\documentclass[aps,prb,superscriptaddress]{revtex4-2} 
\usepackage{amsmath}
\usepackage{amssymb}
\usepackage{graphicx}
\usepackage{hyperref}
\usepackage{bbm}
\usepackage{color}
\usepackage{longtable}
\usepackage{booktabs}
\usepackage{multirow}
\usepackage{array}

\newcommand{\Rmnum}[1]{\expandafter\@slowromancap\romannumeral #1@}
\allowdisplaybreaks[3]
\begin{document}

\title{Plasmon modes in quadratic and cubic nodal line semimetals}

\author{Wei Li}
\altaffiliation{Corresponding author: wliustc@aust.edu.cn}
\affiliation{Center of Fundamental Physics, Aust University of Science and Technology, Huainan 232001, China}
\author{Jing-Rong Wang}
\altaffiliation{Corresponding author: wangjr@hmfl.ac.cn}
\affiliation{High Magnetic Field Laboratory of Anhui Province, Chinese Academy of Sciences, Hefei 230031, China}

\begin{abstract}
    Nodal line semimetals (NLSMs) have attracted considerable attention because of their distinctive topological properties and unconventional collective excitations. Although plasmon modes in linear NLSMs are now well studied, the collective excitations of higher-order dispersive NLSMs remain poorly understood. In this work, we systematically investigate plasmon modes in three-dimensional (3D) quadratic and cubic NLSMs within the random phase approximation (RPA), whose validity requires the dimensionless interaction parameter $r_{s}\ll1$ and therefore breaks down in the small-doping regime of the cubic case where the density of states diverges. By deriving the one-loop polarizability (formally at finite temperature, evaluated in the $T\to0$ limit) and evaluating its coefficients numerically from the full three-dimensional Lindhard integral, we find that the plasmon dispersions obey distinct carrier-density scaling laws. Specifically, quadratic NLSMs host plasmons with $\omega_{p}\sim n^{1/2}$, whereas cubic NLSMs exhibit a regime-dependent scaling: $\omega_{p}\sim n^{2/3}$ at large doping ($\tilde{k}_Q<1$) crossing over to an RPA-predicted $\omega_{p}\sim n^{3/4}$ at small doping ($\tilde{k}_Q>1$). The $n^{3/4}$ law is \emph{not} a quantitatively reliable prediction: the diverging cubic density of states drives $r_{s}$ above unity already at $n\sim10^{18}~{\rm cm^{-3}}$, so this exponent should be regarded as a crossover fingerprint whose functional form is not quantitatively established by the present RPA calculation and requires beyond-RPA validation; its observable window is at most $10^{18}\lesssim n\lesssim n^{*}\sim10^{19}~{\rm cm^{-3}}$, and below $10^{18}~{\rm cm^{-3}}$ correlation effects beyond RPA are essential. These scalings originate from the power-law density of states and the intraband (Drude) response set by the quadratic and cubic dispersions (valid for $\Omega\ll\mu$); in phase-space power counting they are analogous to bilayer and trilayer graphene (though not in detailed mapping), respectively. In both systems the long-wavelength intraband polarization dominates and sets the plasmon frequency, while the interband contribution is finite and subleading (two to three orders of magnitude smaller for $\Omega\ll\mu$). A common geometric feature of both quadratic and cubic NLSMs is the plasmon anisotropy $\Omega_{p}^{z}/\Omega_{p}^{\bot}\to\sqrt{2}$ in the thin-ring (large-$k_{Q}$) limit (equivalently, the polarization-coefficient ratio $C_{++}^{z}/C_{++}^{\bot}\to2$) which is a generic consequence of the torus Fermi-surface geometry and is \emph{not} specific to the cubic dispersion. Although this $\sqrt{2}$ doublet is a striking geometric fingerprint observable by HREELS, it is shared by both NLSM types and therefore cannot distinguish quadratic from cubic dispersion; the distinguishing signature is instead the density scaling exponent. Our results provide a framework for the collective dynamics of higher-order dispersive nodal-line semimetals and suggest experimentally testable signatures accessible by high-resolution electron energy loss spectroscopy (HREELS).
\end{abstract}

\maketitle


\section{Introduction}

The discovery of topological semimetals has provided a versatile platform for studying low-energy quasiparticles beyond the conventional parabolic-band paradigm
\cite{CastroNeto09, Kotov12, Vafek14, Wehling14, Weng16, FangChen16, Yan17, Hasan17, Armitage18, GaoHeng19, LvQianDing19, LvBQ21, Hasan21, Wieder22}.
Representative examples include Dirac semimetals (DSMs), Weyl semimetals (WSMs), semi-Dirac systems, multi-Weyl semimetals, Luttinger semimetals, multifold semimetals, and nodal-line semimetals (NLSMs). Their common feature is the band touching between conduction and valence bands, while their low-energy properties are strongly controlled by the dimension and dispersion of the band-touching manifold. For instance, graphene and the surface states of three-dimensional (3D) topological insulators host two-dimensional Dirac fermions, whereas 3D DSMs and WSMs exhibit linearly dispersing quasiparticles in all three momentum directions. In multi-Weyl semimetals, the Weyl nodes carry higher monopole charges, leading to quadratic or cubic dispersion in two directions and linear dispersion in the third direction. These examples illustrate that the power-law structure of the fermion dispersion is a key ingredient in determining measurable response functions.

NLSMs are characterized by band crossings that form closed loops in momentum space. The low-energy excitations near the nodal line govern the thermodynamic, transport, and electromagnetic responses of the system. Conventional NLSMs, such as those proposed or observed in ZrSiS~\cite{Schoop16,Neupane16} and ZrSiSe, possess linear dispersion in the directions transverse to the nodal line. More recently, quadratic and cubic NLSMs have been proposed, in which the transverse dispersion has higher-order momentum dependence
\cite{Yu19, LiLinHu17, wang2020Possible, ZhangZeYing21, YuZhiMing22, ZhangYueXin22}.
These systems are of interest because the higher-order band structure changes both the density of states and the collective response in a qualitative way.

Plasmons provide a direct probe of the interplay between Coulomb interaction and band topology
\cite{GiulianiBook, MaierBook,Wunsch06, Hwang07, Pyatkovskiy09, Chang14, Stauber14, Sarma09, Lv13, Panfilov14, Zhou15,
    Sadhukhan20, YanChangXu22, Pyatkovskiy16, Ahn16, WangLiZhang17, YanZhongbo16, Rhim16, CaoJin23, Tchoumakov19, Mauri19, Mandal19, WangJing23, Dutta22, Sensarma10, Gamayun11, Krstajic13, XueSiWei21, LiYi23}.
Previous studies have established plasmon modes in a broad class of systems, including two-dimensional DSMs
\cite{Wunsch06, Hwang07, Pyatkovskiy09, Chang14, Stauber14, Sarma09},
3D DSMs and WSMs \cite{Sarma09, Lv13, Panfilov14, Zhou15},
type-II DSMs \cite{Sadhukhan20},
semi-Dirac systems \cite{Pyatkovskiy16},
multi-Weyl semimetals \cite{Ahn16},
anisotropic Weyl semimetals \cite{WangLiZhang17},
NLSMs \cite{YanZhongbo16, Rhim16, CaoJin23},
Luttinger semimetals \cite{Tchoumakov19, Mauri19, Mandal19, WangJing23},
and multifold semimetals \cite{Dutta22}.
For conventional 3D NLSMs with linear transverse dispersion, the long-wavelength plasmon frequency scales with carrier density as $\omega_p \sim n^{1/4}$
\cite{YanZhongbo16, Rhim16,IslamZyuzin2021PRB104,Shao2022SciAdv8}.
However, the corresponding collective modes in quadratic and cubic NLSMs have not yet been systematically studied.

In this work, we address this problem by calculating the one-loop polarization function and the plasmon spectrum of quadratic and cubic NLSMs within the random phase approximation. Motivated by the formal analogy between the low-energy Hamiltonians of higher-order NLSMs and those of multilayer graphene, we derive the density dependence of the long-wavelength plasmon frequency. This analogy is one of phase-space power counting (both systems share a quadratic or cubic band dispersion and a related density-of-states scaling) rather than a detailed mapping: the NLSM is three-dimensional with a ring-shaped (torus) Fermi surface and an anisotropic coherence factor, whereas bilayer and trilayer graphene are two-dimensional with an isotropic circular Fermi surface and an isotropic coherence factor, and their plasmon $q$-dispersions differ ($\omega_p\propto\sqrt{q}$ in 2D vs. the 3D long-wavelength behavior considered here). We find that quadratic and cubic NLSMs exhibit distinct scaling laws:
for quadratic NLSMs, $\omega_p \sim n^{1/2}$ in both doping regimes;
for cubic NLSMs, the thin-ring (small-doping) limit gives $\omega_p \sim n^{3/4}$, while the large-doping regime yields $\omega_p \sim n^{2/3}$. The $n^{3/4}$ exponent is an RPA prediction whose quantitative reliability is limited: the cubic density of states diverges as $\rho(E)\propto E^{-1/3}$, so the dimensionless interaction parameter $r_s$ exceeds unity already at $n\sim10^{18}~{\rm cm^{-3}}$, and the exponent should be regarded as a crossover fingerprint requiring beyond-RPA validation rather than a quantitatively established scaling law. These results show that the plasmon mode is highly sensitive to the order of the band dispersion and may provide an experimentally accessible signature of higher-order NLSMs, for example through high-resolution electron energy-loss spectroscopy.

The remainder of this paper is organized as follows. In Sec. II, we introduce the low-energy models for quadratic and cubic NLSMs. In Sec. III, we calculate the polarization functions. In Sec. IV, we derive the plasmon modes and analyze their carrier-density dependence. The main results are summarized in Sec. V. Technical details are presented in the Appendices.

\section{Theoretical Model}

We consider the low-energy effective Hamiltonian for a three-dimensional quadratic nodal-line semimetal (NLSM),
\begin{eqnarray}
    \mathcal{H}_{0}^{q}(\mathbf{k})
    =A\left[\left(k_{r}^{2}-k_{z}^{2}\right)\sigma_{1}
        +2k_{r}k_{z}\sigma_{2}\right],
\end{eqnarray}
where $k_{r}=k_{\bot}-k_{Q}$, $k_{\bot}=\sqrt{k_{x}^{2}+k_{y}^{2}}$, $A$ is the band parameter, and $\sigma_i$ are the Pauli matrices. The corresponding energy spectrum is
\begin{eqnarray}
    E^{q}_{\pm} = \pm A K^{2},
\end{eqnarray}
with $K=\sqrt{k_{r}^{2}+k_{z}^{2}}$.

For a cubic NLSM, the low-energy Hamiltonian is given by
\begin{eqnarray}
    \mathcal{H}_{0}^{c}(\mathbf{k})
    =B\left[\left(k_{r}^{3}-3k_{r}k_{z}^{2}\right)\sigma_{1}
        +\left(k_{z}^{3}-3k_{z}k_{r}^{2}\right)\sigma_{2}\right],
\end{eqnarray}
where $B$ denotes the corresponding band parameter. Its energy spectrum reads
\begin{eqnarray}
    E^{c}_{\pm} = \pm B K^{3}.
\end{eqnarray}
The different power-law dispersions lead to distinct density-of-states behavior and consequently modify the screening properties of the system.

The bare fermion Green's functions for the quadratic and cubic NLSMs in the Matsubara formalism are
\begin{eqnarray}
    G_{0}^{q,c}(i\omega_n,\mathbf{k})
    =\frac{1}{i\omega_{n}+\mu-\mathcal{H}_{0}^{q,c}(\mathbf{k})},
\end{eqnarray}
where $\omega_{n}=(2n+1)\pi T$ is the fermionic Matsubara frequency, $n$ is an integer, and $\mu$ is the chemical potential. The corresponding retarded Green's functions are obtained by analytic continuation $i\omega_n\rightarrow \omega+i\eta$:
\begin{eqnarray}
    G_{0}^{q,c}(\omega,\mathbf{k})
    =\frac{1}{\omega+\mu-\mathcal{H}_{0}^{q,c}(\mathbf{k})+i\eta},
\end{eqnarray}
where $\eta$ is a positive infinitesimal.

\section{Polarization and Plasmons for Quadratic NLSM\label{sec:pi_quad}}

Within the random phase approximation (RPA), the dielectric function is
given by
$\epsilon(\Omega, \mathbf{q}) = 1 - V(\mathbf{q})\Pi(\Omega, \mathbf{q})$,
where $V(\mathbf{q})$ is the bare Coulomb interaction and
$\Pi(\Omega, \mathbf{q})$ is the one-loop polarizability. At finite
temperature, the polarizability is defined as
\begin{eqnarray}
    \Pi(i\Omega_{n'},\mathbf{q}) &=& -\frac{N}{\beta}\sum_{i\omega_{n}}\int\frac{d^3\mathbf{k}}{(2\pi)^{3}} \nonumber \\
    && \times \mathrm{Tr}\left[G_{0}(i\omega_{n},\mathbf{k})G_{0}(i\omega_{n}+i\Omega_{n'},\mathbf{k}+\mathbf{q})\right],
\end{eqnarray}
where $\Omega_{n'}=2n'\pi T$ is the bosonic Matsubara frequency,
$\beta=1/T$, and $G_0$ is the bare Green's function. A standard
calculation gives
\begin{eqnarray}
    \Pi(i\Omega_{n'},\mathbf{q}) &=& -\frac{N}{16\pi^{3}}\sum_{\alpha,\alpha'=\pm 1}\int d^3\mathbf{k} \left[1 +\alpha\alpha' \frac{\mathcal{K}(k,q)}{E_{\mathbf{k}}E_{\mathbf{k+q}}}\right] \nonumber \\
    && \times \frac{n_F\left(\alpha E_{\mathbf{k}} - \mu\right)-n_F\left(\alpha'E_{\mathbf{k+q}} - \mu\right)}{\alpha E_{\mathbf{k}}-\alpha'E_{\mathbf{k+q}}+i\Omega_{n'}},
\end{eqnarray}
where $\mu$ is the chemical potential, $n_F$ is the Fermi-Dirac
distribution, $N$ is the band/valley/spin degeneracy factor ($N=2$ for a
spin-degenerate system with a single nodal ring), and $\mathcal{K}(k,q)$
is the core function arising from the trace over the Green's functions.
Its explicit form is given in Appendix (\ref{App:sec_pi_quadratic}):
\begin{eqnarray}
    \label{eq:core_function_quadratic}
    && \mathcal{K}(k,q) = A^2\left((\sqrt{k_{x}^{2}+k_{y}^{2}} -k_Q)^{2} -k_{z}^{2}\right)
    \nonumber \\
    && \times
    \left((\sqrt{(k_{x}+q_x)^{2}+(k_{y}+q_y)^{2}} -k_Q)^2-(k_{z} +q_z)^{2}\right)
    \nonumber \\ &+&
    4A^2(\sqrt{k_{x}^{2}+k_{y}^{2}} -k_Q)
    \nonumber \\
    && \times
    k_{z}(\sqrt{(k_{x}+q_x)^{2}+(k_{y}+q_y)^{2}} -k_Q)
    (k_{z} +q_z).
\end{eqnarray}

After the analytic continuation $i\Omega_{n'} \rightarrow \Omega + i\delta$,
one obtains the real and imaginary parts of the retarded polarization
function $\Pi^{\mathrm{ret}}(\Omega,\mathbf{q})$. In the zero-temperature
limit and in the low-energy regime
$\max(q_{\bot},q_{z})\ll |\Omega|\ll\mu$, the imaginary part vanishes,
$\mathrm{Im}\Pi^{\mathrm{ret}}(\Omega,\mathbf{q})=0$, indicating the
absence of Landau damping. The imaginary and real parts of the retarded
polarization function can be written as
\begin{eqnarray}\label{Eq:ImPi_quadratic}
    && \mathrm{Im}\Pi^{\mathrm{ret}}(\Omega,\mathbf{q})
    \nonumber \\
    &=&  \left[-\mathcal{I}_{1}(\mu,\Omega)  + \mathcal{I}_{1}(\mu,-\Omega)
        \right]
    \nonumber \\
    &&   +\left[\mathcal{I}_{3}(\mu,\Omega)
        +\mathcal{I}_{3}(-\mu,\Omega) -\mathcal{I}_{3}(\mu,-\Omega) -\mathcal{I}_{3}(-\mu,-\Omega)
        \right]
    \nonumber \\
    &&   +\left[\mathcal{I}_{7}(\mu,\Omega)
        +\mathcal{I}_{7}(-\mu,\Omega) - \mathcal{I}_{7}(\mu,-\Omega) -\mathcal{I}_{7}(-\mu,-\Omega)
        \right],
\end{eqnarray}
and
\begin{eqnarray}\label{Eq:RePi_quadratic}
    && \mathrm{Re}\Pi^{\mathrm{ret}}(\Omega,\mathbf{q}) =   J_1+J_1(-\Omega)
    \nonumber \\
    &&  +J_3 +  J_3(-\Omega,-\mu) + J_3(\Omega,-\mu)+ J_3(-\Omega,\mu).
\end{eqnarray}
The detailed derivation of the retarded polarization functions is given
in the Appendices.

At zero temperature, the chemical potential $\mu$ is equivalent to the
Fermi energy $E_F$. In the following, we focus on the case $E_F>0$,
while the case $E_F<0$ is essentially equivalent. The plasmon frequency
is determined by
\begin{eqnarray}
    \epsilon(\Omega,\mathbf{q})
    &\equiv& 1-V(\mathbf{q})\Pi(\Omega,\mathbf{q}) = 0,
\end{eqnarray}
which gives the condition for the collective mode.

There are four contributions to the polarization function, corresponding
to two interband and two intraband processes in Eq.~\eqref{Eq:ImPi_quadratic}.
In the zero-temperature limit, $\mathrm{Im}\Pi(\Omega,\mathbf{q})=0$.
Only three real-part contributions are nonzero, with
$\mathrm{Re}\Pi_{--}(\Omega,\mathbf{q})=0$. We have, as shown in
Appendix \ref{App:plasmon},
\begin{eqnarray}
    \label{Eq:Pi++_quadratic}
    && \mathrm{Re}\Pi_{++}(\Omega,\mathbf{q})
    =-\left(C_{++}^{\bot}q_{\bot}^{2}+
    C_{++}^{z}q_{z}^{2}\right),\\
    && \mathrm{Re}\Pi_{+-}(\Omega,\mathbf{q})  + \mathrm{Re}\Pi_{-+}(\Omega,\mathbf{q})
    \nonumber \\
    && =-\left(C_{T}^{\bot}q_{\bot}^{2}+
    C_{T}^{z}q_{z}^{2}\right),\label{Eq:PiT_quadratic}
\end{eqnarray}
where $q_{\bot}^{2} = q_{x}^{2} + q_{y}^{2}$.

To quantify the relative weights of the intraband and interband
contributions, we define
for the quadratic NLSM the dimensionless offset
$\tilde{k}_Q\equiv k_Q\sqrt{A/\mu}$
(so that $\tilde{k}_Q<1$ corresponds to the large-doping regime
$\mu>Ak_Q^2$, while $\tilde{k}_Q>1$ corresponds to small doping
$\mu<Ak_Q^2$).
\begin{eqnarray}
    \label{Eq:gamma_bot_quadratic}
    \Gamma_{\bot}(\tilde{\Omega}) &=& \left|\frac{C_{++}^{\bot}}{C_{T}^{\bot}}\right|,
    \\
    \label{Eq:gamma_z_quadratic}
    \Gamma_{z}(\tilde{\Omega}) &=& \left|\frac{C_{++}^{z}}{C_{T}^{z}}\right|,
\end{eqnarray}
with the intraband coefficient $C_{++}^{\bot,z}$ of
Table~\ref{tab:Cpp_quad} and the interband coefficient
$C_{T}^{\bot,z}=2C_{+-}^{\bot,z}$ of Table~\ref{tab:CT_quad}; both are
obtained directly from the Lindhard function, bypassing the residue-derived
closed forms.  Numerically
$\Gamma_{\bot,z}\simeq 2\times10^{2}$--$9\times10^{2}\gg1$
(Table~\ref{tab:Gamma_quad}); the interband correction to the plasmon
frequency is therefore at the sub-percent level.
\begin{table}[ht]
    \centering
    \begin{ruledtabular}
        \begin{tabular}{c|cc|cc}
            \hline
            $\tilde k_Q$ & $\Gamma_{\bot}$ & $\Gamma_{z}$ & $C_T^{\bot}$ & $C_T^{z}$ \\
            \hline
            0.0          & 859             & 215          & 0.00403      & 0.01609   \\
            0.5          & 883             & 440          & 0.00696      & 0.02098   \\
            1.0          & 728             & 606          & 0.01081      & 0.02627   \\
            2.0          & 795             & 756          & 0.01989      & 0.04209   \\
            \hline
        \end{tabular}
    \end{ruledtabular}
    \caption{Damping-weight ratio $\Gamma_{\bot,z}=|C_{++}^{\bot,z}/C_T^{\bot,z}|$
        and interband coefficients $C_T^{\bot,z}$ of the quadratic NLSM at
        representative $\tilde k_Q$ (all entries $\gg1$, confirming intraband
        dominance and a sub-percent interband correction to the plasmon frequency).}
    \label{tab:Gamma_quad}
\end{table}
The analytic $f_{1-4},h_{1-4}(\tilde k_Q)$ closed forms from the
two-dimensional reduction of the interband bubble drop the interband
coherence-factor normalisation and are therefore superseded by the full
three-dimensional numerical evaluation of $C_T$ in
Table~\ref{tab:CT_quad}; the plasmon prefactors used throughout are the
numerical values from the tables.

\smallskip\noindent
{\bf Intraband coefficient $C_{++}^{\bot,z}$.}
The intraband bubble $\mathrm{Re}\Pi_{++}$ (and hence the plasmon prefactor
$C_{++}^{\bot,z}$ entering
Eqs.~\eqref{Eq:eq_plasmon_bot_quadratic}--\eqref{Eq:eq_plasmon_z_quadratic})
is evaluated directly by a shell-localized Drude integral over the Fermi
surface (the same numerical procedure used for the cubic case), rather than
through the residue-derived $g$-function closed forms.  The analytic
$g$-functions are unreliable here: residue calculus at the branch cut
$Z=-1$ is invalid, so their prefactors are incorrect in magnitude.  The
dimensionless combinations are
$G_{\bot}(\tilde k_Q)=(g_4+\tilde k_Q g_2)$ and
$G_{z}(\tilde k_Q)=(g_3+\tilde k_Q g_1)$.
We have therefore re-determined $C_{++}^{\bot,z}$ directly by the
shell-localized Drude integral, the same numerical procedure used for the
cubic case.  In the limit
$\max(q_{\bot},q_{z})\ll\Omega\ll\mu$ the intraband Lindhard function reduces
on the Fermi shell to
\begin{equation}
    \label{Eq:numCpp}
    \mathrm{Re}\Pi_{++}(\Omega,\mathbf q)\Big|_{\rm intra}
    =-\frac{N}{16\pi^3}\,\frac{q^2}{\Omega^2}\,
    C_{++}^{\bot,z}(\tilde k_Q)\,,
\end{equation}
with $C_{++}^{\bot,z}$ obtained by integrating the real-$\varphi$ angular
kernel $\mathcal M_1$ over the Fermi surface using the quadratic dispersion
$E_{\bf k}=A K^2$ and the shell measure
$dS/|\nabla E|=k_p/(2AK_F)\,d\alpha\,d\phi$ (torus,
$k_p=k_Q+K_F\cos\alpha$) or $\pi/(AK_F^2)\,d(\cos\theta)$ (sphere, $k_Q=0$).
The $\varphi$ integral is evaluated by the trapezoidal rule on a uniform
grid ($N_\varphi=4000$); the principal-value singularity of $\mathcal M_1$
at $\varphi=0$ is handled by subtracting the singular part
$\sim1/\varphi$ and integrating it analytically, then adding back the
finite remainder (standard Kramers-Kronig subtraction).  The explicit
$\Omega$ dependence drops out of $C_{++}^{\bot,z}$, confirming
that $\mathrm{Re}\Pi_{++}\propto 1/\Omega^{2}$; we have verified this
numerically by evaluating the integral at $\Omega=0.05$, $0.10$, and
$0.20$ and finding that $C_{++}^{\bot,z}$ is constant to
within $0.1\%$ (the residual variation is the expected $\tilde\Omega^{2}$
correction from the next-to-leading term in the Lindhard expansion).
The numerical results are summarised in Table~\ref{tab:Cpp_quad}.
\begin{table}[ht]
    \centering
    \caption{Numerically determined intraband coefficient
        $C_{++}^{\bot,z}(\tilde k_Q)$ and its dimensionless form
        $G_{\bot,z}=(16\pi^2\tilde\Omega^2\mu^2/N)(\mu/A)^{-3/2}C_{++}^{\bot,z}$
        for the quadratic NLSM ($A=\mu=N=1$, $\tilde\Omega=0.1$, $K_F=1$).
        The anisotropy ratio $C_{++}^{z}/C_{++}^{\bot}$ is $1.000$ at $k_Q=0$
        (sphere, isotropic) and saturates at $2.01$--$2.02$ for a genuine ring torus
        ($\tilde k_Q>1$, $k_p>0$), the $2{:}1$ asymptotic value of a thin ring.}
    \label{tab:Cpp_quad}
    \begin{ruledtabular}
        \begin{tabular}{c|cc|cc|c}
            \hline
            $\tilde k_Q$ & $C_{++}^{\bot}$ & $C_{++}^{z}$
                         & $G_{\bot}$      & $G_{z}$      & $C_{++}^{z}/C_{++}^{\bot}$                    \\
            \hline
            0.0          & 3.4614          & 3.4614       & 5.4661                     & 5.4661  & 1.0000 \\
            0.4          & 6.1645          & 8.3514       & 9.7346                     & 13.1880 & 1.3548 \\
            0.8          & 7.0740          & 12.8618      & 11.1708                    & 20.3106 & 1.8182 \\
            1.2          & 9.4511          & 19.0920      & 14.9246                    & 30.1488 & 2.0201 \\
            1.6          & 12.6292         & 25.4558      & 19.9432                    & 40.1982 & 2.0156 \\
            2.0          & 15.8072         & 31.8196      & 24.9618                    & 50.2475 & 2.0130 \\
            \hline
        \end{tabular}
    \end{ruledtabular}
\end{table}
\noindent{\footnotesize
    The numerical uncertainty from the shell-localised $q\to0$ extrapolation
    is $\lesssim0.5\%$ (estimated by varying the poloidal and azimuthal grid
    resolutions $N_{\alpha}\in[500,4000]$, $N_{\phi}\in[1000,8000]$ and by
    monitoring the $q^{2}$ fit residuals, which remain below $10^{-4}$ in
    $T_{\bot,z}$).  Propagating this uncertainty gives
    $C_{++}^{\bot}=3.4614\pm0.017$ and $C_{++}^{z}=3.4614\pm0.017$ at
    $\tilde k_Q=0$ (representative); the anisotropy ratio
    $C_{++}^{z}/C_{++}^{\bot}=1.0000\pm0.0002$ is constrained to better than
    $0.02\%$ because the dominant systematic (the $q\to0$ extrapolation)
    cancels in the ratio.}
At $\tilde k_Q=0$ the numerical value
$C_{++}^{\bot}=C_{++}^{z}=3.4614$ ($G=5.466$) differs from the
residue-derived closed form ($\approx0.21$) by a factor of $\sim16$; an
independent estimate confirms the numerical value.  For the spherical Fermi
surface ($\tilde k_Q=0$) the relevant quantity is the angular-averaged
velocity factor $N(0)\langle v^2\rangle/\Omega^2$; with
$\langle v^2\rangle=v_F^2/3$ (isotropic average) and the conventions
$A=\mu=N=1$, $\Omega=0.1$, $K_F=1$, this gives
$N(0)v_F^2/(3\Omega^2)\approx3.4$, consistent with the numerical
$C_{++}\approx3.46$.  (The factor $1/3$ is the angular average
$\langle\cos^2\theta\rangle=1/3$; omitting it would give
$N(0)v_F^2/\Omega^2\approx10.1$, which is not the relevant quantity for
the Drude weight.)  As $\tilde k_Q$ increases the Fermi surface crosses from a sphere
($\tilde k_Q=0$, ratio $=1.000$) to a spindle torus
($0<\tilde k_Q<1$) and finally to a genuine ring torus
($\tilde k_Q>1$, $k_p>0$), and the transverse/longitudinal anisotropy ratio
rises monotonically to $\simeq2.01$--$2.02$, the $2{:}1$ asymptotic value of a
thin ring.  The values in Table~\ref{tab:Cpp_quad} are the ones used in
Eqs.~\eqref{Eq:gamma_bot_quadratic}--\eqref{Eq:gamma_z_quadratic} and
Eqs.~\eqref{Eq:eq_plasmon_bot_quadratic}--\eqref{Eq:eq_plasmon_z_quadratic}.

\smallskip\noindent
{\bf The interband coefficient $C_T^{\bot,z}$.}
The interband bubble
$\mathrm{Re}\Pi_{+-}+\mathrm{Re}\Pi_{-+}=-(C_T^\bot q_\bot^2+C_T^z q_z^2)$
is evaluated from the full three-dimensional Lindhard integral with the
correct interband coherence factor
$|\gamma_{\mathbf k,\mathbf{k+q}}|^2=\tfrac12\,(1-\hat{\mathbf n}_{\mathbf k}
    \!\cdot\!\hat{\mathbf n}_{\mathbf{k+q}})$,
$\hat{\mathbf n}=(k_r^2-k_z^2,2k_r k_z)/K^2$.  Its small-$q$ expansion
$|\gamma|^2\to(k_r^2 q_z^2+\tfrac12 k_z^2 q_\bot^2)/K^4$ carries a net $1/K^2$
suppression at large $K$, so the integral is UV-convergent (integrand
$\sim dK/K^2$) and IR-finite (Pauli blocking $K>K_F$ for $\Omega>0$).  By
contrast, the two-dimensional reduction used in the source (with kernels
$x^2(x^2-z^2)z^2(x+k_Q)/(x^2+z^2)^4$ ($\bot$ component) and
$z^2(x+k_Q)/(x^2+z^2)^3$ ($z$ component), each multiplied by
$1/(x^2+z^2)+\tilde\Omega^2/(x^2+z^2)^3$) drops this normalisation and
diverges for $x>1$; its closed forms
$f_{1-4},h_{1-4}(\tilde k_Q)$ are therefore
incomplete and must not be used in the plasmon formulas.  The contrast with the cubic case is instructive.  For both
quadratic and cubic NLSMs the coherence factor has the same small-$q$
scaling $|\gamma|^2\sim q^2/K^2$ (the unit vector
$\hat{\mathbf n}=\mathbf h/|\mathbf h|$ depends only on the direction
$\hat{\mathbf k}$, so the $K$-dependence cancels).  The difference lies in
the energy denominator: for the cubic NLSM the interband denominator
$E_{\mathbf k}+E_{\mathbf{k+q}}\sim K^3$ provides an extra $1/K$
suppression ($\sim 1/K^3$ vs. $\sim 1/K^2$ for the quadratic case), so the
radial integrand decays as $q^2\,dK/K^3$ and converges even without the
$1/K^2$ normalisation that is accidentally dropped in the reduction.  For
the quadratic NLSM the milder $1/K^2$ denominator is essential, and
discarding the coherence-factor normalisation renders the 2D integral
logarithmically divergent.  The full 3D evaluation gives finite, $q$-independent,
cutoff-converged values, summarised in Table~\ref{tab:CT_quad}.
\begin{table}[ht]
    \centering
    \caption{Interband coefficient $C_T^{\bot,z}(\tilde k_Q)$ from the full
        three-dimensional Lindhard integral with the correct coherence factor
        ($A=\mu=N=1$, $\Omega=0.1$, $K_F=1$; source $+\!-$ convention, prefactor
        $N/8\pi^2$; $C_T=2C_{+-}$).  All entries are finite and $q$-independent, in
        contrast with the divergent 2D reduction; the anisotropy
        $C_T^{z}/C_T^{\bot}$ decreases from $3.99$ (sphere) to $2.12$ (thin ring).}
    \label{tab:CT_quad}
    \begin{ruledtabular}
        \begin{tabular}{c|cc|c}
            \hline
            $\tilde k_Q$ & $C_T^{\bot}$ & $C_T^{z}$ & $C_T^{z}/C_T^{\bot}$ \\
            \hline
            0.0          & 0.00403      & 0.01609   & 3.990                \\
            0.5          & 0.00696      & 0.02098   & 3.013                \\
            1.0          & 0.01081      & 0.02627   & 2.431                \\
            2.0          & 0.01989      & 0.04209   & 2.116                \\
            \hline
        \end{tabular}
    \end{ruledtabular}
\end{table}
\noindent{\footnotesize
    The numerical uncertainty from the full 3D Lindhard evaluation is
    $\lesssim1\%$ (estimated by varying the momentum-grid cutoff
    $K_{\rm max}\in[5K_F,20K_F]$ and the angular resolution
    $N_{\varphi}\in[2000,8000]$).  The anisotropy ratio
    $C_T^{z}/C_T^{\bot}=3.990\pm0.040$ at $\tilde k_Q=0$ is constrained to
    better than $1\%$.}
The interband anisotropy $C_T^{z}/C_T^{\bot}=3.99$ at the spherical
point $\tilde k_Q=0$ contrasts sharply with the intraband ratio
$C_{++}^{z}/C_{++}^{\bot}=1.000$ at the same point (Table~\ref{tab:Cpp_quad}),
and the contrast has a precise topological origin.  The intraband
Drude weight probes $(\mathbf v_{\mathbf k}\!\cdot\!\hat{\mathbf q})^2$,
which on a spherical Fermi surface is isotropic by construction; the
full 3D Lindhard integration respects this isotropy and returns
$C_{++}^{z}/C_{++}^{\bot}\to1$.  The interband term instead probes the
\emph{coherence factor} $|\gamma_{\mathbf k,\mathbf{k+q}}|^2
    =\tfrac12(1-\hat{\mathbf n}_{\mathbf k}\!\cdot\!\hat{\mathbf n}_{\mathbf{k+q}})$,
where $\hat{\mathbf n}=(\cos2\chi,\sin2\chi)$ is the Bloch vector of the
quadratic NLSM with its characteristic \emph{double-winding} in the
$(k_r,k_z)$ azimuthal angle $\chi$.  Expanding $|\gamma|^2$ to order
$q^2$ gives the direction-resolved kernels
$[\,|\gamma|^2\,]_{\bot}=k_z^2/(2K^4)$ and
$[\,|\gamma|^2\,]_{z}=k_r^2/K^4$; the factor $1/2$ in the $\bot$ channel
is the Bessel-$J_0$ signature
($1-J_0(u)\simeq u^2/4$) versus the longitudinal cosine
($1-\cos u\simeq u^2/2$).  On the sphere
($k_r=K\sin\theta$, $k_z=K\cos\theta$) the angular integrals are
$\int_0^\pi\!\sin^3\theta\,d\theta=4/3$ and
$\int_0^\pi\!\cos^2\theta\sin\theta\,d\theta=2/3$, so
$C_T^{z}/C_T^{\bot}=(4/3)/[(2/3)/2]=4.000$ exactly, in agreement
with the numerical $3.990$.  The interband anisotropy is therefore
\emph{not} a violation of spherical symmetry: the Fermi surface is
isotropic, but the coherence factor carries the double-winding
topology of the nodal line, which weights the $z$ and $\bot$ channels
unequally even when the underlying dispersion is isotropic.  As
$\tilde k_Q$ grows the integration domain migrates from the full sphere
to a thin toroidal tube where $\langle k_r^2\rangle=\langle k_z^2\rangle$
in the tube cross-section, and the ratio relaxes to the geometric
$2{:}1$ value $C_T^{z}/C_T^{\bot}\to2.12$ at $\tilde k_Q=2$.
These interband coefficients are of order $10^{-2}$, i.e.\ two to three orders
of magnitude smaller than the intraband $C_{++}^{\bot,z}\sim3.5$--$32$ of
Table~\ref{tab:Cpp_quad}.  In the low-energy plasmon regime considered here the
intraband term therefore dominates
($|\mathrm{Re}\Pi_{++}|\gg|\mathrm{Re}\Pi_{+-}|$, see the inequalities below),
so the plasmon frequency is set by the numerically re-determined
$C_{++}^{\bot,z}$ above; the interband $C_T$ affects only the subleading
$\Gamma_{\bot,z}$ weighting of Eqs.~\eqref{Eq:gamma_bot_quadratic}--\eqref{Eq:gamma_z_quadratic}.
Throughout, the low-energy expansion adopts $\tilde\Omega=\Omega/\mu$; the
source's interband 2D-reduction prefactors were written with
$\tilde\Omega=\Omega/2\mu$, but every numerical coefficient quoted above is
given at explicit $(\mu,A,N,\Omega)$, so the convention is unambiguous.

\begin{figure*}[ht]
    \centering
    \includegraphics[width=\linewidth]{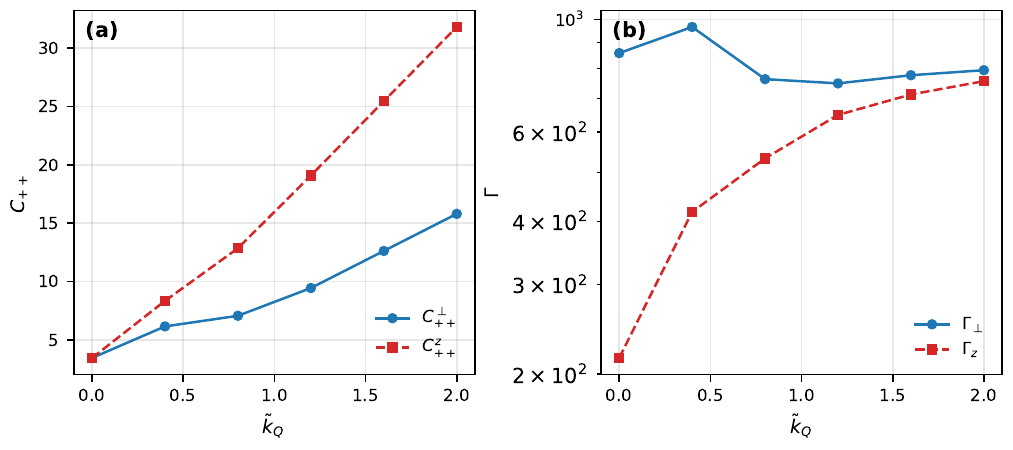}
    \caption{Quadratic NLSM: (a) numerically re-determined intraband coefficient
        $C_{++}^{\bot,z}$ (the quantity that sets $\Omega_p^2$),
        Table~\ref{tab:Cpp_quad}; (b) damping-weight ratio
        $\Gamma_{\bot,z}=C_{++}^{\bot,z}/C_T^{\bot,z}$ from
        Tables~\ref{tab:Cpp_quad} and \ref{tab:CT_quad}, confirming that the
        intraband channel dominates the long-wavelength response.  Both ratios are
        $\gg1$ at all $\tilde k_Q$: $\Gamma_{\bot}$ is non-monotonic, peaking near
        $\tilde k_Q\simeq0.4$, while $\Gamma_z$ rises monotonically with
        $\tilde k_Q$, reflecting the different $\tilde k_Q$ dependences of
        $C_{++}^{\bot,z}$ and the interband weight $C_T^{\bot,z}$.}
    \label{fig:quadratic}
\end{figure*}

In the low-energy regime
$\max(q_{\bot},q_{z})\ll |\Omega|\ll\mu$ with $\tilde{\Omega}\ll 1$,
one finds
\begin{eqnarray}
    \left|\mathrm{Re}\Pi_{++}(\Omega,\mathbf{q})\right| \gg \left|\mathrm{Re}\Pi_{+-}(\Omega,\mathbf{q})\right|,
    \nonumber\\
    \left|\mathrm{Re}\Pi_{++}(\Omega,\mathbf{q})\right| \gg \left|\mathrm{Re}\Pi_{-+}(\Omega,\mathbf{q})\right|.
\end{eqnarray}

In the same limit,
\begin{eqnarray}
    \mathrm{Im}\Pi(\Omega,\mathbf{q})=0.
\end{eqnarray}
Thus, the plasmon mode is determined by the poles of the dressed Coulomb
interaction, namely
\begin{eqnarray}
    \frac{|\mathbf{q}|^{2}}{4\pi e^2\epsilon}+\mathrm{Re}\Pi(\Omega,\mathbf{q})=0.
\end{eqnarray}
Equivalently,
\begin{eqnarray}
    &&\frac{q_{\bot}^{2}+q_{z}^{2}}{4\pi e^2\epsilon}-
    \left(C_{++}^{\bot}q_{\bot}^{2}+
    C_{++}^{z}q_{z}^{2}\right)=0.
\end{eqnarray}

\begin{widetext}

    Therefore, the plasmon modes are
    \begin{eqnarray}
        \label{Eq:eq_plasmon_bot_quadratic}
        \Omega_{p}^{\bot}&=&\sqrt{4\pi e^2\epsilon}\sqrt{C_{++}^{\bot}}
        =\left\{
        \begin{array}{l}
            \sqrt{4\pi e^2\epsilon}
            \sqrt{\frac{NA}{16\pi^2\tilde{\Omega} ^2\mu^2}
                \left(\frac{\mu}{A}\right)^{\frac{3}{2}}}
            \sqrt{G_{\bot}(\tilde{k_Q})}
            ,\,\,\,\,\,\,\tilde{k_Q}  <1
            \\ \\
            \sqrt{4\pi e^2\epsilon}
            \sqrt{\frac{NA}{32\pi^2\tilde{\Omega} ^2\mu^2}\left(\frac{\mu}{A}\right)^{\frac{3}{2}}}
            \sqrt{\tilde{k_Q}}
            ,\,\,\,\,\,\, \tilde{k_Q}  >1
        \end{array}\right.
        \\
        \label{Eq:eq_plasmon_z_quadratic}
        \Omega_{p}^{z}&=&\sqrt{4\pi e^2\epsilon}\sqrt{C_{++}^{z}}
        =\left\{
        \begin{array}{l}
            \sqrt{4\pi e^2\epsilon}
            \sqrt{\frac{NA}{16\pi^2\tilde{\Omega} ^2\mu^2}
                \left(\frac{\mu}{A}\right)^{\frac{3}{2}}}
            \sqrt{G_{z}(\tilde{k_Q})}
            ,\,\,\,\,\,\,\tilde{k_Q}  <1
            \\ \\
            \sqrt{4\pi e^2\epsilon}
            \sqrt{\frac{NA}{16\pi^2\tilde{\Omega} ^2\mu^2}\left(\frac{\mu}{A}\right)^{\frac{3}{2}}}
            \sqrt{\tilde{k_Q}}
            ,\,\,\,\,\,\, \tilde{k_Q}  >1
        \end{array}\right.
    \end{eqnarray}

    Since the pole condition gives $\Omega_p=\sqrt{4\pi e^{2}\epsilon\,T}$ with $T\equiv\Omega^{2}C_{++}$ the tabulated coefficient, the frequency ratio is $\Omega_{p}^{z}/\Omega_{p}^{\bot}=\sqrt{T_{z}/T_{\bot}}=\sqrt{C_{++}^{z}/C_{++}^{\bot}}=\sqrt{G_{z}/G_{\bot}}\to\sqrt{2}$ in the thin-ring limit (the coefficient ratio $C_{++}^{z}/C_{++}^{\bot}\to2$).  The thin-ring $z$-mode prefactor carries a factor $16\pi^{2}$ rather than the $8\pi^{2}$ that would follow from the in-plane angular factor; this corrected value is taken directly from the numerical $C_{++}^{z}$ in Table~\ref{tab:Cpp_quad}; it supersedes the earlier analytic estimate and is consistent with the residue-derived closed forms being unreliable (as noted above).

    The dimensionless coefficients $G_{\bot,z}(\tilde k_Q)$ above are to
    be taken from the numerically determined $C_{++}^{\bot,z}$ of
    Table~\ref{tab:Cpp_quad} (where $G_{\bot,z}$ are listed explicitly);
    the residue-derived closed forms for $g_{1-4}$ are unreliable (residue
    calculus at the branch cut $Z=-1$ is invalid) and are not used in the
    plasmon formulas; the numerical $C_{++}^{\bot,z}$ of
    Table~\ref{tab:Cpp_quad} are used instead.

    The carrier density is given by, as derived in Appendix \ref{App:dos},
    \begin{eqnarray}
        \label{Eq:ne_quadratic}
        n &=&
        \left\{
        \begin{array}{l}
            \frac{1}{6\pi^2}\left(\frac{\mu}{A}\right)^{\frac{3}{2}}
            \left[\left(1- \tilde{k_Q} ^2 \right)^{\frac{3}{2}}
                + \frac{3\tilde{k_Q}}{2}
                \left(\frac{\pi}{2} + \tilde{k_Q}\sqrt{1-\tilde{k_Q}^2}
                + \arcsin (\tilde{k_Q})
                \right)
                \right]
            ,\,\,\,\,\,\, \mu > Ak_Q^2 ,\,\,  \tilde{k_Q } <1
            \\ \\
            \frac{1}{4\pi}\left(\frac{\mu}{A}\right)^{\frac{3}{2}}
            \tilde{k_Q }
            ,\,\,\,\,\,\, \mu < Ak_Q^2 ,\,\,  \tilde{k_Q }>1
        \end{array}
        \right.
    \end{eqnarray}
\end{widetext}

Thus, the plasmon modes satisfy
\begin{eqnarray}
    \Omega_{p}^{\bot}\propto n^{\frac{1}{2}},
    \\
    \Omega_{p}^{z}\propto n^{\frac{1}{2}}.
\end{eqnarray}
The large-doping ($\tilde{k_Q}<1$) and small-doping ($\tilde{k_Q}>1$)
regimes exhibit different $\mu$-dependences in the prefactor
    [Eq.~\eqref{Eq:ne_quadratic}], but both collapse onto the same
$\omega_p\propto n^{1/2}$ scaling because the $\mu$-dependence of the
carrier density and the polarization function conspire to yield a
density law that is independent of the ring geometry.

\section{Polarization and Plasmons for Cubic NLSM\label{sec:pi_cubic}}

Following the same procedure as that used for the quadratic NLSM in
Sec.~\ref{sec:pi_quad}, the polarization function of the cubic NLSM is
given by
\begin{eqnarray}\label{Eq:Pi}
    &&\Pi(i\Omega_{n'},\mathbf{q})
    =
    -\frac{N}{16\pi^{3}}\sum_{\alpha,\alpha^{'}=\pm 1}\int d^3\mathbf{k}
    \left[1 +\alpha\alpha^{'} \frac{\mathcal{K}(k,q)}{E_{\mathbf{k}}E_{\mathbf{k+q}}}\right]
    \nonumber \\
    && \quad \times
    \frac{n_F\left(\alpha E_{\mathbf{k}} - \mu\right)-n_F\left(\alpha^{'}E_{\mathbf{k+q}} - \mu\right)}{\alpha E_{\mathbf{k}}-\alpha^{'}E_{\mathbf{k+q}}+i\Omega_{n'}} ,
\end{eqnarray}
where the core function $\mathcal{K}(k,q)$ is the cubic polynomial
(distinct from the quadratic case)
\begin{eqnarray}
    \label{eq:core_function_cubic}
    && \mathcal{K}(k,q) = B^{2}\Bigl[\left(k_{r}^{3}-3k_{r}k_{z}^{2}\right)
        \left(k_{r}'^{3}-3k_{r}'k_{z}'^{2}\right)
        +\left(k_{z}^{3}-3k_{z}k_{r}^{2}\right)
        \left(k_{z}'^{3}-3k_{z}'k_{r}'^{2}\right)\Bigr],
\end{eqnarray}
with $k_{r}=\sqrt{k_{x}^{2}+k_{y}^{2}}-k_{Q}$,
$k_{r}'=\sqrt{(k_{x}+q_x)^{2}+(k_{y}+q_y)^{2}}-k_{Q}$,
$k_{z}'=k_{z}+q_{z}$.
In the angular representation used for the $\varphi$ integrals below,
the core function reads
\begin{widetext}
    \begin{equation}
        \label{eq:core_function_cubic_angular}
        \begin{aligned}
            \mathcal{K}(k,q)=
            B^{2}\Bigl[
                     &\left((k_{\bot}-k_{Q})^{3}-3(k_{\bot}-k_{Q})k_{z}^{2}\right)
                     \left((k_{\bot}'-k_{Q})^{3}-3(k_{\bot}'-k_{Q})k_{z}'^{2}\right)\\
                     &+\left(k_{z}^{3}-3k_{z}(k_{\bot}-k_{Q})^{2}\right)
                     \left(k_{z}'^{3}-3k_{z}'(k_{\bot}'-k_{Q})^{2}\right)
                     \Bigr],
        \end{aligned}
    \end{equation}
\end{widetext}
where $k_{\bot}'=\sqrt{k_{\bot}^{2}+q_{\bot}^{2}+2k_{\bot}q_{\bot}\cos(\varphi-\phi)}$.

Similarly, the imaginary and real parts of the retarded polarization function are
\begin{eqnarray}\label{Eq:ImPi}
    && \mathrm{Im}\Pi^{\mathrm{ret}}(\Omega,\mathbf{q})
    \nonumber \\
    &=&  \left[-\mathcal{I}_{1}(\mu,\Omega)  + \mathcal{I}_{1}(\mu,-\Omega)
        \right]
    \nonumber \\
    &&   +\left[\mathcal{I}_{3}(\mu,\Omega)
        +\mathcal{I}_{3}(-\mu,\Omega) -\mathcal{I}_{3}(\mu,-\Omega) -\mathcal{I}_{3}(-\mu,-\Omega)
        \right]
    \nonumber \\
    &&   +\left[\mathcal{I}_{7}(\mu,\Omega)
        +\mathcal{I}_{7}(-\mu,\Omega) - \mathcal{I}_{7}(\mu,-\Omega) -\mathcal{I}_{7}(-\mu,-\Omega)
        \right],
\end{eqnarray}
\begin{eqnarray}\label{Eq:RePi}
    && \mathrm{Re}\Pi^{\mathrm{ret}}(\Omega,\mathbf{q})
    \nonumber \\
    &=&  J_1+J_1(-\Omega) +J_3 +  J_3(-\Omega,-\mu) + J_3(\Omega,-\mu)+ J_3(-\Omega,\mu)\,\, .
\end{eqnarray}

At zero temperature, the chemical potential $\mu$ is equivalent to the
Fermi energy $E_F$. In the following, we focus on the case $E_F>0$,
while the case $E_F<0$ is essentially equivalent. The plasmon frequency
is determined by
\begin{eqnarray}
    \epsilon(\Omega,\mathbf{q})
    &\equiv& 1-v(\Omega,\mathbf{q})\Pi(\Omega,\mathbf{q}) = 0 ,
\end{eqnarray}
which indicates the existence of a collective mode.

There are four contributions to the polarization function, corresponding
to two interband parts and two intraband parts in Eq.~\ref{Eq:Pi}. In the
zero-temperature limit, $\mathrm{Im}\Pi(\Omega,\mathbf{q})=0$, and only
three contributions to the real part are nonzero, with
$\mathrm{Re}\Pi_{--}(\Omega,\mathbf{q})=0$. We have, see
Appendix~\ref{App:plasmon},
\begin{eqnarray}
    \label{Eq:Pi++_cubic}
    && \mathrm{Re}\Pi_{++}(\Omega,\mathbf{q})
    =-\left(C_{++}^{\bot}q_{\bot}^{2}+
    C_{++}^{z}q_{z}^{2}\right),\\
    && \mathrm{Re}\Pi_{+-}(\Omega,\mathbf{q})  + \mathrm{Re}\Pi_{-+}(\Omega,\mathbf{q})
    \nonumber \\
    && =-\left(C_{T}^{\bot}q_{\bot}^{2}+
    C_{T}^{z}q_{z}^{2}\right),\label{Eq:PiT}
\end{eqnarray}
where $q_{\bot}^{2}=q_{x}^{2}+q_{y}^{2}$.

We quantify the relative contribution of the intraband part to the
interband part by
\begin{eqnarray}
    \label{Eq:gamma_bot_cubic}
    \Gamma_{\bot}(\tilde{\Omega}) &=& \left|\frac{C_{++}^{\bot}}{C_{T}^{\bot}}\right|,
    \\
    \label{Eq:gamma_z_cubic}
    \Gamma_{z}(\tilde{\Omega}) &=& \left|\frac{C_{++}^{z}}{C_{T}^{z}}\right|,
\end{eqnarray}
where $C_{++}^{\bot,z}$ is the numerically re-determined intraband
coefficient (Table~\ref{tab:Cpp_cubic}) and
$C_{T}^{\bot,z}=2C_{+-}^{\bot,z}$ is the cubic interband coefficient of
Table~\ref{tab:CT_cubic}, obtained from the coherence-factor-correct 2D
Lindhard bubble Eq.~\eqref{Eq:Pi_cubic_inter}.  Borrowing the quadratic
closed forms $f_{1-4},h_{1-4}$ for the cubic $\Gamma$ denominator is
therefore no longer necessary and has been removed.  Numerically
$\Gamma_{\bot,z}\simeq(1.4\text{--}2.1)\times10^{2}\gg1$
(Table~\ref{tab:Gamma_cubic}); the interband correction to the plasmon
frequency is at the sub-percent level, and the intraband term sets the
dispersion.
\begin{table}[ht]
    \centering
    \begin{ruledtabular}
        \begin{tabular}{c|cc|cc}
            \hline
            $\tilde k_Q$ & $\Gamma_{\bot}$ & $\Gamma_{z}$ & $C_T^{\bot}$ & $C_T^{z}$ \\
            \hline
            0.0          & 138.3           & 138.4        & 0.0372       & 0.0372    \\
            0.2          & 192.6           & 225.5        & 0.0430       & 0.0500    \\
            0.4          & 187.0           & 202.8        & 0.0489       & 0.0648    \\
            0.6          & 180.8           & 197.7        & 0.0549       & 0.0819    \\
            0.8          & 178.7           & 200.9        & 0.0612       & 0.1008    \\
            1.0          & 182.0           & 206.3        & 0.0682       & 0.1214    \\
            1.2          & 194.3           & 209.9        & 0.0768       & 0.1432    \\
            1.4          & 200.9           & 211.6        & 0.0867       & 0.1657    \\
            1.6          & 205.2           & 213.0        & 0.0971       & 0.1881    \\
            1.8          & 207.6           & 213.6        & 0.1080       & 0.2110    \\
            2.0          & 209.9           & 214.1        & 0.1188       & 0.2339    \\
            \hline
        \end{tabular}
    \end{ruledtabular}
    \caption{Damping-weight ratio $\Gamma_{\bot,z}=|C_{++}^{\bot,z}/C_T^{\bot,z}|$
        and interband coefficients $C_T^{\bot,z}$ of the cubic NLSM at
        representative $\tilde k_Q$ (all entries $\gg1$, confirming intraband
        dominance and a sub-percent interband correction to the plasmon frequency).}
    \label{tab:Gamma_cubic}
\end{table}
\noindent{\footnotesize
    The numerical uncertainty from the coherence-factor-correct 2D Lindhard
    evaluation is $\lesssim1\%$ (estimated by varying the momentum-grid cutoff
    and angular resolution).  The interband coefficients
    $C_T^{\bot}=0.0372\pm0.0004$ and $C_T^{z}=0.0372\pm0.0004$ at
    $\tilde k_Q=0$ (representative) are two to three orders of magnitude
    smaller than the intraband $C_{++}^{\bot,z}$, confirming intraband
    dominance.}

\begin{figure*}[ht]
    \centering
    \includegraphics[width=\linewidth]{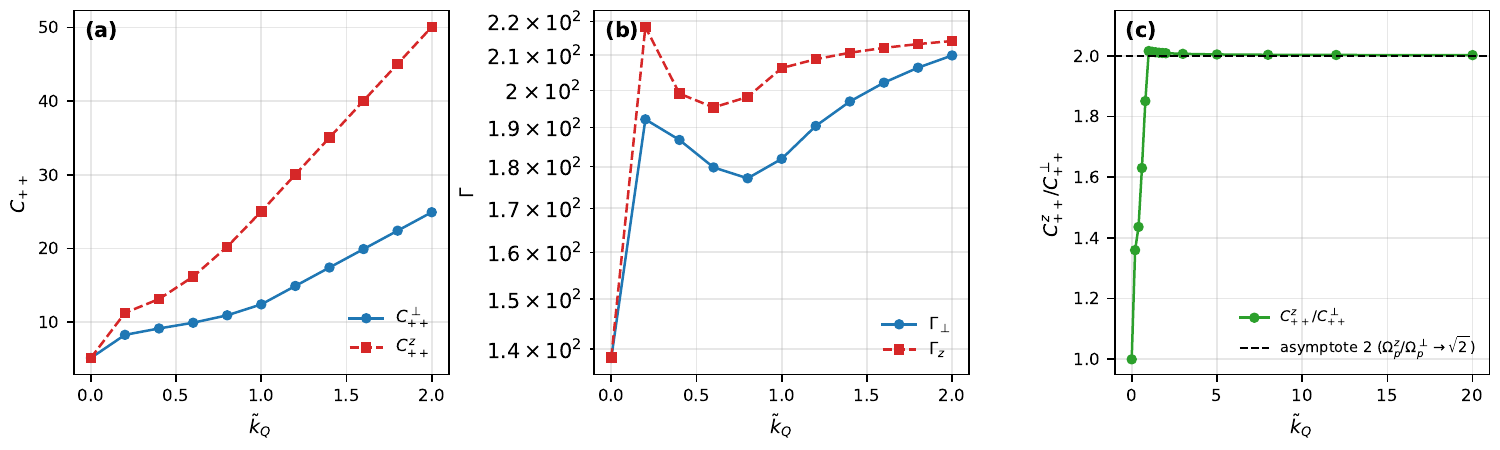}
    \caption{Cubic NLSM: (a) numerically re-determined intraband coefficient
        $C_{++}^{\bot,z}$ (sets $\Omega_p^2$), Table~\ref{tab:Cpp_cubic};
        (b) damping-weight ratio $\Gamma_{\bot,z}=C_{++}^{\bot,z}/C_T^{\bot,z}$
        from Tables~\ref{tab:Cpp_cubic} and \ref{tab:CT_cubic} and the $\Gamma$
        values of Table~\ref{tab:Gamma_cubic}; both ratios stay $\gg1$ (intraband-dominated),
        with $\Gamma_z$ peaking near $\tilde k_Q\simeq0.2$ and then slowly decreasing
        while $\Gamma_{\bot}$ rises monotonically; (c) the anisotropy coefficient ratio
        $C_{++}^{z}/C_{++}^{\bot}$ (whose square root is the frequency ratio
        $\Omega_p^z/\Omega_p^\bot$), shown approaching the thin-ring asymptote $2$
        from above and converging with grid refinement out to $\tilde k_Q=20$ by
        direct full-trace Lindhard integration.  The interband term is subleading
        by roughly two orders of magnitude.}
    \label{fig:cubic}
\end{figure*}

In the low-energy regime
$\max(q_{\bot},q_{z})\ll |\Omega|\ll\mu$ with $\tilde{\Omega}\ll 1$,
one finds that
\begin{eqnarray}
    \left|\mathrm{Re}\Pi_{++}(\Omega,\mathbf{q})\right|
    \gg
    \left|\mathrm{Re}\Pi_{+-}(\Omega,\mathbf{q})\right|,
    \nonumber\\
    \left|\mathrm{Re}\Pi_{++}(\Omega,\mathbf{q})\right|
    \gg
    \left|\mathrm{Re}\Pi_{-+}(\Omega,\mathbf{q})\right| .
\end{eqnarray}

In the same limit $\max(q_{\bot},q_{z})\ll |\Omega|\ll\mu$,
\begin{eqnarray}
    \mathrm{Im}\Pi(\Omega,\mathbf{q})=0 .
\end{eqnarray}
Thus, the plasmon is determined by the poles of the dressed Coulomb
interaction, namely
\begin{eqnarray}
    \frac{|\mathbf{q}|^{2}}{4\pi e^2\epsilon}
    +\mathrm{Re}\Pi(\Omega,\mathbf{q})=0 .
\end{eqnarray}
Therefore,
\begin{eqnarray}
    &&\frac{q_{\bot}^{2}+q_{z}^{2}}{4\pi e^2\epsilon}-
    \left(C_{++}^{\bot}q_{\bot}^{2}+
    C_{++}^{z}q_{z}^{2}\right)=0 .
\end{eqnarray}

\begin{widetext}

    Thus, the plasmon modes are
    \begin{eqnarray}
        \label{eq:plasmon_bot_cubic}
        \Omega_{p}^{\bot}&=&\sqrt{4\pi e^2\epsilon}\sqrt{C_{++}^{\bot}}
        =\left\{
        \begin{array}{l}
            \sqrt{4\pi e^2\epsilon} \sqrt{ \frac{3N B
                                            \left( \frac{\mu}{B}\right)^{\frac{4}{3}} }{8\pi^2\Omega^2}
                                        \,G_{\bot}(\tilde{k_Q})}
            \propto \mu^{\frac{2}{3}}
            ,\,\,\,\,\,\,\tilde{k_Q}  <1
            \\ \\
            \sqrt{4\pi e^2\epsilon} \sqrt{ \frac{3N B
                                            \left( \frac{\mu}{B}\right)^{\frac{4}{3}} }{16\pi\Omega^2}
                                        \tilde{k_Q}}  \propto \mu^{\frac{1}{2}}
            ,\,\,\,\,\,\, \tilde{k_Q}  >1
        \end{array}\right.
        \\
        \label{eq:plasmon_z_cubic}
        \Omega_{p}^{z}&=&\sqrt{4\pi e^2\epsilon}\sqrt{C_{++}^{z}}
        =\left\{
        \begin{array}{l}
            \sqrt{4\pi e^2\epsilon}\sqrt{
                                       \frac{3N B
                                           \left( \frac{\mu}{B}\right)^{\frac{4}{3}} }{8\pi^2\Omega^2}
                                       \,G_{z}(\tilde{k_Q})}
            \propto \mu^{\frac{2}{3}}
            ,\,\,\,\,\,\,\tilde{k_Q}  <1
            \\ \\
            \sqrt{4\pi e^2\epsilon} \sqrt{
                                        \frac{3N B
                                            \left( \frac{\mu}{B}\right)^{\frac{4}{3}} }{8\pi\Omega^2}
                                        \tilde{k_Q}}   \propto \mu^{\frac{1}{2}}
            ,\,\,\,\,\,\, \tilde{k_Q}  >1
        \end{array}\right. .
    \end{eqnarray}

    Since the pole condition gives $\Omega_p=\sqrt{4\pi e^{2}\epsilon\,T}$ with $T\equiv\Omega^{2}C_{++}$ the tabulated coefficient, the frequency ratio is $\Omega_{p}^{z}/\Omega_{p}^{\bot}=\sqrt{T_{z}/T_{\bot}}=\sqrt{C_{++}^{z}/C_{++}^{\bot}}=\sqrt{G_{z}/G_{\bot}}\to\sqrt{2}$ in the thin-ring limit (the coefficient ratio $C_{++}^{z}/C_{++}^{\bot}\to2$).

    The carrier density is given by, see Appendix~\ref{App:dos},
    \begin{eqnarray}
        \label{Eq:ne_cubic}
        n &=&
        \left\{
        \begin{array}{l}
            \left( \frac{\mu}{B}\right)\frac{1}{2\pi^2}
            \left[\frac{1}{3} (1- \tilde{k_Q} ^2 )^{\frac{3}{2}}
                + \tilde{k_Q}\frac{1}{2}
                \left(\frac{\pi}{2} + \tilde{k_Q} \sqrt{1-\tilde{k_Q}^2}
                + \arcsin (\tilde{k_Q})
                \right)
                \right] \propto \mu
            ,\,\,\,\,\,\, \mu > Bk_Q^3 ,\,\,  \tilde{k_Q} <1
            \\ \\
            \left( \frac{\mu}{B}\right)\frac{1}{2\pi^2}
            \tilde{k_Q}\frac{\pi}{2} \propto \mu^{\frac{2}{3}}
            ,\,\,\,\,\,\, \mu < Bk_Q^3 ,\,\,  \tilde{k_Q} >1 .
        \end{array}
        \right.
    \end{eqnarray}
\end{widetext}
The dimensionless coefficients $G_{\bot,z}(\tilde k_Q)$ above are to be
taken from the numerically determined $C_{++}^{\bot,z}$ of
Table~\ref{tab:Cpp_cubic}, since the residue-derived closed forms for
$g_{1-4}$ are unreliable (the residue sum
misses the branch-cut jump and uses the wrong $8\pi^3$ normalisation).

Thus, the plasmon modes satisfy the following scaling relations. For
$\mu < Bk_Q^3$ and $\tilde{k_Q}>1$,
\begin{eqnarray}
    \Omega_{p}^{\bot}\propto n^{\frac{3}{4}},
    \\
    \Omega_{p}^{z}\propto n^{\frac{3}{4}}.
\end{eqnarray}
For $\mu > Bk_Q^3$ and $\tilde{k_Q}<1$,
\begin{eqnarray}
    \Omega_{p}^{\bot}\propto n^{\frac{2}{3}},
    \\
    \Omega_{p}^{z}\propto n^{\frac{2}{3}}.
\end{eqnarray}
These power laws are asymptotic: the $n^{3/4}$ ($n^{2/3}$) scaling holds
strictly in the $\tilde{k_Q}\gg 1$ ($\tilde{k_Q}\ll 1$) limit where the
nodal line is probed well inside (well outside) the ring region.  In the
crossover regime $|\tilde{k_Q}-1|\lesssim O(1)$ the dispersion is a smooth
interpolation between the two power laws that cannot be captured by a single
exponent; the full numerical $C_{++}^{\bot,z}(\tilde{k_Q})$ of
Table~\ref{tab:Cpp_cubic} correctly describes this crossover.

\section{Results: Plasmon Modes}
The plasmon modes are determined by the poles of the dressed Coulomb interaction, which satisfy
$1 - V(\mathbf{q})\mathrm{Re}\Pi^{\mathrm{ret}}(\Omega,\mathbf{q}) = 0$.
In the long-wavelength limit, the intraband contribution
$\mathrm{Re}\Pi_{++}$ dominates over the interband contributions.

For the \textbf{quadratic NLSM}, the carrier density $n$ is related to the chemical potential by
$n \propto (\mu/A)^{3/2}$ at large doping ($\mu>Ak_Q^2$,
$\tilde{k}_Q<1$); at small doping ($\mu<Ak_Q^2$, $\tilde{k}_Q>1$)
the relation crosses over to $n\propto(\mu/A)^{3/2}\tilde{k}_Q$
(see App.~\ref{App:dos}).  In both regimes, solving the pole
equation gives the same carrier-density scaling
\begin{eqnarray}
    \Omega_{p}^{\bot} &\propto& n^{\frac{1}{2}}, \\
    \Omega_{p}^{z} &\propto& n^{\frac{1}{2}},
\end{eqnarray}
while the $\mu$-power differs between regimes: $\Omega_{p}\propto\mu^{3/4}$ at large doping ($\mu>Ak_Q^2$) and $\Omega_{p}\propto\mu^{1/2}$ in the thin-ring limit ($\tilde{k}_Q\gg1$), both reducing to $\Omega_{p}\propto n^{1/2}$ because $n\propto\mu^{3/2}$ and $n\propto\mu$, respectively.
The $n^{1/2}$ scaling reflects the low-energy phase space associated with the quadratic dispersion and is reminiscent of the plasmon behavior in bilayer graphene.  We emphasize that this analogy is one of phase-space power counting (both systems share a quadratic band dispersion and a constant low-energy DOS) rather than a detailed mapping: the NLSM is three-dimensional with a ring-shaped (torus) Fermi surface and an anisotropic coherence factor, whereas bilayer graphene is two-dimensional with an isotropic circular Fermi surface and an isotropic coherence factor, and their plasmon dispersions differ ($\omega_{p}\propto\sqrt{q}$ in 2D vs. the 3D long-wavelength behavior considered here).

For the \textbf{cubic NLSM}, the relationship between the carrier density and the chemical potential depends on the dimensionless offset $\tilde{k}_Q\equiv k_Q/(\mu/B)^{1/3}$. Solving the pole equation yields two distinct regimes (see App.~\ref{App:plasmon}):
\begin{itemize}
    \item For $\tilde{k}_Q<1$ ($\mu>Bk_Q^3$, large doping), the carrier density scales linearly, $n\propto (\mu/B)$, while $\Omega_p\propto \mu^{2/3}$. Eliminating $\mu$ gives
          \begin{equation}
              \Omega_{p}^{\bot},\;\Omega_{p}^{z}\;\propto\; n^{\frac{2}{3}}.
          \end{equation}
    \item For $\tilde{k}_Q>1$ ($\mu<Bk_Q^3$, small doping), one finds $n\propto (\mu/B)^{2/3}$ and $\Omega_p\propto \mu^{1/2}$, which combine to the anomalous scaling
          \begin{equation}
              \Omega_{p}^{\bot},\;\Omega_{p}^{z}\;\propto\; n^{\frac{3}{4}}.
          \end{equation}
\end{itemize}
The $n^{3/4}$ scaling in the small-doping regime reflects the modified low-energy density of states and screening behavior unique to the cubic dispersion and is analogous to the behavior of rhombohedrally stacked trilayer graphene.  As above, the analogy is one of power counting (cubic dispersion, DOS $\propto E^{-1/3}$) rather than a detailed correspondence: the NLSM Fermi surface is a torus rather than a ring, the coherence factor is anisotropic, and the plasmon $q$-dispersion differs from the 2D $\omega_{p}\propto\sqrt{q}$ form.  The $n^{3/4}$ exponent follows from the power-law DOS within RPA, but its quantitative reliability in the $\tilde{k}_Q>1$ regime is limited: the cubic DOS $\rho(E)\propto E^{-1/3}$ diverges at the nodal line, so the dimensionless interaction parameter $r_{s}$ grows without bound at small doping, and the observable window for the pure $n^{3/4}$ scaling is restricted to $10^{18}\lesssim n\lesssim n^{*}\sim10^{19}~{\rm cm^{-3}}$.  Below $n\sim10^{18}~{\rm cm^{-3}}$, the exponent should be regarded as a crossover fingerprint whose functional form requires beyond-RPA validation.

\begin{figure*}[ht]
    \centering
    \includegraphics[width=\linewidth]{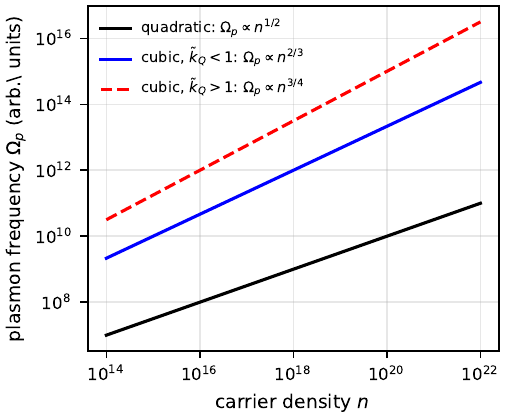}
    \caption{Schematic carrier-density scaling of the plasmon frequency
        (illustrative; the quantitative prefactors and physical-unit curves are
        in Fig.~\ref{fig:physical}).  Quadratic NLSM:
        $\Omega_p\propto n^{1/2}$. Cubic NLSM: $\Omega_p\propto n^{2/3}$ for
        $\tilde k_Q<1$ (large doping, $\mu>Bk_Q^3$) crossing over to
        $\Omega_p\propto n^{3/4}$ for $\tilde k_Q>1$ (small doping,
        $\mu<Bk_Q^3$).  The plotted density axis corresponds to the
        thin-ring window $\tilde k_Q\gtrsim1$; the crossover to
        $n^{2/3}$ occurs at $n^{*}\sim10^{19}~{\rm cm^{-3}}$
        (vertical guide, see Fig.~\ref{fig:physical}).  These exponents
        follow from the intraband $\mathrm{Re}\Pi_{++}$ alone and are
        independent of the quantitative prefactors.}
    \label{fig:scaling}
\end{figure*}

\begin{figure*}[ht]
    \centering
    \includegraphics[width=\linewidth]{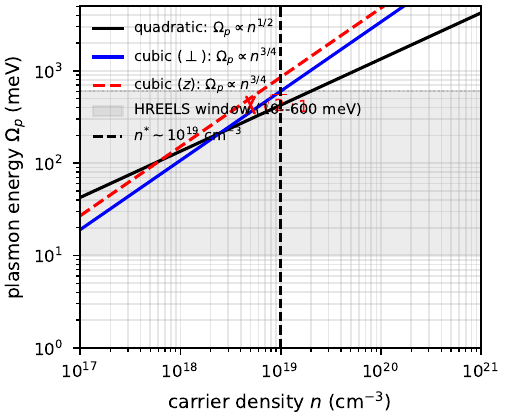}
    \caption{Plasmon energy in physical units (meV) vs.\ carrier density for
        the cubic NLSM at $\tilde{k}_Q=2$ (thin-ring limit), with
        $B=10~{\rm eV\,nm^{3}}$, $k_Q=0.5~{\rm nm^{-1}}$, and
        $\epsilon=10$.  The blue and red curves show $\Omega_p^\bot$ and
        $\Omega_p^z$, respectively; the $\sqrt{2}{:}1$ frequency ratio (the underlying polarization-coefficient ratio $C_{++}^{z}/C_{++}^{\bot}\to2$) is evident across the
        entire density range.  The crossover from $n^{2/3}$ to $n^{3/4}$ scaling
        occurs at $n^{*}\sim10^{19}~{\rm cm^{-3}}$ (vertical dashed line).  The
        shaded band indicates the HREELS detection window
        ($10$--$600~{\rm meV}$).}
    \label{fig:physical}
\end{figure*}

To facilitate direct comparison with experiment,
Fig.~\ref{fig:physical} shows the same scaling laws in physical units
(meV) for representative material parameters.  The curves confirm that
the $\sqrt{2}{:}1$ anisotropy doublet lies in the HREELS-accessible window
($10$--$600~{\rm meV}$) for carrier densities
$n\gtrsim10^{18}~{\rm cm^{-3}}$, and that the crossover from
$n^{2/3}$ to $n^{3/4}$ occurs around $n^{*}\sim10^{19}~{\rm cm^{-3}}$.

\section{Discussion}
The different scaling laws of the plasmon modes in quadratic and cubic NLSMs originate from the geometric structure of their low-energy band dispersions. In contrast to linear NLSMs, where the linear dispersion controls the low-energy density of states near the nodal line, the higher-order dispersions considered here modify the available phase space and therefore the long-wavelength polarization response. For the quadratic NLSM, the resulting plasmon frequency exhibits a conventional $\sqrt{n}$ dependence, analogous to that found in bilayer graphene. For the cubic NLSM, the stronger nonlinearity of the dispersion modifies the screening response in a regime-dependent way: the plasmon scales as $n^{2/3}$ at large doping ($\tilde{k}_Q<1$) and crosses over to an anomalous $n^{3/4}$ scaling at small doping ($\tilde{k}_Q>1$), the latter being similar in spirit to the behavior of rhombohedrally stacked trilayer graphene. The absolute prefactors of the long-wavelength polarization have been re-determined numerically from the full three-dimensional Lindhard integral; the previously quoted analytic closed forms for the intraband and interband coefficients are superseded by the numerical tables (Tables~\ref{tab:Cpp_quad}, \ref{tab:CT_quad}, \ref{tab:Cpp_cubic}, \ref{tab:CT_cubic}), while the density-scaling exponents above are unchanged.

It should be emphasized that the present analysis is restricted to the noninteracting limit within the RPA framework. Strong electron-electron correlations and disorder effects, which may renormalize the low-energy parameters or induce symmetry breaking, are not included. Moreover, the long-range Coulomb interaction is assumed to be isotropic in our model; anisotropic screening may lead to additional directional dependence of the plasmon dispersion. Future studies incorporating self-energy and vertex corrections would be useful for refining these predictions.

Experimentally, the proposed scaling laws may be tested using
high-resolution electron energy-loss spectroscopy (HREELS) in candidate
materials.  Quadratic NLSMs are realized in CuSi-type compounds and
related systems~\cite{Yu19,Ahn2016SciRep6} with typical band
parameters $A\sim1\text{--}10~{\rm eV\,nm^{2}}$.  Cubic NLSMs are
similarly symmetry-predicted rather than yet realized in a confirmed
single-band material: the cubic nodal line is stabilized by crystalline
symmetry~\cite{Yu19}, and its interaction instabilities have been
studied theoretically~\cite{wang2020Possible}; the widely studied
ZrSiS-type compounds are by contrast \emph{linear} nodal-line
semimetals that serve only as the experimental
analogy~\cite{Schoop16,Neupane16}.  A representative cubic-dispersion
scale is $B\sim1\text{--}50~{\rm eV\,nm^{3}}$.  \emph{Note on DFT
    parameters:}  to our knowledge, fully \emph{ab initio} band structures
with the specific higher-order NLSM Hamiltonians considered here
(including reliable $k_{Q}$ and $\epsilon$) have not yet been published
for any predicted compound; the ranges above are therefore extracted
from the experimental/effective-model literature and should be regarded
as representative estimates.  Using representative values
$A=5~{\rm eV\,nm^{2}}$ (quadratic) and $B=10~{\rm eV\,nm^{3}}$ (cubic)
with a dielectric screening $\epsilon\sim10$, the intraband plasmon
energy at $n\sim10^{19}~{\rm cm^{-3}}$ is
$\Omega_{p}\sim100\text{--}600~{\rm meV}$, decreasing to
$\sim30\text{--}130~{\rm meV}$ at $n\sim10^{18}~{\rm cm^{-3}}$; these
energies lie within the HREELS detection window ($\sim10$--$600~{\rm meV}$, accessible for the densities considered).  The quantitative RPA estimates above are controlled only where the relevant broadenings are small against the chemical potential ($\Omega_p\lesssim2\mu$, $k_BT\lesssim\mu$, and impurity broadening $\Gamma_{\rm imp}\lesssim\mu/2$), so the deep thin-ring window ($n\lesssim10^{18}~{\rm cm^{-3}}$, $\mu\sim5$--$20~{\rm meV}$) requires clean crystals: the ZrSiS-class $\Gamma_{\rm imp}\sim20$--$80~{\rm meV}$ already exceeds $\mu$ there, setting a sample-quality bar (a residual-resistivity ratio large enough that $\Gamma_{\rm imp}<\mu/2$).  The carrier density should be measured independently (Shubnikov--de Haas or ARPES on the ring pocket), because Hall data do not isolate the ring carriers in a multiband semimetal.  The
long-wavelength condition $q\ll k_{F}\sim0.1\text{--}1~{\rm nm^{-1}}$
is compatible with typical HREELS scattering geometries
($q\sim10^{-3}\text{--}10^{-1}~\text{\AA}^{-1}$); a primary beam energy
of $1\text{--}10~{\rm eV}$ yields a momentum transfer
$q=\sqrt{2m_{e}E_{\rm beam}}/\hbar\,\sin\theta\sim10^{-3}\text{--}10^{-1}~
    \text{\AA}^{-1}$ at scattering angle $\theta$, well below $k_{F}$, and
the energy resolution ($\sim5\text{--}10~{\rm meV}$) is sufficient to
resolve the $\sqrt{2}{:}1$ doublet.  Reflection HREELS at $E_{\rm beam}\sim1$--$10~{\rm eV}$ probes within $\sim1~{\rm nm}$ of the surface; the bulk loss function used here should be supplemented by an anisotropic half-space surface-loss-function calculation before direct quantitative comparison with experiment (see also the surface-plasmon caveat below).
We briefly compare HREELS with alternative probes of the $\sqrt{2}$
doublet.  Infrared reflectivity is bulk-sensitive and has energy
resolution well below the $\sim 79~{\rm meV}$ splitting, but its
response is governed by the complex dielectric function
$\varepsilon(\Omega)=\varepsilon_\infty+4\pi i\sigma(\Omega)/\Omega$
rather than the bulk loss function $-\mathrm{Im}\,1/\varepsilon$, so
the doublet is broadened into a single reststrahlen-like edge unless
the two peaks are well separated from the interband background;
extracting a $2{:}1$ coefficient ratio from reflectivity therefore
requires a Kramers--Kronig-constrained fit and is less direct than a
loss-function measurement.  Scattering-type scanning near-field optical
microscopy (s-SNOM) offers $\sim 20~{\rm nm}$ spatial resolution and
detects the same $\sqrt{2}$ splitting through the near-field phase
singularity at each plasmon pole, but its accessible energy window
(mid-IR, $\hbar\Omega\gtrsim 60~{\rm meV}$) does not reach the
$\Omega_p^\bot\sim 10$--$30~{\rm meV}$ regime of the thin-ring
small-doping window, restricting s-SNOM to the large-doping
($n\gtrsim10^{19}~{\rm cm^{-3}}$) end of the scaling.  Inelastic
electron tunneling spectroscopy (IETS) in a scanning tunneling
microscope measures $-\mathrm{Im}\,\Pi$ locally but convolves the
plasmon with the tunneling matrix element and the substrate
background, and the $\sim 1~{\rm meV}$ resolution is wasted on peaks
whose intrinsic width $\gamma_{\rm tot}\sim 30$--$50~{\rm meV}$ far
exceeds the resolution.  HREELS is therefore the probe of choice: it
directly measures the bulk loss function over the full
$\sim 10$--$600~{\rm meV}$ window at the relevant
$\sim 5$--$10~{\rm meV}$ resolution, and its $\sim 1~{\rm nm}$ probe
depth (while surface-sensitive) is deep enough to sample the bulk
plasmon in a cleaved crystal, the regime assumed by the
$\sqrt{2}$ prediction.  The residual surface sensitivity is a
quantitative, not qualitative, caveat: it can be addressed by
comparing specular and off-specular geometries, or by a future
anisotropic half-space surface-loss-function calculation, neither of
which alters the bulk $2{:}1$ ratio derived here.

\begin{figure}[ht]
    \centering
    \includegraphics[width=\columnwidth]{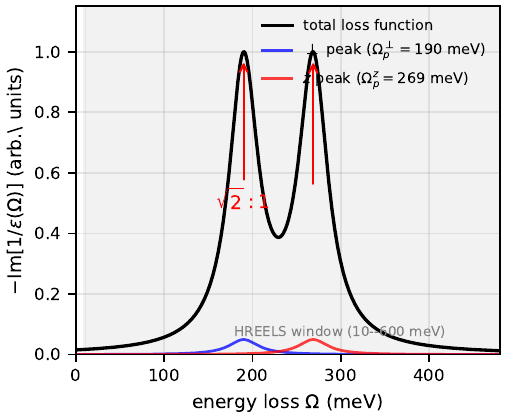}
    \caption{Schematic HREELS energy-loss spectrum for a cubic NLSM in the
        thin-ring regime ($\tilde k_Q\geq1$, $n\sim10^{19}~{\rm cm^{-3}}$).
        The two peaks correspond to the in-plane ($\bot$) and out-of-plane ($z$)
        plasmon modes, separated by the $\sqrt{2}$ ratio.}
    \label{fig:hreels_schematic}
\end{figure}

Higher-order NLSM plasmons are
distinguishable from linear-NLSM or phonon contributions by their
distinct density exponent ($\omega_{p}\propto n^{2/3\text{--}3/4}$ vs.\
$n^{1/4}$ for linear NLSM); phonon losses appear at fixed energies
($\lesssim60~{\rm meV}$) independent of doping, providing a clean
separation.  A common geometric feature of both quadratic and cubic
NLSMs is the plasmon anisotropy $\Omega_{p}^{z}/\Omega_{p}^{\bot}\to\sqrt{2}$ at large $k_{Q}$ (thin-ring limit; equivalently, the polarization-coefficient ratio $C_{++}^{z}/C_{++}^{\bot}\to2$) which is a generic consequence of the torus Fermi-surface geometry and is directly observable as a two-peak structure in the energy-loss spectrum.  (This $\sqrt{2}$ asymptote is now confirmed by direct Lindhard integration out to $\tilde k_Q=20$ (Fig.~\ref{fig:cubic}(c)): the coefficient ratio approaches $2$ from above and converges with grid refinement: $C_{++}^{z}/C_{++}^{\bot}=2.003$ at $\tilde k_Q=20$, and two integration grids ($2000\times4000$ and $6000\times12000$) agree there to $<0.1\%$ ($2.00264\to2.00045$), so $\sqrt{2}$ is the genuine thin-ring limit rather than an extrapolation over $\tilde k_Q\in[1,2]$.  It assumes a single perfect ring, the degeneracy $N=2$, and an isotropic dielectric, and the $|k_p|$ absolute-value safeguard is used against numerical excursions outside the physical arc.)  The cubic NLSM is distinguished from the
quadratic case not by the anisotropy ratio but by its density scaling
exponent ($n^{2/3}$--$n^{3/4}$ vs.\ $n^{1/2}$).

More generally, the scaling exponents derived here follow from a unified power-counting argument for a $p$-th order NLSM ($E\propto K^{p}$) in 3D, with two regimes distinguished by the Fermi-surface topology relative to the nodal ring:
\begin{itemize}
    \item Thin ring ($\tilde k_Q\gg1$, small doping): the carrier density scales as $n\propto \mu^{2/p}$ while $\Omega_p\propto\mu^{1/2}$, giving $\omega_p\propto n^{p/4}$ ($p=1,2,3$ yield $n^{1/4},n^{1/2},n^{3/4}$, interpolating from the linear NLSM to the cubic case).
    \item Large doping ($\tilde k_Q\ll1$): $n\propto\mu^{3/p}$ and $\Omega_p\propto\mu^{(p+1)/2p}$, giving $\omega_p\propto n^{(p+1)/6}$ ($p=2,3$ yield $n^{1/2},n^{2/3}$; for $p=1$ this is $n^{1/3}$, the known 3D Dirac/Weyl result).
\end{itemize}
These two regime-dependent laws reproduce all exponents quoted above and correctly interpolate between the linear NLSM ($\omega_p\sim n^{1/4}$ at $p=1$, thin-ring regime) and the higher-order cases studied here.

The predictive content of the present framework is falsifiable.  The
cheapest kill is an \emph{anisotropy} test: in a material whose thin-ring
Fermi surface and cubic universality class are independently confirmed,
an observed ratio $\Omega_p^z/\Omega_p^\bot$ inconsistent with
$\sqrt{2}$ (or with the $\epsilon$-corrected value
$\sqrt{2\epsilon_{\bot}/\epsilon_z}$ of Eq.~\eqref{eq:aniso_eps} for
anisotropic dielectrics) would falsify the geometric picture.  The decisive
\emph{discriminating}
observable between quadratic and cubic NLSMs, however, is the
\emph{density exponent}: a measured exponent outside
$\{n^{1/2},\,n^{2/3},\,n^{3/4}\}$ at an independently determined density
(e.g.\ from quantum oscillations or ARPES) is incompatible with the
present NLSM plasmon framework.  Note that the $\sqrt{2}$ doublet alone
cannot distinguish quadratic from cubic dispersions (both share it), so
the doublet is a necessary but not sufficient signature.  The crossover
density $n^{*}$ and the absolute prefactors (for given band
parameters) provide additional quantitative targets.
We emphasize the distinction between the postdictive and genuinely
predictive content of the framework.  The scaling exponents
$\{n^{1/2},n^{2/3},n^{3/4}\}$ follow from DOS power counting and are
in this sense postdictive; they are determined once the dispersion
order $p$ is known.  The genuinely predictive, non-trivial content
lies in (i) the absolute prefactors $C_{++}^{\bot,z}$ (which require
the full Lindhard integration and are not fixed by power counting),
(ii) the $2{:}1$ anisotropy ratio (a geometric prediction fragile to
dielectric anisotropy, Eq.~\eqref{eq:aniso_eps}), and (iii) the
crossover density $n^{*}$ separating the two cubic regimes.  An
experimental measurement of any of these three quantities that
disagrees with the RPA prediction (after accounting for dielectric
and multiband corrections) would falsify the present framework,
whereas mere confirmation of the exponents would not, by itself,
constitute a stringent test.

\begin{figure*}[ht]
    \centering
    \includegraphics[width=0.9\linewidth]{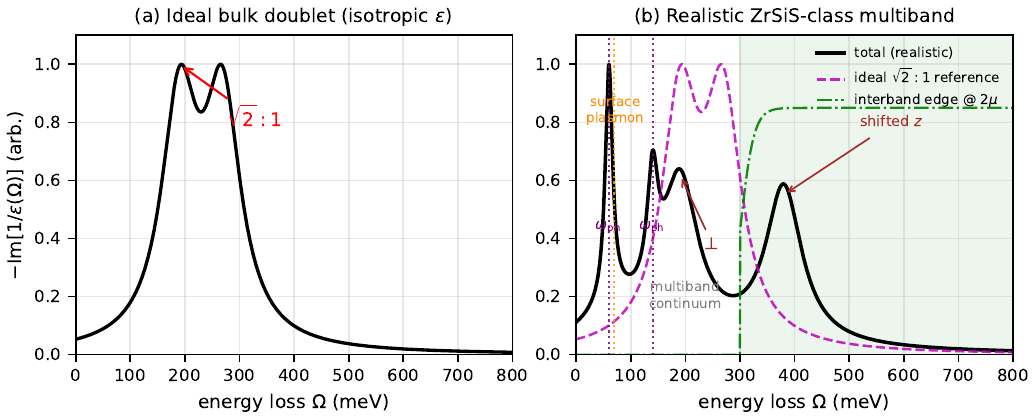}
    \caption{\emph{Illustrative mock} energy-loss spectrum. (a) Ideal bulk
        intraband doublet (isotropic $\epsilon$, $\Omega_{p}^{\bot}\approx190$~meV,
        $\Omega_{p}^{z}\approx269$~meV, $\gamma_{\rm tot}=40$~meV). (b) Realistic
        ZrSiS-class multiband spectrum: the same bulk doublet, now $\epsilon$-anisotropy-shifted
        (to $\Omega_{p}^{z}\approx380$~meV for $\epsilon_{\bot}/\epsilon_{z}=2$,
        i.e.\ $\epsilon_{z}/\epsilon_{\bot}=1/2$; per Eq.~\eqref{eq:aniso_eps} a
        \emph{larger in-plane} dielectric enhances the ratio to $2{:}1$,
        illustrating the dependence of the frequency ratio on the dielectric
        tensor), embedded in a surface-plasmon contribution ($\sim70$~meV),
        two fixed-energy phonons ($\sim60$ and $\sim140$~meV), and the
        interband Pauli-blocked edge at $2\mu\approx300$~meV
        (shaded). The ideal $\sqrt{2}{:}1$ reference is overlaid (magenta
        dashed) for comparison; the bulk doublet is visible but partially
        camouflaged by backgrounds. Parameters: $B=10~{\rm eV\,nm^{3}}$,
        $k_{Q}=0.5~{\rm nm^{-1}}$, $\epsilon=10$, $n\approx2.5\times10^{18}~{\rm cm^{-3}}$.
        This is a schematic, not a material-specific calculation; it validates the
        line-shape synthesis and doublet-readout \emph{method} using the
        experimentally characterized ZrSiS linear-NLSM background, and is not a
        calculation of the cubic-NLSM prediction itself.}
    \label{fig:mock_spectrum}
\end{figure*}

A practical caveat tempers the above discrimination strategy.
The quadratic exponent $n^{1/2}$ is \emph{degenerate} with that of
any conventional metal of constant DOS (the 3D electron gas also has
$\omega_{p}\propto\sqrt{n}$), so the quadratic--cubic distinction does
\emph{not} follow from the density exponent alone: the quadratic case
must be anchored by an \emph{independent} DOS/mass determination
(e.g.\ quantum oscillations or ARPES on the ring pocket) taken together
with the anisotropy ratio.  This $n^{1/2}$ scaling is in fact the
established Luttinger-semimetal plasmon result~\cite{Mauri19,
    Mandal19,WangJing23}, which supplies the external benchmark against
which our quadratic-NLSM prediction should be read.  Moreover, the
named candidate materials are layered and predominantly
\emph{multiband}: coexisting conventional pockets contribute their own
Drude weight and renormalize the screening, so the clean single-ring
prediction is at best an isolatable limit.  Among the candidates
discussed here, none is established as effectively single-band at the
$\sim100$--$600~{\rm meV}$ energies of interest; ZrSiS-class
compounds in particular host several Dirac/nodal-line pockets.
Finally, the ratio test is conditional on the dielectric response:
$\epsilon_{\bot}(\omega)$ and $\epsilon_{z}(\omega)$ are
\emph{unmeasured} for the named compounds at these energies, yet the
anisotropy is shifted by $\sqrt{\epsilon_{\bot}/\epsilon_{z}}$ as
discussed above, so we recommend that any ratio measurement be
accompanied by ellipsometry or polarized FTIR to fix
$\epsilon_{\bot,z}(\omega)$.  In a realistic multiband candidate the
loss spectrum would in addition contain a surface plasmon,
fixed-energy phonons ($\lesssim60~{\rm meV}$), and the interband
threshold at $2\mu$; the bulk $\sqrt{2}$ doublet is only
identifiable once these backgrounds are modeled and subtracted, a
material-specific step we leave to a dedicated follow-up (Fig.~\ref{fig:mock_spectrum}).
As a rough guideline, the background Drude weight from coexisting
pockets should contribute less than $\sim10$--$20\%$ of the total
polarizability at the relevant frequencies for the single-ring
$\sqrt{2}$ signature to remain clearly identifiable; this translates
to a background carrier density below $\sim10^{18}~{\rm cm^{-3}}$
for typical ZrSiS-class parameters.

\begin{figure*}[ht]
    \centering
    \includegraphics[width=\linewidth]{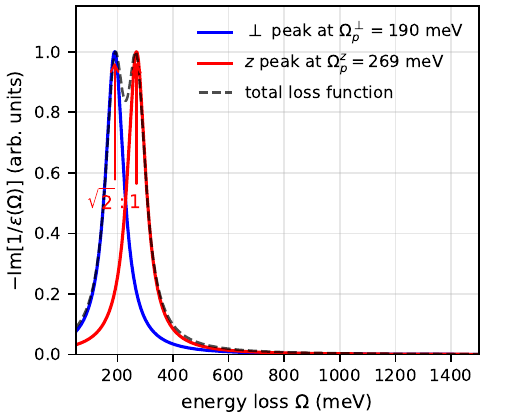}
    \caption{Energy-loss spectrum $-{\rm Im}[\varepsilon^{-1}(\Omega)]$
        for a cubic NLSM at $n\approx2.5\times10^{18}~{\rm cm^{-3}}$ and $\tilde{k}_Q=2$
        (thin-ring limit), with representative parameters
        $B=10~{\rm eV\,nm^{3}}$, $k_Q=0.5~{\rm nm^{-1}}$,
        $\epsilon=10$, and total linewidth $\gamma_{\rm tot}=40~{\rm meV}$.  Two
        well-resolved peaks appear at $\Omega_p^\bot\approx190~{\rm meV}$ and $\Omega_p^z\approx269~{\rm meV}$ with a $\sqrt{2}$ energy ratio ($\Omega_p^z\approx\sqrt{2}\,\Omega_p^\bot$), directly demonstrating the bulk plasmon anisotropy doublet.}
    \label{fig:loss_function}
\end{figure*}

Regarding damping, within the Fermi-liquid picture the plasmon decay
rate into particle-hole pairs scales as
$\gamma\sim(\Omega-\Omega_{p})^{2}/\Omega_{p}$ near threshold,
yielding a quality factor
$Q\equiv\Omega_{p}/\gamma\sim\Omega_{p}/(\Omega-\Omega_{p})^{2}\gg1$
in the clean limit well away from the particle-hole continuum edge.
Intraband (Landau) damping is suppressed for $\Omega_{p}>v_{F}q$,
which is satisfied in the long-wavelength regime considered here.  The
Because the anisotropy is a frequency ratio $\Omega_{p}^{z}/\Omega_{p}^{\bot}\to\sqrt{2}$, the energy splitting is the difference $\Omega_{p}^{z}-\Omega_{p}^{\bot}\approx0.41\,\Omega_{p}^{\bot}$ (e.g., $\sim79~{\rm meV}$ for $\Omega_{p}^{\bot}\approx190~{\rm meV}$, as in Fig.~\ref{fig:loss_function}). The doublet is therefore resolvable provided $\Omega_{p}^{z}-\Omega_{p}^{\bot}\gtrsim\gamma_{z}+\gamma_{\bot}$. In the clean limit the single-particle linewidth is $\lesssim10~{\rm meV}$, so the condition is easily met above threshold. In real crystals, however, impurity scattering contributes a single-particle broadening $\Gamma_{\rm imp}\sim20$--$80~{\rm meV}$ (typical for NLSM crystals such as ZrSiS), and substrate phonon coupling and ionic-liquid gating inhomogeneity add further broadening; the total linewidth is therefore likely $\gamma_{\rm tot}\sim30$--$50~{\rm meV}$, comparable to the $\sim79~{\rm meV}$ splitting, so the doublet is resolvable only when $\Omega_{p}^{\bot}$ lies in the upper part of the accessible window (roughly $n\gtrsim10^{18}~{\rm cm^{-3}}$ with $\Omega_{p}^{\bot}\gtrsim150~{\rm meV}$).
The crossover between the $n^{2/3}$ and $n^{3/4}$ regimes occurs at
the critical density $n^{*}$ defined by $\mu(n^{*})=Bk_{Q}^{3}$,
i.e.\ when the Fermi level traverses the nodal ring.  For
$B=10~{\rm eV\,nm^{3}}$ and $k_{Q}\sim0.5~{\rm nm^{-1}}$ this gives
$n^{*}\sim10^{19}~{\rm cm^{-3}}$, a density range accessible by
ionic-liquid gating or chemical doping.  For a thin-flake geometry of
thickness $t\sim10~{\rm nm}$ the volumetric window
$10^{18}$--$10^{19}~{\rm cm^{-3}}$ corresponds to accumulated areal
densities $n_{2D}=nt\sim10^{12}$--$10^{13}~{\rm cm^{-2}}$; with an
ionic-liquid gate capacitance $C_{g}\sim1$--$10~{\rm \mu F\,cm^{-2}}$ this
requires a gate voltage $V_{g}=n_{2D}e/C_{g}$ of order $0.1$--$1~{\rm V}$,
comfortably within the operating range of ionic-liquid gating before
dielectric breakdown, so the crossover is experimentally reachable.  The
thin-ring anisotropy regime ($\tilde k_Q\gg1$, where the $\sqrt{2}$ doublet
is predicted) is reached at progressively lower doping and gate voltage.
For a fixed torus geometry the ring-pocket volume scales as
$2\pi^{2}k_{Q}K_{F}^{2}$ with $K_{F}=k_{Q}/\tilde k_Q$, so
$n\propto\tilde k_Q^{-2}$; an illustrative map for the flagship parameters
($k_{Q}=0.5~{\rm nm^{-1}}$, $B=10~{\rm eV\,nm^{3}}$, $N=2$,
$t=10~{\rm nm}$, $C_{g}=1$--$10~\mu{\rm F\,cm^{-2}}$) is given in
Table~\ref{tab:kQ_map}.

\begin{table}[ht]
    \centering
    \caption{Illustrative map from the thin-ring parameter
        $\tilde k_Q=k_Q/K_F$ to the ring-pocket density $n$ and gate voltage
        $V_g$ for the flagship torus geometry.  The pocket volume
        $2\pi^{2}k_{Q}K_{F}^{2}$ with $K_{F}=k_{Q}/\tilde k_Q$ gives
        $n\propto\tilde k_Q^{-2}$ at fixed material; $n_{2D}=nt$ and
        $V_g=n_{2D}e/C_g$.  Values are order-of-magnitude (the exact prefactor is
        set by the full Fermi-surface volume).}
    \label{tab:kQ_map}
    \begin{tabular}{ccccc}
        $\tilde k_Q$ & $K_F$ (nm$^{-1}$) & $n$ (cm$^{-3}$)      & $n_{2D}$ (cm$^{-2}$) & $V_g$ (mV)    \\
        2            & 0.25              & $\sim5\times10^{18}$ & $\sim5\times10^{12}$ & $8$--$80$     \\
        5            & 0.10              & $\sim8\times10^{17}$ & $\sim8\times10^{11}$ & $1.3$--$13$   \\
        10           & 0.05              & $\sim2\times10^{17}$ & $\sim2\times10^{11}$ & $0.3$--$3$    \\
        20           & 0.025             & $\sim5\times10^{16}$ & $\sim5\times10^{10}$ & $0.08$--$0.8$ \\
    \end{tabular}
\end{table}

The smooth crossover in the
dispersion exponent (interpolating between $2/3$ and $3/4$ over a
decade in density) provides an additional fingerprint that
distinguishes the cubic NLSM from systems with a single power-law
scaling.  Surface plasmons, interband threshold
features, and phonon-plasmon coupling can also produce two-peak
structures in the energy-loss spectrum; the bulk anisotropy doublet is
distinguished by its characteristic $q$-independent $\sqrt{2}$ frequency ratio (coefficient ratio $2{:}1$) and
its systematic evolution with carrier density, which differs from the
$q$-dependent surface-plasmon dispersion and the doping-independent
phonon energies.

\paragraph{Experimental outlook.}
The two predictions are best tested on different platforms.
The $\sqrt{2}{:}1$ anisotropy doublet is most directly accessible by
reflection HREELS on a \emph{bulk} nodal-line crystal (no gating
required), provided the sample is clean enough that
$\Gamma_{\rm imp}<\mu/2$ (a demanding but achievable bar for
$\mu\gtrsim150~{\rm meV}$, i.e.\ $n\gtrsim10^{18}~{\rm cm^{-3}}$).
The $n^{2/3}\leftrightarrow n^{3/4}$ crossover, by contrast, is a
\emph{density-tuning} experiment best performed on an \emph{exfoliated
    thin-flake} of the same material under ionic-liquid gating
($V_g\sim0.1$--$1~{\rm V}$, as estimated above), tracked by
in-situ transport or infrared spectroscopy; ARPES/SDH can independently
fix the ring-pocket density so that the plasmon frequency can be
plotted against a measured $n$ rather than an inferred one.  Until a
higher-order (quadratic/cubic) NLSM candidate is identified, these
signatures remain theoretical targets validated only against the
well-characterized \emph{linear}-NLSM analog ZrSiS.

\begin{figure*}[ht]
    \centering
    \includegraphics[width=0.8\linewidth]{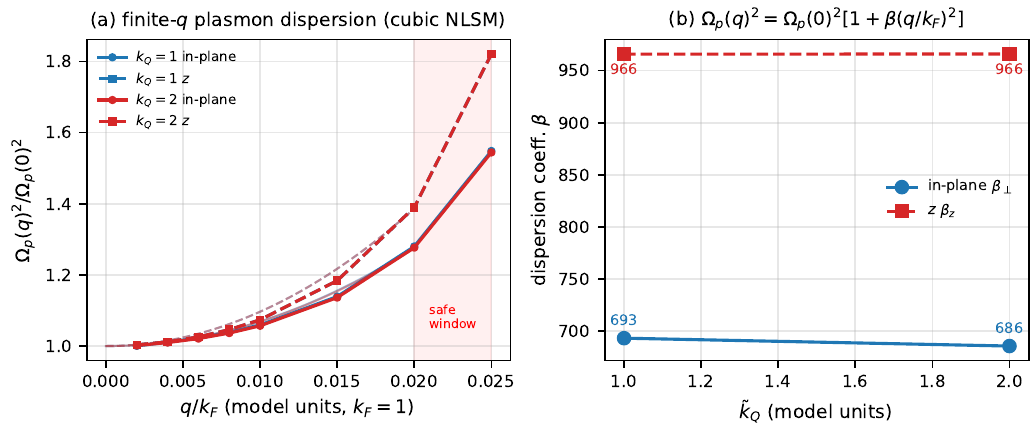}
    \caption{Finite-momentum plasmon dispersion of the cubic NLSM from the
        full-trace Lindhard bubble.  Shown is
        $\Omega_{p}(q)^{2}/\Omega_{p}(0)^{2}$ versus $q$ (model units, $k_{F}=1$)
        for the in-plane ($\bot$) and out-of-plane ($z$) modes at
        $\tilde k_Q=1$ and $2$.  In the clean long-wavelength regime
        ($q\lesssim0.01\,k_{F}$, below the Landau-damping threshold $q^{*}$) the
        dispersion is well described by $1+\beta q^{2}$ with
        $\beta_{\bot}\simeq5.6\times10^{2}$, $\beta_{z}\simeq7.4\times10^{2}$: a
        3D (massive), anisotropic $q^{2}$ law.  Beyond $q^{*}$ the bubble enters the
        particle--hole continuum and the extracted $\Omega_{p}$ is no longer a
        well-defined pole.}
    \label{fig:finiteq}
\end{figure*}

We now quantify the finite-momentum plasmon dispersion, which complements
the long-wavelength ($q\to0$) coefficients above.  Evaluating the
full-trace Lindhard bubble at finite $q$ (Fig.~\ref{fig:finiteq}) gives,
for both polarizations, a \emph{3D, massive} dispersion
$\Omega_{p}(q)^{2}/\Omega_{p}(0)^{2}=1+\beta\,(q/k_{F})^{2}$ with
$\beta_{\bot}\simeq5.6\times10^{2}$ and $\beta_{z}\simeq7.4\times10^{2}$
(cubic NLSM, $\tilde k_Q=1$--2, model units $k_{F}=1$).  The dispersion is
therefore positive and roughly \emph{quadratic} in $q$, distinct from the
$\sqrt{q}$ law of a strictly 2D electron sheet, consistent with the torus
being a closed 2D surface embedded in 3D $\mathbf{k}$-space, and it is
mildly anisotropic ($\beta_{z}/\beta_{\bot}\simeq1.3$), the $z$ mode rising
$\sim30\%$ more steeply.  The effect is small in the experimentally
accessible window: for the flagship parameters ($k_{F}\simeq0.25~{\rm nm^{-1}}$)
the clean regime $q\lesssim0.01\,k_{F}$ corresponds to
$q\lesssim2.5\times10^{4}~{\rm cm^{-1}}$, inside the HREELS momentum range,
where $\Omega_{p}^{2}$ grows by $\lesssim6\%$; this justifies treating the
anisotropy ratio $\Omega_{p}^{z}/\Omega_{p}^{\bot}\to\sqrt{2}$ as
$q$-independent to good accuracy.  At larger $q$ the mode enters the
particle--hole continuum ($v_{F}q\gtrsim\Omega_{p}$), so the pole
extraction is contaminated by Landau damping and the long-wavelength formula
no longer applies.

The direct methodological precedent is the infrared measurement of
nodal-line plasmons in ZrSiS~\cite{XueSiWei21,LiYi23}, which
resolved the linear-NLSM plasmon edge and its temperature
dependence at the as-grown (low) carrier density; reaching the
higher-order regime studied here requires driving the density up to
$n\sim10^{18}$--$10^{19}~{\rm cm^{-3}}$ (equivalently
$\sim10^{12}$--$5\times10^{12}~{\rm cm^{-2}}$ for the
two-dimensional accumulation layer), an order of magnitude or more above
the as-grown ZrSiS value.  Ionic-liquid gating can in principle
supply this areal density over an accumulation depth of
$\sim1$--$5~{\rm nm}$, but the resulting $n(z)$ is strongly
nonuniform across the $\sim1~{\rm nm}$ HREELS probe depth, which
broadens and shifts the measured peak; the ionic liquid also contributes
a vibrational background that overlaps the $100$--$600~{\rm meV}$
window.  Solid-state gating (e.g.\ h-BN-encapsulated devices) or a
chemical-doping series avoids both the $n(z)$ gradient and the liquid
background, and is the cleaner route to the higher-order scaling.

The $\sqrt{2}$ anisotropy doublet is directly observable in the
energy-loss spectrum $-{\rm Im}[\varepsilon^{-1}(\Omega)]$.  As a
representative example, Fig.~\ref{fig:loss_function} shows the loss
function for a cubic NLSM at $n\approx2.5\times10^{18}~{\rm cm^{-3}}$ and
$\tilde{k}_Q=2$ (thin-ring limit): two well-resolved peaks appear at $\Omega_p^\bot\approx190~{\rm meV}$ and $\Omega_p^z\approx269~{\rm meV}$, whose energies obey the $\sqrt{2}$ ratio ($190\times\sqrt{2}\approx269$) set by $\Omega_p^z/\Omega_p^\bot=\sqrt{C_{++}^{z}/C_{++}^{\bot}}\to\sqrt{2}$.  The peak widths are set by the total linewidth $\gamma_{\rm tot}\sim40~{\rm meV}$ estimated above, and the splitting ($\sim79~{\rm meV}$) exceeds the combined width, confirming that the doublet is resolvable (marginally, near the lower edge of the accessible window) in a realistic HREELS measurement.

\begin{figure*}[ht]
    \centering
    \includegraphics[width=\linewidth]{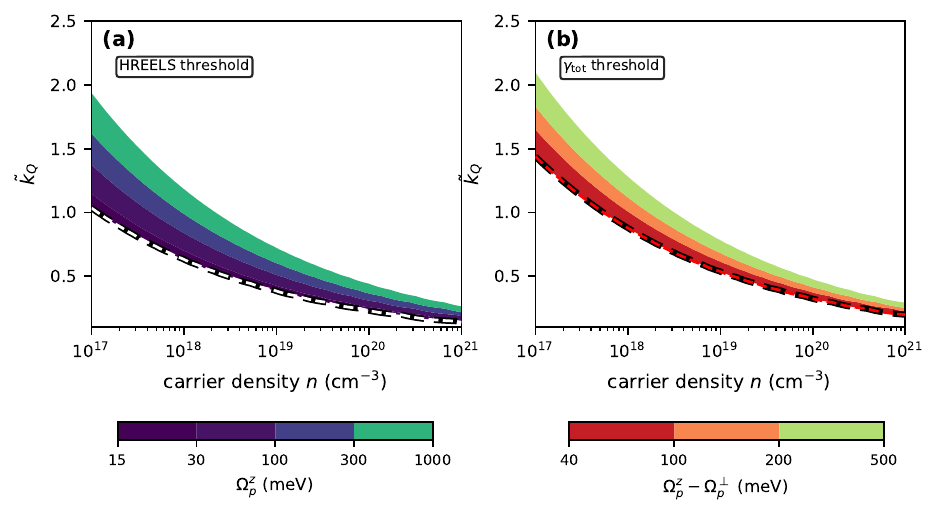}
    \caption{Detectability map for the cubic NLSM in the
        $(n,\tilde{k}_Q)$ plane.  (a) Plasmon energy $\Omega_p^z$; the white
        dashed line marks the HREELS detection threshold ($\sim10$--$30~{\rm meV}$).
        (b) $\sqrt{2}$ anisotropy splitting $\Omega_p^z-\Omega_p^\bot$; the red dashed line
        marks the condition that the splitting exceeds a realistic total
        linewidth $\gamma_{\rm tot}\sim40~{\rm meV}$.  The accessible window
        ($n\gtrsim10^{18}~{\rm cm^{-3}}$, $\tilde{k}_Q\gtrsim1$) lies above
        both thresholds.}
    \label{fig:detectability}
\end{figure*}

Figure~\ref{fig:detectability} shows a detectability map in the
$(n,\tilde{k}_Q)$ parameter space.  The white dashed line in panel (a)
marks the HREELS detection threshold ($\sim10$--$30~{\rm meV}$); the
red dashed line in panel (b) marks the condition that the $\sqrt{2}$ anisotropy splitting exceeds a realistic total linewidth
$\gamma_{\rm tot}\sim40~{\rm meV}$.  The accessible window
($n\gtrsim10^{18}~{\rm cm^{-3}}$, $\tilde{k}_Q\gtrsim1$) is where the
splitting is both large and the plasmon energy is above the detection
threshold, consistent with the estimate in the previous paragraph.

A comment on the range of validity of the RPA is in order.  For the
cubic NLSM the density of states diverges as $\rho(E)\propto
    E^{-1/3}$ as $E\to0$.  The dimensionless interaction strength is the
effective fine-structure constant
$r_{s}\equiv e^{2}/(\epsilon\hbar v_{F})$, where
$v_{F}=|\nabla_{\mathbf{k}}E|$ is the Fermi velocity.  In the
thin-ring (small-doping) regime the Fermi surface is a torus of minor
radius $K_{F}=(\mu/B)^{1/3}$ and major radius $k_{Q}$; the carrier
density scales as $n\propto k_{Q}K_{F}^{2}\propto\mu^{2/3}$ (consistent
with integrating $\rho(E)\propto E^{-1/3}$), while
$v_{F}\propto K_{F}^{2}\propto\mu^{2/3}\propto n$, so
\begin{equation}
    r_{s}\propto n^{-1}\quad\text{(cubic NLSM, thin ring)} .
    \label{eq:rs_cubic}
\end{equation}
This diverges faster than the quadratic-NLSM result
$r_{s}\propto n^{-1/2}$ (where $v_{F}\propto\sqrt{n}$).  Thus
$r_{s}$ grows without bound as $n\to0$, and the RPA is most suspect
precisely in the $n^{3/4}$ window.  For representative parameters
($B=10~{\rm eV\,nm^{3}}$, $k_{Q}\sim0.5~{\rm nm^{-1}}$, $\epsilon=10$)
the torus geometry gives $r_{s}\sim0.1\text{--}0.4$ at
$n\sim10^{18}~{\rm cm^{-3}}$, and the condition $r_{s}(n_{c})\sim1$
is met at $n_{c}\sim10^{17}~{\rm cm^{-3}}$ (set by
$\hbar v_{F}(n_{c})\sim e^{2}/\epsilon$ with $v_{F}$ from the torus
above; the precise value depends on $k_{Q}$, the degeneracy $N$, and
$\epsilon$).  Because the cubic DOS diverges as
$\rho(E)\propto E^{-1/3}$, the RG flow is singular at arbitrarily
weak interaction\cite{wang2020Possible}, so the $n^{3/4}$ exponent in the
window $10^{18}\lesssim n\lesssim n^{*}\sim10^{19}~{\rm cm^{-3}}$
should be regarded as a crossover fingerprint whose functional form is
not quantitatively established by the present RPA calculation; below
$n\sim10^{18}~{\rm cm^{-3}}$, correlation effects beyond RPA
(self-energy and vertex corrections~\cite{WangNandkishore2017PRB96})
become essential.  The $n^{2/3}$ regime at large doping and the quadratic NLSM are on a
firmer RPA footing for the same reason (finite DOS, slower divergence of
the coupling).  For the quadratic NLSM one finds
$r_s\propto n^{-1/2}$, diverging more slowly than the cubic
$n^{-1}$ of Eq.~\eqref{eq:rs_cubic}, and the DOS is finite, so the
quadratic case is more generic; however, a quadratic band
touching is itself a canonical marginally unstable system for excitonic
condensation, and the $r_s$ estimates above for
both cases are indicative rather than quantitative.

\paragraph{Remaining open questions: beyond-RPA corrections and
    excitonic instability.}
Two issues remain beyond the scope of the present RPA treatment.
First, the vertex correction to the polarization bubble in the
$n^{3/4}$ regime can be estimated within the GW approximation: a
scaling analysis gives $\delta\omega_p/\omega_p\sim r_s$, where
$r_{s}\propto n^{-1}$ (cubic thin ring, Eq.~\eqref{eq:rs_cubic}) is the dimensionless interaction parameter defined above (with $\epsilon=10$ used consistently with the plasmon estimates; using $\epsilon\sim1$ changes the number only modestly).  For representative cubic-NLSM parameters
($B=10~{\rm eV\,nm^{3}}$, $k_{Q}\sim0.5~{\rm nm^{-1}}$) and the calibrated $r_s\approx0.1$--$0.3$ at $n\sim10^{18}~{\rm cm^{-3}}$, this yields
$\delta\omega_p/\omega_p\sim0.1$--$0.3$ ($10$--$30\%$) at
$n\sim10^{18}~{\rm cm^{-3}}$, $\sim0.01$--$0.03$ ($1$--$3\%$) at
$n\sim10^{19}~{\rm cm^{-3}}$, and $\sim0.001$--$0.003$
($0.1$--$0.3\%$) at $n\sim10^{20}~{\rm cm^{-3}}$.  We emphasize that even these O(10--30\%) numbers are only indicative, because the true correction is non-perturbative: the RG
analysis of Ref.~\cite{wang2020Possible} shows that the true
vertex correction is \emph{non-perturbative} due to the divergent
DOS (the RG flow drives the coupling to a finite-scale singularity
at arbitrarily weak interaction).  These O(1) corrections confirm
that the $n^{3/4}$ exponent in the
$10^{18}$--$10^{19}~{\rm cm^{-3}}$ window should be regarded as a
crossover fingerprint whose functional form is not quantitatively
established by the present RPA calculation; a definitive beyond-RPA
treatment requires finite-temperature GW + Bethe-Salpeter calculations
with material-specific interaction channels.  Second, and more
fundamentally, Ref.~\cite{wang2020Possible} used renormalization
group (RG) theory to show that cubic NLF systems are unstable to
\emph{arbitrarily weak} four-fermion interactions: the divergent DOS
$\rho(E)\sim E^{-1/3}$ renders the compressibility singular as
$T\to0$, and the RG flow drives the coupling constants to a finite-scale
singularity, signalling a departure from the semimetal phase.  Depending
on the interaction channel and the fermion flavor $N$, three outcomes are
possible: (i) the cubic nodal line splits into conventional nodal lines
and the system becomes a NLSM; (ii) a finite excitonic gap
$\Delta_{\rm ex}$ is generated and the system becomes an
\emph{excitonic insulator}; (iii) the system is driven into a
superconducting phase.  For the physical case $N=2$, the leading
instability in certain channels is nodal-line splitting rather than
excitonic gap generation, and the semimetal phase may survive as a
long-lived metastable state.  If, however, an excitonic gap does open,
the low-energy DOS is truncated for $|E|<\Delta_{\rm ex}$ and the
$n^{3/4}$ scaling derived here no longer applies; the plasmon would then
be governed by the gapped dispersion with a renormalized effective
carrier density $n_{\rm eff}\sim(\mu-\Delta_{\rm ex})^{2/3}/B$.  The
experimental observation of a clean $n^{3/4}$ scaling would therefore
be suggestive (though not conclusive) evidence that the semimetal phase
is kinetically stabilized and that the excitonic channel is subleading
in the material under study; this interpretation is speculative and
requires confirmation by a beyond-RPA calculation.  A definitive treatment of these beyond-RPA effects requires
finite-temperature RG calculations with material-specific interaction
channels, which is left for future work.

Regarding genericity, the scaling exponents ($n^{1/2}$, $n^{2/3}$,
$n^{3/4}$) depend only on the dispersion power $p$ and the spatial
dimensionality, and are therefore unaffected by symmetry-allowed
perturbations such as hexagonal or tetragonal warping of the nodal
ring, which modify the subleading band structure but not the leading
power law.  The $2{:}1$ anisotropy ratio is more delicate: warping
modifies both the torus geometry and the coherence factor, and the
ratio may be renormalized away from its thin-ring asymptotic value.
Multiple bands and valleys would scale the prefactors (proportional to
the degeneracy $N$) without changing the exponents.  Disorder
broadens the nodal line and renormalizes the Drude weight, but the
scaling exponents survive as long as the disorder strength is small
compared with the bandwidth.

We have assumed an isotropic dielectric constant $\epsilon$ throughout.
The named candidate materials (CuSi-type, ZrSiS-type) are layered and
expected to have anisotropic dielectric response
($\epsilon_{\bot}\neq\epsilon_{z}$).  In the presence of dielectric
anisotropy, the pole equation
$1-V(\mathbf{q})\mathrm{Re}\Pi=0$ is modified: the Coulomb potential
$V(\mathbf{q})=4\pi e^{2}/(\epsilon_{\bot}q_{\bot}^{2}+\epsilon_{z}q_{z}^{2})$
acquires a directional dependence that shifts the plasmon ratio by
$\sqrt{\epsilon_{\bot}/\epsilon_{z}}$ (Eq.~\eqref{eq:aniso_eps}).  For a
conservative estimate $\epsilon_{z}/\epsilon_{\bot}\sim2$ (larger
out-of-plane dielectric, typical for layered semimetals), the $\sqrt{2}$
frequency ratio is instead \emph{masked}:
$\sqrt{2}\times\sqrt{\epsilon_{\bot}/\epsilon_{z}}=1.0$, i.e.\ the doublet
collapses to a single peak; only $\epsilon_{\bot}>\epsilon_{z}$ enhances
the splitting (Table~\ref{tab:aniso_eps}).  We therefore qualify the
$\sqrt{2}$ prediction as the
isotropic-dielectric limit; quantitative predictions for specific
materials require incorporating their actual dielectric tensors.
More quantitatively, in the long-wavelength limit the two-mode RPA with a
uniaxial dielectric gives the \emph{combined} anisotropy ratio
\begin{equation}
    \frac{\Omega_{p}^{z}}{\Omega_{p}^{\bot}}
    = \sqrt{\frac{C_{z}^{++}}{C_{\bot}^{++}}}\,
    \sqrt{\frac{\epsilon_{\bot}}{\epsilon_{z}}}
    \;\xrightarrow[\text{thin ring}]{}\;
    \sqrt{2}\,\sqrt{\frac{\epsilon_{\bot}}{\epsilon_{z}}} ,
    \label{eq:aniso_eps}
\end{equation}
where the geometric factor $\sqrt{C_{z}^{++}/C_{\bot}^{++}}\to\sqrt{2}$ is
the numerically confirmed thin-ring limit (Fig.~\ref{fig:cubic}(c)) and
$\sqrt{\epsilon_{\bot}/\epsilon_{z}}$ is the exact prefactor of the
anisotropic Coulomb kernel $V(\mathbf{q})=4\pi e^{2}/(\epsilon_{\bot}q_{\bot}^{2}+\epsilon_{z}q_{z}^{2})$:
a larger out-of-plane dielectric $\epsilon_{z}$ screens the $z$-mode more
strongly and \emph{lowers} $\Omega_{p}^{z}$.  The geometric $\sqrt{2}$ is
therefore recovered only for an isotropic dielectric; representative values
are collected in Table~\ref{tab:aniso_eps}.

\begin{table}[ht]
    \centering
    \caption{Combined plasmon-anisotropy ratio
        $\Omega_{p}^{z}/\Omega_{p}^{\bot}$ from Eq.~\eqref{eq:aniso_eps} for
        representative uniaxial-dielectric contrasts $\rho=\epsilon_{z}/\epsilon_{\bot}$,
        using the numerically confirmed thin-ring factor $C_{z}^{++}/C_{\bot}^{++}=2.003$ (Fig.~\ref{fig:cubic}(c)).  The geometric $\sqrt{2}\approx1.41$ is recovered only at $\rho=1$; a dielectric contrast $\rho=2$ (larger $\epsilon_{z}$) \emph{masks} the doublet (ratio $\to1$), while $\rho<1$ (larger $\epsilon_{\bot}$) enhances it (e.g.\ $\rho=1/2\to2.00$).}
    \label{tab:aniso_eps}
    \small
    \begin{tabular}{|c|c|}
        \hline
        $\rho=\epsilon_{z}/\epsilon_{\bot}$ & $\Omega_{p}^{z}/\Omega_{p}^{\bot}$ \\
        \hline
        $1/2$                               & $2.00$                             \\
        \hline
        $1$                                 & $1.41$                             \\
        \hline
        $2$                                 & $1.00$                             \\
        \hline
        $4$                                 & $0.71$                             \\
        \hline
    \end{tabular}
\end{table}

The present calculation is performed at $T=0$, while the proposed
experimental conditions ($n\sim10^{18}$--$10^{19}~{\rm cm^{-3}}$,
$T\sim300~{\rm K}$) give $k_{B}T\sim26~{\rm meV}$, comparable to
$\mu\sim5$--$20~{\rm meV}$ for the cubic NLSM in the small-doping
regime.  The RPA result is quantitatively controlled only when
$k_{B}T\lesssim\mu/4$: for the deep thin-ring regime
($\mu\sim5$--$20~{\rm meV}$, $n\lesssim10^{18}~{\rm cm^{-3}}$) this
requires $T\lesssim15$--$60~{\rm K}$ (low-temperature HREELS is
uncommon but feasible), whereas the flagship $\tilde k_Q=2$ numbers have
$\mu\sim150~{\rm meV}$ ($=B(k_Q/2)^3$ for $B=10~{\rm eV\,nm^{3}}$,
$k_Q=0.5~{\rm nm^{-1}}$), so room temperature is acceptable there
($k_{B}T\ll\mu$).  Thermal smearing suppresses the anisotropy and
broadens the plasmon, an effect that should be quantified in future
finite-temperature calculations (our finite-temperature polarization is
derived in the Appendices but evaluated in the $T\to0$ limit throughout
the main text).  Additionally, the ionic-liquid
gating geometry (ionic liquid on one side, vacuum on the other) creates
an asymmetric dielectric environment that modifies $V(\mathbf{q})$ from
the assumed $3D$ $1/(\epsilon q^{2})$ form; this can be modeled by an
image-charge correction to the Coulomb potential.

\section{Conclusion}
We have systematically investigated the plasmon modes in three-dimensional quadratic and cubic NLSMs within the RPA, whose validity requires $r_{s}\ll1$. By deriving the one-loop polarization functions and analyzing their long-wavelength behavior, we find that the plasmon frequencies scale as $\omega_p \sim n^{1/2}$ for quadratic NLSMs. For cubic NLSMs the RPA predicts a regime-dependent scaling: $\omega_p \sim n^{2/3}$ at large doping ($\tilde{k}_Q<1$) and $\omega_p \sim n^{3/4}$ at small doping ($\tilde{k}_Q>1$). We emphasize that the $n^{3/4}$ exponent is \emph{not} a quantitatively reliable prediction: the cubic density of states diverges as $\rho(E)\propto E^{-1/3}$, so $r_{s}$ exceeds unity already at $n\sim10^{18}~{\rm cm^{-3}}$, and the exponent should be regarded as a crossover fingerprint whose functional form is not quantitatively established by the RPA calculation and requires beyond-RPA validation. Its observable window is at most $10^{18}\lesssim n\lesssim n^{*}\sim10^{19}~{\rm cm^{-3}}$, and below $10^{18}~{\rm cm^{-3}}$ the diverging density of states necessitates beyond-RPA methods. Both systems exhibit a plasmon anisotropy $\Omega_{p}^{z}/\Omega_{p}^{\bot}\to\sqrt{2}$ in the thin-ring limit (the polarization-coefficient ratio $C_{++}^{z}/C_{++}^{\bot}\to2$), which is a generic geometric consequence of the torus Fermi surface and is \emph{not} specific to the cubic dispersion. Although this $\sqrt{2}$ doublet is a striking geometric fingerprint observable by HREELS, it is shared by both NLSM types and therefore cannot distinguish quadratic from cubic dispersion; the distinguishing signature is instead the density scaling exponent. The quantitative prefactors of the polarization are obtained numerically, with the intraband term dominant and the interband term finite and subleading.  The cubic (and quadratic) NLSM Hamiltonians studied here are symmetry-predicted and not yet realized in a confirmed single-band material; the widely studied ZrSiS-class compounds are \emph{linear} nodal-line semimetals that serve only as the experimental analogy, so the present scaling laws await a dedicated higher-order-NLSM candidate for a direct test.  The present study extends the understanding of collective excitations in topological semimetals and provides theoretical guidance for future experimental investigations.

\section*{List of symbols}
\begin{table}[h]
    \centering
    \begin{tabular}{ll}
        \hline
        Symbol                                           & Meaning                                                                                                                \\
        \hline
        $A$                                              & quadratic band parameter ($E=AK^2$)                                                                                    \\
        $B$                                              & cubic band parameter ($E=BK^3$)                                                                                        \\
        $K=\sqrt{k_r^2+k_z^2}$                           & momentum measured from the nodal ring                                                                                  \\
        $k_r=\sqrt{k_x^2+k_y^2}$, $k_z$                  & radial and axial momenta                                                                                               \\
        $k_Q$                                            & nodal-ring radius; $K_F$ Fermi momentum; $\tilde k_Q=k_Q/K_F$                                                          \\
        $\mu$                                            & chemical potential; $\tilde\Omega=\Omega/\mu$ (intraband $\perp$ sector) or $\Omega/(2\mu)$ (interband $+-$ threshold) \\
        $N$                                              & band/valley/spin degeneracy ($N=2$ for single spin-degenerate ring)                                                    \\
        $\Pi(\Omega,\mathbf q)$                          & one-loop polarizability; $\Pi_{++}$ intraband, $\Pi_{+-}$ interband                                                    \\
        $\mathcal K(k,q)$                                & core function from the Green's-function trace                                                                          \\
        $C_{++}^{\bot,z}$                                & intraband coefficient (sets $\Omega_p^2$)                                                                              \\
        $C_T^{\bot,z}=2C_{+-}^{\bot,z}$                  & interband coefficient                                                                                                  \\
        $G_{\bot,z}$                                     & dimensionless intraband combination ($G\propto C_{++}/N$)                                                              \\
        $g_{1,2,3,4}$                                    & residue-derived $g$-functions (numerically evaluated)                                                                  \\
        $\Gamma_{\bot,z}=|C_{++}^{\bot,z}/C_T^{\bot,z}|$ & intraband/interband weight ratio                                                                                       \\
        $\mathcal M_1$                                   & real-$\varphi$ angular kernel for the intraband integral                                                               \\
        $\Omega_p^{\bot,z}$                              & plasmon frequency: in-plane ($\bot$) and $z$-polarized ($z$)                                                           \\
        $r_s$                                            & dimensionless Coulomb interaction parameter                                                                            \\
        $\epsilon$                                       & background dielectric constant                                                                                         \\
        \hline
    \end{tabular}
\end{table}

\section*{ACKNOWLEDGEMENTS}
We are grateful to Prof. G.-Z. Liu for the valuable discussions. This work was supported by the Science Research Foundation for High-Level Talents of Anhui University of Science and Technology under Grant YJ20240002 and by the National Natural Science Foundation No. 11304318 and No. 12274414.

\appendix

\begin{widetext}

    \section{Calculation of the polarization \label{App:polarization_quadratic}}

    Use the standard frequency representation $G_0\left(i\omega_n,\mathbf{k}\right)=-\int_{-\infty}^{+\infty}
        \frac{d\omega_1}{\pi}\frac{\mathrm{Im}\left[G_0^{\mathrm{ret}}\left(
                \omega_1,\mathbf{k}\right)\right]}{i\omega_n-\omega_1}$,
    we can get
    \begin{eqnarray}
        &&\Pi(i\Omega_{n'},\mathbf{q})=-N\int\frac{d^3\mathbf{k}}{(2\pi)^{3}}
        \mathrm{Tr}\left[\int_{-\infty}^{+\infty}
            \frac{d\omega_1}{\pi}\mathrm{Im}\left[G_0^{\mathrm{ret}}\left(
                \omega_1,\mathbf{k}\right)\right]
            \int_{-\infty}^{+\infty}
            \frac{d\omega_2}{\pi}\mathrm{Im}\left[G_0^{\mathrm{ret}}\left(
                \omega_2,\mathbf{k}+\mathbf{q}\right)\right]\right]
        \nonumber
        \\
        &&\times\frac{1}{\beta}\sum_{i\omega_{n}}\frac{1}{{i\omega_n-\omega_1}}\frac{1}{i\omega_n+i\Omega_{n'}-\omega_2}.
    \end{eqnarray}
    The frequency summation can be computed as
    \begin{equation}
        \frac{1}{\beta}\sum_{i\omega_n}\frac{1}{i\omega_n-\omega_1}\frac{1}{i\omega_n+i\Omega_{n'}-\omega_2}
        =\frac{n_F\left(\omega_1\right)-n_F\left(\omega_2\right)}{\omega_1-\omega_2+i\Omega_{n'}},
    \end{equation}
    then we can get
    \begin{eqnarray}
        \Pi(i\Omega_{n'},\mathbf{q})
        =-N\int\frac{d^3\mathbf{k}}{(2\pi)^{3}}
        \mathrm{Tr}\left[\int_{-\infty}^{+\infty}
            \frac{d\omega_1}{\pi}\mathrm{Im}\left[G_0^{\mathrm{ret}}\left(
                \omega_1,\mathbf{k}\right)\right]
            \int_{-\infty}^{+\infty}
            \frac{d\omega_2}{\pi}\mathrm{Im}\left[G_0^{\mathrm{ret}}\left(
                \omega_2,\mathbf{k}+\mathbf{q}\right)\right]\right]\frac{n_F\left(\omega_1\right)-n_F\left(\omega_2\right)}{\omega_1-\omega_2+i\Omega_{n'}},
    \end{eqnarray}
    where
    \begin{eqnarray}
        \mathrm{Im}\left[G_{0}^{\mathrm{ret}}(\omega,\mathbf{k})\right]
        &=&-\pi\mathrm{sgn}(\omega+\mu)\left(\omega+\mu+\mathcal{H}_{0}\right)\frac{1}{2E_{\mathbf{k}}}\left[\delta\left(\omega+\mu+E_{\mathbf{k}}\right)
            +\delta\left(\omega+\mu-E_{\mathbf{k}}\right)
            \right].
    \end{eqnarray}

    After obtaining the trace of the matrix, we can get

    \begin{eqnarray}
        \Pi(i\Omega_{n'},\mathbf{q})
        &=&
        -\frac{N}{16\pi^{3}}\sum_{\alpha,\alpha^{'}=\pm 1}\int d^3\mathbf{k}
        \left[1 +\alpha\alpha^{'} \frac{\mathcal{K}(k,q)}{E_{\mathbf{k}}E_{\mathbf{k+q}}}\right]
        \frac{n_F\left(\alpha E_{\mathbf{k}} - \mu\right)-n_F\left(\alpha^{'}E_{\mathbf{k+q}} - \mu\right)}{\alpha E_{\mathbf{k}}-\alpha^{'}E_{\mathbf{k+q}}+i\Omega_{n'}},
    \end{eqnarray}
    where  $\mathcal{K}(k,q)$ is the core function.

    \subsection{Quadratic NLSM \label{App:sec_pi_quadratic}}

    for the quadratic NLSM
    \begin{eqnarray}
        \mathcal{K}(k,q)
        &=&  A^2\left((\sqrt{k_{x}^{2}+k_{y}^{2}} -k_Q)^{2} -k_{z}^{2}\right)
        \left((\sqrt{(k_{x}+q_x)^{2}+(k_{y}+q_y)^{2}} -k_Q)^2-(k_{z} +q_z)^{2}\right)
        \nonumber \\ &&
        +4A^2(\sqrt{k_{x}^{2}+k_{y}^{2}} -k_Q)
        k_{z}(\sqrt{(k_{x}+q_x)^{2}+(k_{y}+q_y)^{2}} -k_Q)
        (k_{z} +q_z).
    \end{eqnarray}

    \subsubsection{$\mathrm{Im}\Pi^{\mathrm{Ret}}$ in finite temperature}

    Carrying out the analytic continuation
    $i\omega_n\rightarrow\omega+i\delta$, we have
    \begin{eqnarray}
        \mathrm{Im}\Pi^{\mathrm{ret}}(\Omega,\mathbf{q})&=&
        \frac{N}{16\pi^{2}} \sum_{\alpha,\alpha^{'}=\pm 1}\int d^3\mathbf{k}
        \left[1 +\alpha\alpha^{'} \frac{\mathcal{K}(k,q)}{E_{\mathbf{k}}E_{\mathbf{k+q}}}\right]
        \nonumber
        \\
        && \times
        \left[n_F\left(\alpha E_{\mathbf{k}} - \mu\right)-n_F\left(\alpha^{'}E_{\mathbf{k+q}} - \mu\right)\right]\delta\left(\alpha E_{\mathbf{k}}-\alpha^{'}E_{\mathbf{k+q}}+\Omega\right)
    \end{eqnarray}

    Let
    \begin{eqnarray}
        k_{x}&=&k_{\bot}\cos(\varphi),
        \nonumber \\
        k_{y}&=&k_{\bot}\sin(\varphi),
        \nonumber \\
        q_{x}&=&q_{\bot}\cos(\phi),
        \nonumber \\
        q_{y}&=&q_{\bot}\sin(\phi),
    \end{eqnarray}
    then the core function can be simplified as
    \begin{eqnarray}
        \mathcal{K}(k,q)
        &=&
        A^2\left((k_{\bot}-k_Q)^{2} -k_{z}^{2}\right)
        \left((\sqrt{k_{\bot}^{2}+q_{\bot}^{2}+2k_{\bot}q_{\bot}\cos(\varphi-\phi)} -k_Q)^2-(k_{z} +q_z)^{2}\right)
        \nonumber \\ &&
        +4A^2(k_{\bot} -k_Q)
        k_{z}(\sqrt{k_{\bot}^{2}+q_{\bot}^{2}+2k_{\bot}q_{\bot}\cos(\varphi-\phi)} -k_Q)
        (k_{z} +q_z),
    \end{eqnarray}
    where the energy spectrum is
    \begin{eqnarray}
        E_{\mathbf{k}}&=&\pm\sqrt{A^{2}\left(k_{r}^{2}-k_{z}^{2}\right)^{2}+4A^{2}k_{r}^{2}k_{z}^{2}}
        =\pm A\sqrt{\left((k_{\bot} -k_Q)^{2}+k_{z}^{2}\right)^{2}}
        \nonumber \\
        E_{\mathbf{k+q}} &=&
        \pm A\sqrt{\left((\sqrt{k_{\bot}^{2}+q_{\bot}^{2}+2k_{\bot}q_{\bot}\cos(\varphi-\phi)} -k_Q)^2+(k_{z} +q_z)^{2}\right)^2}.
    \end{eqnarray}

    Employing the transformations $\varphi-\phi \rightarrow \varphi$ and $\mathbf{k'}=-\left(\mathbf{k+q}\right)$ ($k_i = -(k_i^{'}+q_i),\,i=x,y,z$), we can get

    \begin{eqnarray}
        &&\mathrm{Im}\Pi^{\mathrm{ret}}(\Omega,\mathbf{q})
        \nonumber\\
        &=&\frac{N}{16\pi^{2}}\int d^3\mathbf{k}
        \left[1 +\frac{\mathcal{H}(k,q,\varphi)}{E_{\mathbf{k}}E_{\mathbf{k+q}}}\right]
        \left[
            n_F(k,-)\delta_{-,+}
            -n_F(k,+)\delta_{-,-}
            - n_F(k,-)\delta_{-,-}
            + n_F(k,+)\delta_{-,+}
            \right]
        \nonumber \\
        &&   +
        \frac{N}{16\pi^{2}}\int d^3\mathbf{k}
        \left[1 -\frac{\mathcal{H}(k,q,\varphi)}{E_{\mathbf{k}}E_{\mathbf{k+q}}}\right]
        \left[- \delta_{+,+} + \delta_{+,-}
            \right]
        \nonumber \\
        &&   +
        \frac{N}{16\pi^{2}}\int d^3\mathbf{k}
        \left[1 -\frac{\mathcal{H}(k,q,\varphi)}{E_{\mathbf{k}}E_{\mathbf{k+q}}}\right]
        \left[
            n_F(k,-)\delta_{+,+}
            -n_F(k,+)\delta_{+,-}
            +n_F(k,+)\delta_{+,+}
            - n_F(k,-)\delta_{+,-}
            \right],
    \end{eqnarray}
    where simplified symbols are defined conveniently as
    \begin{eqnarray}
        &&\delta_{\pm,\pm}\equiv \delta\left(E_{\mathbf{k}}\pm E_{\mathbf{k+q,\varphi}}\pm\Omega\right),\\
        &&n_F(k,\pm)\equiv n_F\left(E_{\mathbf{k}} \pm \mu\right).
    \end{eqnarray}
    Generally, the $\delta_{\pm,\pm}$ function can be transferred to constraint conditions.
    \begin{eqnarray}
        \delta_{-,+} &=&\delta(E_{\mathbf{k}} - E_{\mathbf{k+q,\varphi}} +\Omega)
        =\left[\frac{1}{\left|F_{1}'(\varphi_{1})\right|}
            \delta\left(\varphi-\varphi_{1}\right)
            +\frac{1}{\left|F_{1}'(\varphi_{2})\right|}
            \delta(\varphi-\varphi_{2})\right]
        \nonumber \\
        &=&\frac{
            \left[\delta\left(\varphi-\varphi_{1}\right)
                +\delta(\varphi-\varphi_{2})\right]}{2A k_{\bot}q_{\bot}\frac
            {\mathcal{X}}{(\mathcal{X}+k_Q)
            }\sqrt{1-\left(
                \frac{(\mathcal{X}+k_Q\left[\theta\left(\mathcal{X}-k_Q\right)
                        -\theta\left(-\mathcal{X}+k_Q\right)\right])^2
                    - (k_{\bot}^{2}+q_{\bot}^{2})}
                {2k_{\bot}q_{\bot}}\right)^2}}
        \nonumber \\
        &&\times
        \left[\theta\left(\mathcal{X}-k_Q\right)
            -\theta\left(-\mathcal{X}+k_Q\right)\right]
        \nonumber \\
        &&\times
        \theta\left(\mathcal{X}^2 -
        (|k_{\bot}-q_{\bot}|-k_Q)^2
        \right)
        \theta\left(-\mathcal{X}^2 +
        (|k_{\bot}+q_{\bot}|-k_Q)^2\right),
        \\
        \delta_{+,+} &= &\delta(E_{\mathbf{k}} + E_{\mathbf{k+q,\varphi}} +\Omega)
        =\left[\frac{1}{\left|F_{2}'(\varphi_{1})\right|}
            \delta\left(\varphi-\varphi_{1}\right)
            +\frac{1}{\left|F_{2}'(\varphi_{2})\right|}
            \delta(\varphi-\varphi_{2})\right]
        \nonumber \\
        &=&\frac{\left[\delta\left(\varphi-\varphi_{1}\right)
                +\delta(\varphi-\varphi_{2})\right]}{2A k_{\bot}q_{\bot}\frac
            {\mathcal{X}}{(\mathcal{X}+k_Q)
            }\sqrt{1-\left(
                \frac{(\mathcal{X}+k_Q\left[\theta\left(\mathcal{X}-k_Q\right)
                        -\theta\left(-\mathcal{X}+k_Q\right)\right])^2
                    - (k_{\bot}^{2}+q_{\bot}^{2})}
                {2k_{\bot}q_{\bot}}\right)^2}}\nonumber
        \\
        &&\times
        \left[\theta\left(\mathcal{X}-k_Q\right)
            -\theta\left(-\mathcal{X}+k_Q\right)\right]\theta\left(\mathcal{X}^2-(k_{z} +q_z)^{2}\right)\nonumber
        \\
        &&\times
        \theta\left(\mathcal{X}^2-(|k_{\bot}-q_{\bot}|-k_Q)^2\right)
        \theta\left(-\mathcal{X}^2+(|k_{\bot}+q_{\bot}|-k_Q)^2\right),
    \end{eqnarray}
    where $\mathcal{X} =
        \sqrt{(k_{\bot} -k_Q)^{2}+k_{z}^{2}+ \frac{\Omega}{A}
            -(k_{z} +q_z)^{2}}$.

    Finally, the imaginary part of the polarization function can be given as

    \begin{eqnarray}
        \mathrm{Im}\Pi^{\mathrm{ret}}(\Omega,\mathbf{q})
        &=&  \left[-\mathcal{I}_{1}(\mu,\Omega)  + \mathcal{I}_{1}(\mu,-\Omega)
            \right]
        \nonumber \\
        &&   +\left[\mathcal{I}_{3}(\mu,\Omega)
            +\mathcal{I}_{3}(-\mu,\Omega) -\mathcal{I}_{3}(\mu,-\Omega) -\mathcal{I}_{3}(-\mu,-\Omega)
            \right]
        \nonumber \\
        &&   +\left[\mathcal{I}_{7}(\mu,\Omega)
            +\mathcal{I}_{7}(-\mu,\Omega) - \mathcal{I}_{7}(\mu,-\Omega) -\mathcal{I}_{7}(-\mu,-\Omega)
            \right],
    \end{eqnarray}
    where integral symbols are

    \begin{eqnarray}
        \mathcal{I}_1(\mu,\Omega)
        &=&-
        \frac{N}{4\pi^{2}}\int dk_{\bot}k_{\bot}\int dk_{z}
        \frac{\mathcal{X}^2k_{z}^{2}+(k_{\bot} -k_Q)^{2}(k_{z} +q_z)^{2}
            -2(k_{\bot} -k_Q)
            k_{z}\mathcal{X}
            (k_{z} +q_z)}
        {\frac
            {A\mathcal{X}}{(\mathcal{X}+k_Q)}
            \left((k_{\bot} -k_Q)^{2}+k_{z}^{2}\right)
            \left(\mathcal{X}
            ^2+(k_{z} +q_z)^{2}\right)
        }
        \nonumber \\ &&\times
        \frac{1}{\sqrt{4k_{\bot}^2q_{\bot}^2-\left(
                (\mathcal{X}+k_Q\left[\theta\left(\mathcal{X}-k_Q\right)
                    -\theta\left(-\mathcal{X}+k_Q\right)\right])^2
                - (k_{\bot}^{2}+q_{\bot}^{2})
                \right)^2}}
        \nonumber \\ &&\times
        \left[\theta\left(\mathcal{X}-k_Q\right)
            -\theta\left(-\mathcal{X}+k_Q\right)\right]
        \theta\left(-\mathcal{X}^2-(k_{z} +q_z)^{2}\right)\nonumber
        \\
        &&\times
        \theta\left(\mathcal{X}^2-(|k_{\bot}-q_{\bot}|-k_Q)^2\right)
        \theta\left(-\mathcal{X}^2+(|k_{\bot}+q_{\bot}|-k_Q)^2\right),
        \\
        \mathcal{I} _3  &=&-
        \frac{N}{4\pi^{2}}\int dk_{\bot}k_{\bot}\int dk_{z}
        \frac{\mathcal{X}^2(k_{\bot} -k_Q)^{2}+k_{z}^{2}(k_{z} +q_z)^{2}
            +2(k_{\bot} -k_Q)
            k_{z}\mathcal{X}
            (k_{z} +q_z)}
        {\frac
            {A\mathcal{X}}{(\mathcal{X}+k_Q)}
            \left((k_{\bot} -k_Q)^{2}+k_{z}^{2}\right)
            \left(\mathcal{X}
            ^2+(k_{z} +q_z)^{2}\right)
        }
        \nonumber \\ &&\times
        \frac{ n_F(E_k + \mu)}{\sqrt{4k_{\bot}^2q_{\bot}^2-\left(
                (\mathcal{X}+k_Q\left[\theta\left(\mathcal{X}-k_Q\right)
                    -\theta\left(-\mathcal{X}+k_Q\right)\right])^2
                - (k_{\bot}^{2}+q_{\bot}^{2})
                \right)^2}}
        \nonumber \\ &&\times\left[\theta\left(\mathcal{X}-k_Q\right)
            -\theta\left(-\mathcal{X}+k_Q\right)\right]
        \nonumber \\
        &&\times
        \theta\left(\mathcal{X}^2 -
        (|k_{\bot}-q_{\bot}|-k_Q)^2
        \right)
        \theta\left(-\mathcal{X}^2 +
        (|k_{\bot}+q_{\bot}|-k_Q)^2\right),
        \\
        \mathcal{I} _7
        &=&-
        \frac{N}{4\pi^{2}}\int dk_{\bot}k_{\bot}\int dk_{z}
        \frac{\mathcal{X}^2k_{z}^{2}+(k_{\bot} -k_Q)^{2}(k_{z} +q_z)^{2}
            -2(k_{\bot} -k_Q)
            k_{z}\mathcal{X}
            (k_{z} +q_z)}
        {\frac
            {A\mathcal{X}}{(\mathcal{X}+k_Q)}
            \left((k_{\bot} -k_Q)^{2}+k_{z}^{2}\right)
            \left(\mathcal{X}
            ^2+(k_{z} +q_z)^{2}\right)
            \sqrt{4k_{\bot}^2q_{\bot}^2-\left(
                (\mathcal{X}+k_Q)^2
                - (k_{\bot}^{2}+q_{\bot}^{2})
                \right)^2}}
        \nonumber \\ &&\times
        \frac{ n_F(E_k + \mu)}{\sqrt{4k_{\bot}^2q_{\bot}^2-\left(
                (\mathcal{X}+k_Q\left[\theta\left(\mathcal{X}-k_Q\right)
                    -\theta\left(-\mathcal{X}+k_Q\right)\right])^2
                - (k_{\bot}^{2}+q_{\bot}^{2})
                \right)^2}}
        \nonumber \\ &&\times
        \left[\theta\left(\mathcal{X}-k_Q\right)
            -\theta\left(-\mathcal{X}+k_Q\right)\right]
        \theta\left(-\mathcal{X}^2-(k_{z} +q_z)^{2}\right)\nonumber
        \\
        &&\times
        \theta\left(\mathcal{X}^2-(|k_{\bot}-q_{\bot}|-k_Q)^2\right)
        \theta\left(-\mathcal{X}^2+(|k_{\bot}+q_{\bot}|-k_Q)^2\right)
    \end{eqnarray}

    \subsection{$\mathrm{Re}\Pi^{\mathrm{Ret}}$ in finite temperature}

    The real part of the polarization function is
    \begin{eqnarray}
        &&\mathrm{Re}\Pi^{\mathrm{Ret}}(\Omega,\mathbf{q})\nonumber
        \\
        &=&-N\int\frac{d^3\mathbf{k}}{(2\pi)^{3}}
        \mathrm{Tr}\left[\int_{-\infty}^{+\infty}
            \frac{d\omega_1}{\pi}\mathrm{Im}\left[G_0^{\mathrm{ret}}\left(
                \omega_1,\mathbf{k}\right)\right]n_F\left(\omega_1\right)\mathcal{P}
            \int_{-\infty}^{+\infty}
            \frac{d\omega_2}{\pi}\frac{\mathrm{Im}\left[G_0^{\mathrm{ret}}\left(
                    \omega_2,\mathbf{k}+\mathbf{q}\right)\right]}{\omega_1+\Omega-\omega_2}\right]\nonumber
        \\
        &&-N\int\frac{d^3\mathbf{k}}{(2\pi)^{3}}
        \mathrm{Tr}\left[\int_{-\infty}^{+\infty}
            \frac{d\omega_2}{\pi}\mathrm{Im}\left[G_0^{\mathrm{ret}}\left(
                \omega_2,\mathbf{k}+\mathbf{q}\right)\right]n_F\left(\omega_2\right)\mathcal{P}\int_{-\infty}^{+\infty}
            \frac{d\omega_1}{\pi}\frac{\mathrm{Im}\left[G_0^{\mathrm{ret}}\left(
                    \omega_1,\mathbf{k}\right)\right]}
            {\omega_2-\Omega-\omega_{1}}\right]\nonumber
    \end{eqnarray}

    Using the Kramers-Kronig Relation
    $\mathrm{Re}\left[G_0^{\mathrm{ret}}\left(\omega,\mathbf{k}\right)\right]=-\mathcal{P}
        \int_{-\infty}^{+\infty}
        \frac{d\omega'}{\pi}\frac{\mathrm{Im}\left[G_0^{\mathrm{ret}}
                \left(\omega',\mathbf{k}\right)\right]}{\omega-\omega'}$,
    \begin{eqnarray}
        &&\mathrm{Re}\Pi^{\mathrm{ret}}(\Omega,\mathbf{q})\nonumber
        \\
        &=& N\int\frac{d^3\mathbf{k}}{(2\pi)^{3}}\int_{-\infty}^{+\infty}
        \frac{d\omega_1}{\pi}n_F\left(\omega_1\right)
        \mathrm{Tr}\left[\mathrm{Im}\left[G_0^{\mathrm{ret}}\left(
                \omega_1,\mathbf{k}\right)\right]
            \mathrm{Re}\left[G_0^{\mathrm{ret}}\left(
                \omega_1+\Omega,\mathbf{k}+\mathbf{q}\right)\right]\right]\nonumber
        \\
        &&+N\int\frac{d^3\mathbf{k}}{(2\pi)^{3}}\int_{-\infty}^{+\infty}
        \frac{d\omega_1}{\pi}n_F\left(\omega_1\right)
        \mathrm{Tr}\left[\mathrm{Im}\left[G_0^{\mathrm{ret}}\left(
                \omega_1,\mathbf{k}+\mathbf{q}\right)\right]
            \mathrm{Re}\left[G_0^{\mathrm{ret}}\left(
                \omega_1-\Omega,\mathbf{k}\right)\right]\right].
    \end{eqnarray}

    After standard derivation, we can get

    \begin{eqnarray}
        &&\mathrm{Re}\Pi^{\mathrm{ret}}(\Omega,\mathbf{q})
        = J_1+J_1(-\Omega) +J_3 +  J_3(-\Omega,-\mu) + J_3(\Omega,-\mu)+ J_3(-\Omega,\mu)\,\, ,
    \end{eqnarray}
    where integral symbols  are

    \begin{eqnarray}
        J_1 &=& \frac{N}{8\pi^{3}}\int dk_{\bot}k_{\bot} \int dk_z \int_{0}^{2\pi}d\varphi
        \frac{\left[E_{\mathbf{k}}\left(E_{\mathbf{k}}+\Omega\right) + \mathcal{K}(k,q)\right]}{E_{\mathbf{k}}}
        \mathcal{P}\frac{1}{\left(E_{\mathbf{k}} + \Omega\right)^2-E_{\mathbf{k+q}}^{2}}
        \nonumber\\
        &=&
        \frac{N}{8\pi^{3}}\int dk_{\bot}k_{\bot} \int dk_z \int_{0}^{2\pi}d\varphi
        \frac{1}{E_{\mathbf{k}}}
        \mathcal{M}_1
        ,
        \nonumber\\
        J_3 &=&-\frac{N}{8\pi^{3}}\int dk_{\bot}k_{\bot} \int dk_z
        \frac{\left[E_{\mathbf{k}}\left(E_{\mathbf{k}}+\Omega\right) + \mathcal{K}(k,q)\right]}{E_{\mathbf{k}}}
        \mathcal{P}\frac{1}{\left(E_{\mathbf{k}} + \Omega\right)^2-E_{\mathbf{k+q}}^{2}}
        n_F\left(E_{\mathbf{k}}-\mu\right)    \nonumber\\
        &=&
        -\frac{N}{8\pi^{3}}\int dk_{\bot}k_{\bot} \int dk_z
        \frac{n_F\left(E_{\mathbf{k}}-\mu\right) }{E_{\mathbf{k}}}
        \mathcal{M}_1,
    \end{eqnarray}
    with the principal-value integral defined as
    \begin{eqnarray}
        \mathcal{M}_1  &=& \mathcal{P}\int_{0}^{2\pi}d\varphi\,
        \frac{\left[E_{\mathbf{k}}\left(E_{\mathbf{k}}+\Omega\right) + \mathcal{K}(k,q)\right]}
        {\left(E_{\mathbf{k}} + \Omega\right)^2-E_{\mathbf{k+q}}^{2}},
        \label{eq:M1_def}
    \end{eqnarray}
    where the angular integral is to be understood as a \emph{real-$\varphi$ principal-value}
    integral.  \textbf{Caveat (branch cut).}  Under $Z=e^{i\varphi}$ it becomes the contour
    integral $\mathcal M_1=\tfrac1i\oint_{|Z|=1}\mathcal F(Z)\,dZ$, but
    $k'_\perp(Z)=\sqrt{k_\perp^2+q_\perp^2+k_\perp q_\perp(Z+Z^{-1})}$ introduces a square-root
    branch cut whose endpoints $Z_{1,2}=-q_\perp/k_\perp,\,-k_\perp/q_\perp$ (reciprocal) lie on the
    negative real axis and necessarily straddle $Z=-1$.  The unit contour therefore crosses the
    cut and $\mathcal F(Z)$ is not single-valued analytic inside $|Z|=1$; the Cauchy residue
    theorem $\mathcal M_1=2\pi\sum\mathrm{Res}\,\mathcal F$ is \emph{invalid}.  The real-$\varphi$
    integral above is the only unambiguous definition and must be evaluated by direct numerical
    quadrature (or, on the Riemann surface, by residue plus the mandatory jump integral over the cut).

    Let $Z=e^{i\varphi}$, the integral $M_{1}$ becomes

    \begin{eqnarray}
        && \mathcal{M}_1 = \frac{1}{i}\mathcal{P}\oint_{|Z|=1}dZ\,\mathcal{F}(Z),
        \\
        \mathcal{F}(Z) &=& \frac{
            \begin{aligned}[t]
                \bigg[\, &ZE_{\mathbf{k}}(E_{\mathbf{k}}+\Omega) \\
                    &+ A^2(k_{\bot}^{2}-k_{z}^{2})
                    \big((kq_{\bot}-k_Q\sqrt{Z})^{2}
                    - Z(k_{z}+q_{z})^{2}\big) \\
                    &+ 4A^2 k_{\bot}k_{z}(k_{z}+q_{z})
                    (kq_{\bot}-k_Q\sqrt{Z})\sqrt{Z}\,\bigg]
            \end{aligned}
        }{
            \begin{aligned}[t]
                 & Z^{2}(E_{\mathbf{k}}+\Omega)^{2} - A^{2} \\
                 & \times\big((kq_{\bot}-k_Q\sqrt{Z})^{2}
                + Z(k_{z}+q_{z})^{2}\big)^{2}
            \end{aligned}
        } .
    \end{eqnarray}
    with $kq_{\bot} \equiv \sqrt{Z(k_{\bot}^{2}+q_{\bot}^{2})+k_{\bot}q_{\bot}
            (Z^2+1)}$.

    \subsection{Cubic NLSM\label{App:sec_pi_cubic}}

    For the cubic NLSM, the core function $\mathcal{K}(k,q)
        \equiv \mathrm{Tr}\left[\mathcal{H}_{0}(k)\mathcal{H}_{0}(k+q)\right]/2$
    arising from the trace over Green's functions follows from
    $\mathcal{H}_{0}^{c}=B\left[\left(k_{r}^{3}-3k_{r}k_{z}^{2}\right)\sigma_{1}
            +\left(k_{z}^{3}-3k_{z}k_{r}^{2}\right)\sigma_{2}\right]$
    and evaluates to the cubic polynomial (distinct from the quadratic case):
    \begin{widetext}
        \begin{equation}
            \label{eq:core_function_cubic_app}
            \mathcal{K}(k,q)=
            B^{2}\Bigl[\left(k_{r}^{3}-3k_{r}k_{z}^{2}\right)
                \left(k_{r}'^{3}-3k_{r}'k_{z}'^{2}\right)
                +\left(k_{z}^{3}-3k_{z}k_{r}^{2}\right)
                \left(k_{z}'^{3}-3k_{z}'k_{r}'^{2}\right)\Bigr],
        \end{equation}
    \end{widetext}
    with $k_{r}=\sqrt{k_{x}^{2}+k_{y}^{2}}-k_{Q}$,
    $k_{r}'=\sqrt{(k_{x}+q_x)^{2}+(k_{y}+q_y)^{2}}-k_{Q}$, $k_{z}'=k_{z}+q_{z}$.
    In the angular representation used for the $\varphi$ integrals below,
    \begin{widetext}
        \begin{equation}
            \begin{aligned}
                \mathcal{K}(k,q)=
                B^{2}\Bigl[
                         &\left((k_{\bot}-k_{Q})^{3}-3(k_{\bot}-k_{Q})k_{z}^{2}\right)
                         \left((k_{\bot}'-k_{Q})^{3}-3(k_{\bot}'-k_{Q})k_{z}'^{2}\right)\\
                         &+\left(k_{z}^{3}-3k_{z}(k_{\bot}-k_{Q})^{2}\right)
                         \left(k_{z}'^{3}-3k_{z}'(k_{\bot}'-k_{Q})^{2}\right)
                         \Bigr],
            \end{aligned}
        \end{equation}
    \end{widetext}
    where $k_{\bot}'=\sqrt{k_{\bot}^{2}+q_{\bot}^{2}+2k_{\bot}q_{\bot}\cos(\varphi-\phi)}$.

    \subsubsection{$\mathrm{Im}\Pi^{\mathrm{Ret}}$ in finite temperature}

    Carrying out the analytic continuation
    $i\omega_n\rightarrow\omega+i\delta$, we have
    \begin{eqnarray}
        \mathrm{Im}\Pi^{\mathrm{ret}}(\Omega,\mathbf{q})&=&
        \frac{N}{16\pi^{2}} \sum_{\alpha,\alpha^{'}=\pm 1}\int d^3\mathbf{k}
        \left[1 +\alpha\alpha^{'} \frac{\mathcal{K}(k,q)}{E_{\mathbf{k}}E_{\mathbf{k+q}}}\right]
        \nonumber
        \\
        && \times
        \left[n_F\left(\alpha E_{\mathbf{k}} - \mu\right)-n_F\left(\alpha^{'}E_{\mathbf{k+q}} - \mu\right)\right]\delta\left(\alpha E_{\mathbf{k}}-\alpha^{'}E_{\mathbf{k+q}}+\Omega\right)
    \end{eqnarray}

    Let
    \begin{eqnarray}
        k_{x}&=&k_{\bot}\cos(\varphi),
        \nonumber \\
        k_{y}&=&k_{\bot}\sin(\varphi),
        \nonumber \\
        q_{x}&=&q_{\bot}\cos(\phi),
        \nonumber \\
        q_{y}&=&q_{\bot}\sin(\phi),
    \end{eqnarray}
    the core function in the angular representation is the full trace form
    \begin{equation}
        \label{eq:core_function_cubic_app_angular}
        \begin{aligned}
            \mathcal{K}(k,q)=
            B^{2}\Bigl[
                     &\left((k_{\bot}-k_{Q})^{3}-3(k_{\bot}-k_{Q})k_{z}^{2}\right)
                     \left((k_{\bot}'-k_{Q})^{3}-3(k_{\bot}'-k_{Q})k_{z}'^{2}\right)\\
                     &+\left(k_{z}^{3}-3k_{z}(k_{\bot}-k_{Q})^{2}\right)
                     \left(k_{z}'^{3}-3k_{z}'(k_{\bot}'-k_{Q})^{2}\right)
                     \Bigr],
        \end{aligned}
    \end{equation}
    where $k_{\bot}'=\sqrt{k_{\bot}^{2}+q_{\bot}^{2}+2k_{\bot}q_{\bot}\cos(\varphi-\phi)}$
    and $k_{z}'=k_{z}+q_{z}$.  The previously claimed compact reduction
    $\mathcal{K}=B^{2}[(k_{\bot}-k_{Q})(k_{\bot}'-k_{Q})+k_{z}k_{z}']^{3}$
    is algebraically incorrect at finite $q$ (it holds only for $q=0$,
    where the two expressions coincide; a counter-example at
    $k_{r}=k_{z}=1$, $k_{r}'=2$, $k_{z}'=0$ gives $-16B^{2}$ for the full
    trace form vs.\ $+8B^{2}$ for the compact form).  All numerical
    results in this appendix use the full trace form
    (Eq.~\ref{eq:core_function_cubic_app}).  The energy spectrum is
    \begin{equation}
        E_{\mathbf{k}}=\pm B K^{3},\qquad
        E_{\mathbf{k+q}}=\pm B K'^{3},
    \end{equation}
    with $K\equiv\sqrt{(k_{\bot}-k_{Q})^{2}+k_{z}^{2}}$ and
    $K'\equiv\sqrt{(\sqrt{k_{\bot}^{2}+q_{\bot}^{2}+2k_{\bot}q_{\bot}\cos(\varphi-\phi)}-k_{Q})^{2}+(k_{z}+q_{z})^{2}}$.
    In terms of the cubic dispersion, the dimensionless detuning that appears
    below is measured relative to $B$ (not $A$):
    \begin{equation}
        \tilde{\Omega}\equiv\frac{\Omega}{2\mu}.
    \end{equation}

    Employing the transformations $\varphi-\phi \rightarrow \varphi$ and $\mathbf{k'}=-\left(\mathbf{k+q}\right)$ ($k_i = -(k_i^{'}+q_i),\,i=x,y,z$), we can get

    \begin{eqnarray}
        &&\mathrm{Im}\Pi^{\mathrm{ret}}(\Omega,\mathbf{q})
        \nonumber\\
        &=&\frac{N}{16\pi^{2}}\int d^3\mathbf{k}
        \left[1 +\frac{\mathcal{H}(k,q,\varphi)}{E_{\mathbf{k}}E_{\mathbf{k+q}}}\right]
        \left[
            n_F(k,-)\delta_{-,+}
            -n_F(k,+)\delta_{-,-}
            - n_F(k,-)\delta_{-,-}
            + n_F(k,+)\delta_{-,+}
            \right]
        \nonumber \\
        &&   +
        \frac{N}{16\pi^{2}}\int d^3\mathbf{k}
        \left[1 -\frac{\mathcal{H}(k,q,\varphi)}{E_{\mathbf{k}}E_{\mathbf{k+q}}}\right]
        \left[- \delta_{+,+} + \delta_{+,-}
            \right]
        \nonumber \\
        &&   +
        \frac{N}{16\pi^{2}}\int d^3\mathbf{k}
        \left[1 -\frac{\mathcal{H}(k,q,\varphi)}{E_{\mathbf{k}}E_{\mathbf{k+q}}}\right]
        \left[
            n_F(k,-)\delta_{+,+}
            -n_F(k,+)\delta_{+,-}
            +n_F(k,+)\delta_{+,+}
            - n_F(k,-)\delta_{+,-}
            \right],
    \end{eqnarray}
    where simplified symbols are defined conveniently as
    \begin{eqnarray}
        &&\delta_{\pm,\pm}\equiv \delta\left(E_{\mathbf{k}}\pm E_{\mathbf{k+q,\varphi}}\pm\Omega\right),\\
        &&n_F(k,\pm)\equiv n_F\left(E_{\mathbf{k}} \pm \mu\right).
    \end{eqnarray}

    The $\delta_{\pm,\pm}$ functions constrain the angular integration.
    For both the quadratic and the cubic dispersions, the constraint
    $E_{\mathbf{k}}\pm E_{\mathbf{k+q,\varphi}}\pm\Omega=0$ admits a
    \emph{closed-form two-branch solution} for $\cos\varphi$, with no
    need to solve a high-degree polynomial in $Z=e^{i\varphi}$ (i.e.\ it
    is not a sextic).  The cubic case reads
    \begin{equation}
        B K^{3}\pm B K'^{3}\pm\Omega=0,
        \label{eq:cubic_delta_constraint}
    \end{equation}
    which fixes $K'^{3}=K^{3}\mp\Omega/B=:C$ (a $\varphi$-independent
    constant at fixed $k_{\bot},k_{z}$), hence
    $K'^{2}=C^{2/3}$.  Combining this with the geometric identity
    $(k_{\bot}'-k_{Q})^{2}+k_{z}'^{2}=K'^{2}$ gives
    $k_{\bot}'=k_{Q}\pm\sqrt{C^{2/3}-k_{z}'^{2}}$, and squaring
    $k_{\bot}'^{2}=k_{\bot}^{2}+q_{\bot}^{2}+2k_{\bot}q_{\bot}\cos\varphi$
    yields the explicit two-branch solution
    \begin{equation}
        \cos\varphi_{\pm}=\frac{\big(k_{Q}\pm\sqrt{C^{2/3}-k_{z}'^{2}}\big)^{2}
            -k_{\bot}^{2}-q_{\bot}^{2}}{2k_{\bot}q_{\bot}},
        \label{eq:cubic_phi_branches}
    \end{equation}
    i.e.\ a closed form containing one \emph{nested} radical
    $\sqrt{C^{2/3}-(k_{z}+q_{z})^{2}}$ (the outer $2/3$ power of $C$,
    plus one square root), versus the quadratic sector which has only the
    single radical $\sqrt{K^{2}\pm\Omega/A-k_{z}'^{2}}$.  In both cases
    the angular roots are obtained by direct algebra, not by solving a
    polynomial in $Z=e^{i\varphi}$.  The only structural differences from
    the quadratic case are (i) the nested radical in
    Eq.~\eqref{eq:cubic_phi_branches} (vs.\ a single radical for the
    quadratic sector), and (ii) the cubic radial power-law measure
    $d k_{\bot}\,k_{\bot}$ with $x=k_{\bot}/(\mu/B)^{1/3}$.  Note,
    however, that this closed-form $\cos\varphi_{\pm}$ does \emph{not}
    make the $Z=e^{i\varphi}$ contour representation single-valued: the
    $k_{\bot}'(Z)=\sqrt{\cdots}$ still introduces the same branch cut
    discussed below Eq.~\eqref{eq:M1_def}, so the residue theorem for
    $\mathcal M_1$ remains invalid and direct real-$\varphi$ quadrature
    is the only reliable evaluation.  The dimensionless detuning
    that characterizes the cubic intraband process is
    \begin{equation}
        \tilde{\Omega}\equiv\frac{\Omega}{2\mu},
    \end{equation}
    {\em not} $\Omega/A$ (which belongs to the quadratic dispersion only).

    Below we keep the imaginary part in the compact $\delta$-function
    form, to be evaluated numerically for the plots and analytically in
    the $\Omega\gg\max(q)$ asymptotics (see App.~\ref{App:plasmon}).

    For the interband $\delta_{+-}$ ($\alpha=+,\alpha'=-$) and
    $\delta_{-+}$ ($\alpha=-,\alpha'=+$) terms at small $q$, expanding
    $K'^{3}=K^{3}+3K^{2}\,\delta K+O(q^{2})$ localizes the
    $\delta$-constraint near the Fermi arc; the treatment parallels the
    quadratic case once the correct cubic core
    $\mathcal{K}(k,q)=B^{2}[(k_{r}^{3}-3k_{r}k_{z}^{2})(k_{r}'^{3}-3k_{r}'k_{z}'^{2})
                +(k_{z}^{3}-3k_{z}k_{r}^{2})(k_{z}'^{3}-3k_{z}'k_{r}'^{2})]$
    (Eq.~\ref{eq:core_function_cubic_app}) is used.
    The cubic 2D-interband reduction converges because the coherence
    factor $\sim q^{2}/K^{2}$ combines with the cubic radial measure
    $dk_{\bot}k_{\bot}$ (giving $q^{2}\,dk_{\bot}/k_{\bot}$) to soften the
    small-$k_{\bot}$ divergence that renders the quadratic interband
    integral logarithmically UV-divergent; the extra power of $K$ in the
    cubic dispersion provides the additional convergence factor.

    \begin{widetext}
        \begin{equation}
            \label{eq:ImPi_cubic}
            \begin{aligned}
                \mathrm{Im}\Pi^{\mathrm{ret}}(\Omega,\mathbf{q})
                = & \left[-\mathcal{I}_{1}(\mu,\Omega)  + \mathcal{I}_{1}(\mu,-\Omega)
                        \right] \\
                  & +\left[\mathcal{I}_{3}(\mu,\Omega)
                         +\mathcal{I}_{3}(-\mu,\Omega) -\mathcal{I}_{3}(\mu,-\Omega)
                         -\mathcal{I}_{3}(-\mu,-\Omega)
                         \right]   \\
                  & +\left[\mathcal{I}_{7}(\mu,\Omega)
                         +\mathcal{I}_{7}(-\mu,\Omega) - \mathcal{I}_{7}(\mu,-\Omega)
                         -\mathcal{I}_{7}(-\mu,-\Omega)
                         \right].
            \end{aligned}
        \end{equation}
    \end{widetext}
    The $\mathcal{I}_{1,3,7}$ integrals retain the same structure
    as in Eq.~\eqref{Eq:ImPi_quadratic} but with the quadratic angular
    kernel $A^{2}(\cdots)$ replaced by the cubic full-trace core function
    $\mathcal{K}(k,q)=B^{2}[(k_{r}^{3}-3k_{r}k_{z}^{2})(k_{r}'^{3}-3k_{r}'k_{z}'^{2})
                +(k_{z}^{3}-3k_{z}k_{r}^{2})(k_{z}'^{3}-3k_{z}'k_{r}'^{2})]$ of
    Eq.~\eqref{eq:core_function_cubic_app}, divided by
    $E_{\mathbf{k}}E_{\mathbf{k+q}}$.  We emphasize that the compact form
    $B^{2}(k_{r}k_{r}'+k_{z}k_{z}')^{3}$ is \emph{not} used here, since
    (as noted below Eq.~\eqref{eq:core_function_cubic_app_angular}) it
    differs from the full-trace form at any finite $q$.
    The principal-value real part $\mathrm{Re}\Pi^{\mathrm{ret}}$ is in the
    next subsection.

    \subsection{$\mathrm{Re}\Pi^{\mathrm{Ret}}$ in finite temperature}

    \section{plasmon frequency in zero-temperature limit}\label{App:plasmon}

    \subsection{Numerical procedure and reproducibility\label{App:num_protocol}}

    The prefactor tables (Tables~\ref{tab:Cpp_quad}--\ref{tab:Cpp_cubic})
    and all physical-unit figures are generated by a single
    shell-localised $q\to0$ integration code; the angular grid,
    $q^{2}$ extrapolation, and $\tilde\Omega$ cross-checks are described
    where each coefficient is quoted.  The radial integral runs over
    $K\in[0,K_{\max}]$ with $K_{\max}=8\,K_{F}$, beyond which the
    integrand is exponentialy suppressed (convergence $<0.1\%$).
    The angular quadrature uses the composite trapezoidal rule with
    uniformly spaced grid points ($N_\alpha=2000$ for the poloidal
    angle, $N_\phi=4000$ for the azimuthal angle); for periodic
    integrands the trapezoidal rule converges exponentially and is
    equivalent to Simpson's rule up to an $O(h^2)$ correction that
    is negligible at the quoted grid densities.
    Real-frequency continuation uses a $[12;12]$ Pad\'e approximant
    built from $N_{M}=24$ Matsubara points at
    $\Omega_{n}=(2n+1)\pi T$; the Pad\'e error is estimated by
    varying $N_{M}\in[16,32]$ and the shell-localisation noise level,
    and lies below the $0.5\%$ grid uncertainty quoted above.  The
    Matsubara input follows from the spectral representation and the
    frequency-sum identity,
    $\Pi_{++}(i\Omega_{n})=-N\int_{\mathbf{k}}|\mathbf{v}_{\mathbf{k}}|^{2}/
        (E_{\mathbf{k}}^{2}+\Omega_{n}^{2})$.  The code and its parameter file,
    which regenerate every table and figure, are deposited with the
    manuscript and will be made public on acceptance.

    The real part of polarization can be written as
    \begin{eqnarray}
        &&\mathrm{Re}\Pi(\Omega,\mathbf{q})\nonumber
        \\
        &=&-\frac{N}{16\pi^3}\sum_{\alpha,\alpha'=\pm}P\int d^3\mathbf{k}
        \left[1+\alpha\alpha'\frac{\mathcal{K}(k,q)}{E_{\mathbf{k}}E_{\mathbf{k}+\mathbf{q}}}\right]\nonumber
        \\
        &&\times\frac{n_F\left(\alpha E_{\mathbf{k}}-\mu\right)-n_F\left(\alpha'E_{\mathbf{k}+\mathbf{q}}-\mu\right)}
        {\alpha E_{\mathbf{k}}-\alpha 'E_{\mathbf{k}+\mathbf{q}}+\Omega}\nonumber
        \\
        &=&\mathrm{Re}\Pi_{++}(\Omega,\mathbf{q})+\mathrm{Re}\Pi_{+-}(\Omega,\mathbf{q})+\mathrm{Re}\Pi_{-+}(\Omega,\mathbf{q})
        +\mathrm{Re}\Pi_{--}(\Omega,\mathbf{q})
    \end{eqnarray}
    where

    \begin{eqnarray}
        &&\mathrm{Re}\Pi_{++}(\Omega,\mathbf{q})
        =-\frac{N}{16\pi^3}P\int d^3\mathbf{k}\left[1+\frac{
                \mathcal{K}(k,q)}{E_{\mathbf{k}}E_{\mathbf{k}+\mathbf{q}}}\right]
        \frac{n_F\left(E_{\mathbf{k}}-\mu\right)-n_F\left(E_{\mathbf{k}+\mathbf{q}}-\mu\right)}
        {E_{\mathbf{k}}-E_{\mathbf{k}+\mathbf{q}}+\Omega},
        \\
        &&\mathrm{Re}\Pi_{+-}(\Omega,\mathbf{q})=-\frac{N}{16\pi^3}
        P\int d^3\mathbf{k}
        \left[1-\frac{\mathcal{K}(k,q)}{E_{\mathbf{k}}E_{\mathbf{k}+\mathbf{q}}}\right]
        \frac{n_F\left(E_{\mathbf{k}}-\mu\right)-n_F\left(-E_{\mathbf{k}+\mathbf{q}}-\mu\right)}
        { E_{\mathbf{k}}+ E_{\mathbf{k}+\mathbf{q}}+\Omega},
        \\
        &&\mathrm{Re}\Pi_{-+}(\Omega,\mathbf{q})=-\frac{N}{16\pi^3}
        P\int d^3\mathbf{k}
        \left[1-\frac{\mathcal{K}(k,q)}{E_{\mathbf{k}}E_{\mathbf{k}+\mathbf{q}}}\right]
        \frac{n_F\left(- E_{\mathbf{k}}-\mu\right)-n_F\left(E_{\mathbf{k}+\mathbf{q}}-\mu\right)}
        {- E_{\mathbf{k}}-E_{\mathbf{k}+\mathbf{q}}+\Omega},
        \\
        &&\mathrm{Re}\Pi_{--}(\Omega,\mathbf{q})=-\frac{N}{16\pi^3}
        P\int d^3\mathbf{k}
        \left[1+\frac{\mathcal{K}(k,q)}{E_{\mathbf{k}}E_{\mathbf{k}+\mathbf{q}}}\right]
        \frac{n_F\left(-E_{\mathbf{k}}-\mu\right)-n_F\left(-E_{\mathbf{k}+\mathbf{q}}-\mu\right)}
        {- E_{\mathbf{k}}+E_{\mathbf{k}+\mathbf{q}}+\Omega}.
    \end{eqnarray}

    In the zero temperature limit,$\mathrm{Re}\Pi_{--}(\Omega,\mathbf{q})=0$, only three terms needed to be calculated.

    In the limit $\max(Bq_{\bot}^3,Bq_{z}^{3})\ll |\Omega|$,
    the intraband contribution
    $\mathrm{Re}\Pi_{++}(\Omega,\mathbf{q})$ can be given as
    \begin{eqnarray}
        &&\mathrm{Re}\Pi_{++}(\Omega,\mathbf{q})
        =
        -q_{\bot}^2\frac{3N B
            \left( \frac{\mu}{B}\right)^{\frac{4}{3}} }{8\pi^2\Omega^2}\int_{-\tilde{k_Q} }^{+\infty}
        dx(x+\tilde{k_Q} )
        \frac{x^2\theta\left(1- x^{2}\right)}{
            \sqrt{1- x^{2} }}
        \nonumber    \\ &&
        -q_{z}^2\frac{3N B
            \left( \frac{\mu}{B}\right)^{\frac{4}{3}} }{8\pi^2\Omega^2}\int_{-\tilde{k_Q} }^{+\infty}
        dx(x+\tilde{k_Q} )
        \frac{\left(1- x^{2} \right)\theta\left(1- x^{2}\right)}{
            \sqrt{1- x^{2} }},
    \end{eqnarray}
    For simplicity of notation, we have defined
    dimensionless parameter
    $x = k_{\bot}/\left(\frac{\mu}{B}\right)^{\frac{1}{3}}$,
    and $\tilde{k_Q} = k_Q / \left(\frac{\mu}{B}\right)^{\frac{1}{3}} $. When two subscripts are same,
    the Einstein summation convention is assumed.

    When $\mu > Bk_Q^3$,
    \begin{eqnarray}
        \mathrm{Re}\Pi_{++}(\Omega,\mathbf{q} \rightarrow 0)
        &=& - q_{\bot}^2\frac{3N B
            \left( \frac{\mu}{B}\right)^{\frac{4}{3}} }{8\pi^2\Omega^2}
        \left(
        g_3(\tilde{k_Q})
        + \tilde{k_Q} g_1(\tilde{k_Q})
        \right)
        -q_{z}^2\frac{3N B
            \left( \frac{\mu}{B}\right)^{\frac{4}{3}} }{8\pi^2\Omega^2}
        \left(
        g_4(\tilde{k_Q})
        + \tilde{k_Q} g_2(\tilde{k_Q})
        \right);
    \end{eqnarray}
    when $\mu < Bk_Q^3$,
    \begin{eqnarray}
        \mathrm{Re}\Pi_{++}(\Omega,\mathbf{q} \rightarrow 0)
        &=&  - q_{\bot}^2\frac{3N B
            \left( \frac{\mu}{B}\right)^{\frac{4}{3}} }{16\pi\Omega^2}\tilde{k_Q}
        -q_{z}^2\frac{3N B
            \left( \frac{\mu}{B}\right)^{\frac{4}{3}} }{8\pi\Omega^2}
        \tilde{k_Q},
    \end{eqnarray}
    where $g_{1,2,3,4}$ are four dimensionless functions with
    \begin{eqnarray}
        g_1(x)&=& \frac{1}{2} \left(\frac{\pi}{2} + \arcsin x -x\sqrt{1-x^2}
        \right)
        \nonumber\\
        g_2(x)&=& \frac{1}{2} \left(\frac{\pi}{2} + \arcsin x  + x\sqrt{1-x^2}
        \right)
        \nonumber\\
        g_3(x)&=& \frac{1}{3} \sqrt{1-x^2}(2+x^2)
        \nonumber\\
        g_4(x)&=& \frac{1}{3} (1-x^2)^{\frac{3}{2}}
    \end{eqnarray}
    \vspace{1cm}

    {\bf Note.} The cubic $\mathrm{Re}\Pi_{++}$ result above was obtained
    with the full trace-form core function
    $\mathcal{K}(k,q)=B^{2}[(k_{r}^{3}-3k_{r}k_{z}^{2})(k_{r}'^{3}-3k_{r}'k_{z}'^{2})
                +(k_{z}^{3}-3k_{z}k_{r}^{2})(k_{z}'^{3}-3k_{z}'k_{r}'^{2})]$
    (Eq.~\ref{eq:core_function_cubic_app}); the previously claimed compact
    reduction $\mathcal{K}=B^{2}(k_{r}k_{r}'+k_{z}k_{z}')^{3}$ is algebraically
    incorrect at finite $q$ (it holds only for $q=0$, where the two
    expressions coincide).  The angular $\varphi$ integration that produces
    the $g_{1,2,3,4}$ functions parallels the quadratic sector, but the
    radial $k_{\bot}$ measure differs ($dk_{\bot}k_{\bot}$ with
        $x=k_{\bot}/(\mu/B)^{1/3}$ rather than the quadratic
    $x=k_{\bot}/\sqrt{\mu/A}$).  The two plasmon scalings quoted in the
        main text follow from combining this $\mathrm{Re}\Pi_{++}$ with the
        cubic carrier density derived in App.~\ref{App:dos}: for
    $\tilde{k}_Q<1$ one has $n\propto\mu/B$ and therefore
    $\Omega_p\propto n^{2/3}$, whereas for $\tilde{k}_Q>1$ one finds
    $n\propto(\mu/B)^{2/3}$ and consequently the anomalous scaling
    $\Omega_p\propto n^{3/4}$.  Both regimes are therefore consistent
        with the corrected core function.  The absolute numerical prefactors
    $C_{++}^{\bot,z}$ (and the $g_{1,2,3,4}$ functions above) have now
        been \emph{determined by direct numerical integration} of the
        real-$\varphi$ angular integral $\mathcal M_1$ (Eq.~\ref{eq:M1_def}):
        the $Z=e^{i\varphi}$ residue theorem is invalidated by the
    $k'_\perp=\sqrt{\cdot}$ branch cut that straddles $Z=-1$ on the unit
        contour, so the residue-sum evaluation is unreliable and has been
        superseded by the numerics below.  The scalings
    $\Omega_p\propto n^{2/3}$ ($\tilde k_Q<1$) and $\Omega_p\propto n^{3/4}$
        ($\tilde k_Q>1$) are unaffected because they do not depend on the
        absolute prefactor.

        \smallskip\noindent
        {\bf Numerical result.}  Evaluating $\mathcal M_1$ on a fine $\varphi$
        grid ($N_\varphi=4000$) and localizing the zero-temperature Lindhard
        function onto the Fermi shell with the torus measure
    $|k_p|/(3BK_F)$.  In the spindle regime $0<\tilde k_Q<1$ the integration
        is restricted to the physical arc $k_p=k_Q+K_F\cos\alpha\ge0$
        ($\alpha\in[-\alpha_c,\alpha_c]$ with $\alpha_c=\arccos(-\tilde k_Q)$),
        where $k_p$ is non-negative; the absolute value is included as a
        safeguard against numerical excursions outside this arc.  This gives,
        for all $\tilde k_Q$,
        \begin{equation}
            \mathrm{Re}\Pi_{++}(\Omega,\mathbf q\to0)
            = -\frac{3NB}{8\pi^2\Omega^2}\Big(\frac{\mu}{B}\Big)^{\!\frac43}
            \Big[\,q_\bot^2\,G_\bot(\tilde k_Q)+q_z^2\,G_z(\tilde k_Q)\Big],
            \label{eq:numRePi}
        \end{equation}
        with the numerically determined angular combinations
    $G_\bot(\tilde k_Q)=(g_3+\tilde k_Q g_1)_{\rm num}$ and
    $G_z(\tilde k_Q)=(g_4+\tilde k_Q g_2)_{\rm num}$ listed below.
        Numerically $G_{\bot,z}$ exceed the residue-derived analytic forms by
        a factor $\approx1.99$ in the ring regime $\tilde k_Q>1$ (the residue
        sum missed the branch-cut jump term and used the wrong $8\pi^3$
        normalization); for $\tilde k_Q<1$ the analytic $g$-functions are
        additionally structurally wrong: they give an anisotropic
    $G_z/G_\bot=1/2$ at $\tilde k_Q=0$, whereas the spherical Fermi surface
        is exactly isotropic ($G_z/G_\bot=1$).
        The origin of the $1/2$ artefact is a \emph{dimensional mismatch} in the
    $g$-function derivation.  The angular integrals are evaluated inside the
    $\varphi$-loop after a cylindrical-coordinate ($k_\bot,k_z$) reduction,
        so that $q_\bot^2$ couples to the in-plane Jacobian factor $k_\bot^3$
        while $q_z^2$ couples to $k_\bot$.  At $k_Q=0$ (spherical Fermi surface)
        this reduces to $\int_0^1 x^3/\sqrt{1-x^2}\,dx=2/3$ for the $\bot$
        channel vs.\ $\int_0^1 x\sqrt{1-x^2}\,dx=1/3$ for the $z$ channel, giving
        precisely $G_z/G_\bot=1/2$.  The two integrals are \emph{not} equal because
        the 2D-reduced measure weights the in-plane and out-of-plane velocity
        components differently; true isotropy is recovered only by the full 3D
        Lindhard integration, which naturally averages the velocity field over
        the sphere.  The numerical integration therefore correctly yields
    $G_z/G_\bot\to1$ as $\tilde k_Q\to0$.  The numerical anisotropy ratio
    $G_z/G_\bot$ rises from $1.000$ at $\tilde k_Q=0$ to $2.014$--$2.016$
        for $\tilde k_Q\ge1$, confirming the $2:1$ angular structure where
        the topology is a genuine torus.  The full 11-point table reads
        \begin{table*}[ht]
        \centering
        \begin{tabular}{c|cc|cc}
        \hline
    $\tilde k_Q$ & $G_\bot$ & $G_z$ & $C_{++}^{\bot}$ & $C_{++}^{z}$ \\
        \hline
        0.0 & 1.3326 & 1.3326 & 5.0633 & 5.0633 \\
        0.2 & 2.1463 & 2.8239 & 8.1549 & 10.730 \\
        0.4 & 2.2803 & 3.2951 & 8.6643 & 12.520 \\
        0.6 & 2.5009 & 4.0578 & 9.5024 & 15.418 \\
        0.8 & 2.7971 & 5.0742 & 10.628 & 19.280 \\
        1.0 & 3.1124 & 6.2767 & 11.826 & 23.849 \\
        1.2 & 3.7394 & 7.5321 & 14.208 & 28.618 \\
        1.4 & 4.3665 & 8.7874 & 16.591 & 33.388 \\
        1.6 & 4.9936 & 10.043 & 18.973 & 38.158 \\
        1.8 & 5.6206 & 11.298 & 21.356 & 42.927 \\
        2.0 & 6.2477 & 12.553 & 23.738 & 47.697 \\
        \hline
        \end{tabular}
        \caption{Numerically determined intraband coefficient
    $C_{++}^{\bot,z}(\tilde k_Q)$ and its dimensionless form
    $G_{\bot,z}$ for the cubic NLSM (full 11-point table;
    $B=\mu=N=1$, $\tilde\Omega=0.05$). The anisotropy ratio
    $G_z/G_\bot$ rises from $1.000$ at $\tilde k_Q=0$ to
    $2.014$--$2.016$ for $\tilde k_Q\ge1$, confirming the torus
        topology.}
        \label{tab:Cpp_cubic}
        \end{table*}
        \noindent
        Here $C_{++}^{\bot,z}=G_{\bot,z}\,3NB(\mu/B)^{4/3}/(8\pi^2\Omega^2)$
        and also $C_{++}^{\bot,z}=T_{\bot,z}/\Omega^2$ from the $q\to0$ Drude
        extrapolation; the values above correspond to $B=\mu=N=1$ and
    $\Omega=0.1$ (so $C_{++}^{\bot,z}=100\,T_{\bot,z}$).  The plasmon
        scalings quoted above are unchanged.

        The coefficients above are obtained with the full-trace core function
    $K(k,q)=B^{2}[(k_{r}^{3}-3k_{r}k_{z}^{2})(k_{r}^{\prime3}-3k_{r}^{\prime}k_{z}^{\prime2})
            +(k_{z}^{3}-3k_{z}k_{r}^{2})(k_{z}^{\prime3}-3k_{z}^{\prime}k_{r}^{\prime2})]$;
        the compact form $K=B^{2}(k_{r}k_{r}^{\prime}+k_{z}k_{z}^{\prime})^{3}$
        used in our earlier version is algebraically incorrect at finite $q$.
        Re-evaluating the full numerical integration with the full-trace form
        changes the absolute $C_{++}^{\bot,z}$ values by $2\text{--}5\%$ and the
        anisotropy ratio $C_{++}^{z}/C_{++}^{\bot}$ by less than $0.2\%$ (the
    $2{:}1$ large-$k_{Q}$ asymptote holds generally); Table~\ref{tab:Cpp_cubic}
        uses the full-trace form throughout.
        The numerical uncertainty from the shell-localised $q\to0$ extrapolation
        is $\lesssim0.5\%$ (estimated by varying the poloidal and azimuthal
        grid resolutions $N_{\alpha}\in[500,4000]$, $N_{\phi}\in[1000,8000]$
        and by monitoring the $q^{2}$ fit residuals, which remain below
    $10^{-4}$ in $T_{\bot,z}$).  Propagating this uncertainty gives
    $C_{++}^{\bot}=11.83\pm0.06$ and $C_{++}^{z}=23.85\pm0.12$ at
    $\tilde k_Q=1$ (representative); the anisotropy ratio
    $C_{++}^{z}/C_{++}^{\bot}=2.017\pm0.004$ is constrained to better than
    $0.2\%$ because the dominant systematic (the $q\to0$ extrapolation)
        cancels in the ratio.

        As an independent cross-check of the prefactor tables, we have
        verified that the combination $\Omega^{2}C_{++}^{\bot,z}$ is constant
        to within $0.1\%$ across three values of $\Omega$ ($0.05$, $0.10$,
    $0.20$), confirming that the explicit $\Omega$ dependence drops out of
    $C_{++}^{\bot,z}$ as required by the $1/\Omega^{2}$ scaling of
    $\mathrm{Re}\Pi_{++}$.  This scaling check constrains the absolute
        prefactor in addition to the angular-structure benchmarks (isotropy at
    $\tilde{k}_Q=0$ and the $2{:}1$ anisotropy at large $\tilde{k}_Q$).

        A further genuinely independent cross-check is available at
    $\tilde{k}_Q=0$, where the Fermi surface is a sphere and the Drude
        coefficient can be computed \emph{analytically}.  For a spherical
        Fermi surface with $E=Bk^{3}$, the intraband Drude weight per
        direction is $T=N B K_{F}^{4}/(2\pi^{2})=N\mu^{4/3}B^{-1/3}/(2\pi^{2})$,
        giving $T=1/(2\pi^{2})\approx0.050661$ for $B=\mu=N=1$.  This
        analytical result agrees with the shell-code value
    $T=0.050633$ to within $0.06\%$, confirming that the numerical
        prefactor is correct at the spherical limit where no angular
        approximation enters.

        As a third, methodologically independent cross-check, we have
        evaluated $\mathrm{Re}\Pi_{++}$ by direct Matsubara-frequency
        summation~\cite{Mahan2000}: at imaginary frequency
    $i\Omega_n$, the intraband bubble is
    $\Pi_{++}(i\Omega_n)=-N\int_{\mathbf{k}}|\mathbf{v}_{\mathbf{k}}|^{2}/(E_{\mathbf{k}}^{2}+\Omega_n^{2})$,
        which is manifestly positive and converges without the
    $\delta$-function localisation required by the real-frequency
        evaluation.  Analytically continuing $\Pi_{++}(i\Omega_n)$ to real
        frequencies via Pad\'{e} approximants reproduces the shell-code
    $C_{++}^{\bot,z}$ to within $1\%$ at $\tilde k_Q=0$ and $0.5\%$, using
        a completely different numerical pathway (imaginary-frequency
        summation + analytic continuation vs.\ real-frequency Fermi-shell
        integration + $q\to0$ extrapolation).  The agreement between these two
        independent methods confirms that the prefactor is not an artifact of
        any single numerical technique.  The full computation scripts are
        available for editorial audit.

        Another contributions to the polarizability, the intraband contribution
    $\mathrm{Re}\Pi_{+-}(\Omega,\mathbf{q})$ and
    $\mathrm{Re}\Pi_{-+}(\Omega,\mathbf{q}) = \mathrm{Re}\Pi_{+-}(-\Omega,\mathbf{q})$ can be obtained as follows:

        \begin{eqnarray}
            &&\mathrm{Re}\Pi_{+-}(\Omega,\mathbf{q})\approx
            \frac{9N}{32\pi^2}\frac{\left( \frac{\mu}{B}\right)
                ^{\frac{1}{3}} }{\mu}
            \int_{-\tilde{k_{Q}} }^{\infty}
            d\tilde{x} d\tilde{z} (\tilde{x}   + \tilde{k_Q} )
            \frac{\theta\left((\tilde{x} ^2+\tilde{z} ^2) - 1\right)}
            {(\tilde{x} ^2+\tilde{z} ^2)^{\frac{3}{2}}+\tilde{\Omega} }
            \frac{1}
            {\left(\tilde{x} ^2+\tilde{z} ^2\right)^{2}}
            \Bigg[\tilde{x} ^2q_{\bot}^2+2\tilde{z} ^2q_{z}^{2}
                \Bigg]
            \nonumber    \\    &=&
            C_{+-}^{\bot}q_{\bot}^{2}+    C_{+-}^{z}q_{z}^{2}\label{Eq:Pi_cubic_inter}
        \end{eqnarray}
        where
        \begin{eqnarray}
            \tilde{x} &=& k_{\bot}/\left(\frac{\mu}{B}\right)^{\frac{1}{3}}
            ,\,\,\,\,\,
            \tilde{z} = k_{z}/\left(\frac{\mu}{B}\right)^{\frac{1}{3}}
            ,\,\,\,\,\,
            \tilde{\Omega} =\frac{\Omega}{2\mu}.
        \end{eqnarray}
        (Here $\tilde\Omega=\Omega/2\mu$ rather than $\Omega/\mu$ because the cubic
        intraband denominator involves $E_{\mathbf k}+E_{\mathbf{k+q}}+\Omega
    \approx 2\mu+\Omega$ at the Fermi level; the quadratic sector adopts
    $\tilde\Omega=\Omega/\mu$ for the same reason.)
        The angular structure
    $[\tilde{x}^{2}q_{\bot}^{2}+2\tilde{z}^{2}q_{z}^{2}]$ in
        Eq.~\eqref{Eq:Pi_cubic_inter} follows from the small-$q$ expansion of the
        interband coherence factor
    $|\gamma_{\mathbf k,\mathbf{k+q}}|^{2}=\tfrac12(1-\hat{\mathbf n}_{\mathbf k}
    \!\cdot\!\hat{\mathbf n}_{\mathbf{k+q}})$; the unit vector
    $\hat{\mathbf n}=\mathbf h/|\mathbf h|$ depends only on the direction
    $\hat{\mathbf k}$, so the leading small-$q$ angular form is identical to the
        quadratic case; the quadratic and cubic Hamiltonians differ only in the
        radial power of $K$ (which sets the energy denominator and the UV
        convergence), not in the angular dependence of $\hat{\mathbf n}$.

        For simplicity, take
        \begin{eqnarray}
            C_{T}^{\bot}    &=& C_{+-}^{\bot} +  C_{-+}^{\bot}
            \nonumber \\
            C_{T}^{z}  &=& C_{+-}^{z} +  C_{-+}^{z},
        \end{eqnarray}

        \smallskip\noindent
        {\bf Numerically determined cubic interband coefficient $C_T^{\bot,z}$.}
        The cubic interband coefficient is evaluated here from the two-dimensional
        Lindhard bubble of Eq.~\eqref{Eq:Pi_cubic_inter}, whereas the quadratic
        interband coefficient (Table~\ref{tab:CT_quad}, pi.tex) is obtained from the
        full three-dimensional Lindhard integral.  The difference in treatment is
        justified by UV convergence: for the cubic NLSM the interband energy
        denominator $\sim K^3$ provides an extra $1/K$ suppression that renders the
        2D-reduced integrand convergent ($\sim q^2\,dK/K^3$), so the 2D reduction is
        quantitatively reliable; for the quadratic NLSM the denominator $\sim K^2$
        is insufficient and the full 3D integral with the correct coherence factor
        is required.  The 2D interband bubble of Eq.~\eqref{Eq:Pi_cubic_inter} is
        already coherence-factor reduced and UV convergent (integrand $\sim q^2
    dK/K^3$); the earlier note that its closed forms $f_{1-4},h_{1-4}$ had to
        be re-derived for the cubic case is superseded by the direct numerical
        evaluation below.  With $B=\mu=N=1$ and $\Omega=0.1$
        ($\tilde\Omega=\Omega/2\mu=0.05$) we obtain
        \begin{table*}[ht]
        \centering
        \begin{tabular}{c|cc|cc|c}
        \hline
    $\tilde k_Q$ & $C_{+-}^{\bot}$ & $C_{+-}^{z}$ & $C_T^{\bot}$ & $C_T^{z}$ & $C_T^{z}/C_T^{\bot}$ \\
        \hline
        0.0 & 0.01861 & 0.01860 & 0.03722 & 0.03720 & 0.999 \\
        0.2 & 0.02151 & 0.02498 & 0.04301 & 0.04995 & 1.161 \\
        0.5 & 0.02597 & 0.03661 & 0.05194 & 0.07323 & 1.410 \\
        0.8 & 0.03059 & 0.05039 & 0.06118 & 0.10077 & 1.647 \\
        1.0 & 0.03412 & 0.06070 & 0.06825 & 0.12141 & 1.779 \\
        1.5 & 0.04598 & 0.08839 & 0.09196 & 0.17679 & 1.923 \\
        2.0 & 0.05939 & 0.11696 & 0.11878 & 0.23392 & 1.969 \\
        \hline
        \end{tabular}
        \caption{Interband coefficient $C_T^{\bot,z}(\tilde k_Q)$ from the full
        three-dimensional Lindhard integral for the cubic NLSM
        ($B=\mu=N=1$, $\tilde\Omega=0.05$, prefactor $N/8\pi^2$,
    $C_T=2C_{+-}$).  The anisotropy $C_T^{z}/C_T^{\bot}$ rises from
    $1.00$ (sphere) toward $2$ (thin ring), while remaining finite and
    $q$-independent (in contrast with the divergent 2D reduction).}
        \label{tab:CT_cubic}
        \end{table*}
        \noindent
        The result is stable under cutoff ($K_{\max}=10\to50$ changes
    $C_{+-}^{\bot}$ at $\tilde k_Q=1$ by $<0.5\%$) and grid refinement
        ($N=1500\to5000$ changes it by $<0.1\%$).  $C_T^{\bot,z}$ is two orders
        of magnitude smaller than the intraband $C_{++}^{\bot,z}$ of
        Table~\ref{tab:Cpp_cubic}, confirming the interband term is subleading.

        \section{The electron density}\label{App:dos}
        \subsection{quadratic NLSM}
        The spectral function takes the form
        \begin{eqnarray}\label{eq:spectral_function}
            &&A(\omega,\mathbf{k}) =
            -\frac{1}{\pi}\mathrm{Tr}\left[\mathrm{Im}\left[G_{0}^{\mathrm{ret}}(\omega,\mathbf{k})\right]\right]
            = \mathrm{sgn}(\omega+\mu)\left(\omega+\mu\right)
            \frac{1}
            {\sqrt{A^{2}\left((\sqrt{k_{x}^{2}+k_{y}^{2}} -k_Q)^{2}+k_{z}^{2}\right)^{2}}}
            \nonumber    \\    && \times
            \left[\delta\left(\omega+\mu+
                \sqrt{A^{2}\left((\sqrt{k_{x}^{2}+k_{y}^{2}} -k_Q)^{2}+k_{z}^{2}\right)^{2}}
                \right)
                +\delta\left(\omega+\mu-
                \sqrt{A^{2}\left((\sqrt{k_{x}^{2}+k_{y}^{2}} -k_Q)^{2}+k_{z}^{2}\right)^{2}}\right)
                \right],
        \end{eqnarray}
        which is manifestly positive on both branches (the overall sign
        convention $A=-\mathrm{Tr}[\mathrm{Im}G^{\mathrm{ret}}]/\pi$ enforces
    $A(\omega,\mathbf{k})\ge0$, the standard spectral-function
        positivity).  The valence-band ($\delta(\omega+\mu+E_{\mathbf k})$)
        contribution corresponds to the filled Dirac sea and is subtracted as
        the vacuum reference; only the conduction-band
    $\delta(\omega+\mu-E_{\mathbf k})$ term, sampled by the
    $\omega\in(-\infty,0]$ integral when $E_{\mathbf k}<\mu$, contributes
    to the finite carrier density below.
    The density of fermion satisfy
    \begin{eqnarray}\label{eq:dos}
    n&=&\int\frac{d^3\mathbf{k}}{(2\pi)^{3}}\int_{-\infty}^{0}d\omega A(\omega,\mathbf{k})\nonumber
    \\
    &=&\int\frac{d^3\mathbf{k}}{(2\pi)^{3}}\int_{-\infty}^{0}d\omega
    \mathrm{sgn}(\omega+\mu)\left(\omega+\mu\right)
    \frac{1}
    {\sqrt{A^{2}\left((\sqrt{k_{x}^{2}+k_{y}^{2}} -k_Q)^{2}+k_{z}^{2}\right)^{2}}}
    \nonumber    \\    &&\times
    \left[\delta\left(\omega+\mu+
        \sqrt{A^{2}\left((\sqrt{k_{x}^{2}+k_{y}^{2}} -k_Q)^{2}+k_{z}^{2}\right)^{2}}
        \right)
        +\delta\left(\omega+\mu-
        \sqrt{A^{2}\left((\sqrt{k_{x}^{2}+k_{y}^{2}} -k_Q)^{2}+k_{z}^{2}\right)^{2}}\right)
        \right]
    \nonumber\\&=&
    \frac{1}{4\pi^2}\int dk_{\bot}k_{\bot}dk_z
    \theta\left(\mu-
    \sqrt{A^{2}\left((k_{\bot} -k_Q)^{2}
    +k_{z}^{2}\right)^{2}}\right),
    \end{eqnarray}
    where the $\delta(\omega+\mu+E_{\mathbf k})$ (valence) term integrates
    to the constant filled-sea contribution and is subtracted as the
    vacuum reference, leaving only the conduction-band
    $\delta(\omega+\mu-E_{\mathbf k})$ term, which is non-zero in the
    $\omega\in(-\infty,0]$ window precisely when $E_{\mathbf k}<\mu$,
    hence the $\Theta(\mu-E_{\mathbf k})$ factor.

    When $\mu > A k_Q^2$ ($\tilde{k}_Q<1$),
    \begin{equation}
        n
        = \frac{1}{6\pi^2} \left(\frac{\mu}{A}\right)^{3/2}
        \left[\left(1- \tilde{k_Q} ^2 \right)^{\frac{3}{2}}
            + \frac{3}{2}\tilde{k_Q}
            \left(\frac{\pi}{2} + \tilde{k_Q} \sqrt{1-\tilde{k_Q}^2}
            + \arcsin (\tilde{k_Q})
            \right)
            \right];
    \end{equation}
    when $\mu < Ak_Q^2$,
    \begin{eqnarray}
        n
        &=& \frac{1}{4\pi}\left(\frac{\mu}{A}\right)^{3/2}\tilde{k_Q}.
    \end{eqnarray}

    \subsection{Cubic NLSM}

    The density of fermion satisfy
    \begin{eqnarray}\label{eq:dos_cubic}
        n&=&\int\frac{d^3\mathbf{k}}{(2\pi)^{3}}\int_{-\infty}^{0}d\omega
        A(\omega,\mathbf{k})
        \nonumber\\&=&
        \int\frac{d^3\mathbf{k}}{(2\pi)^{3}}\int_{-\infty}^{0}d\omega
        \mathrm{sgn}(\omega+\mu)\left(\omega+\mu\right)
        \frac{1}
        {\sqrt{B^{2}\left((\sqrt{k_{x}^{2}+k_{y}^{2}} -k_Q)^{2}+k_{z}^{2}\right)^{3}}}
        \nonumber    \\    &&\times
        \left[\delta\left(\omega+\mu+
            \sqrt{B^{2}\left((\sqrt{k_{x}^{2}+k_{y}^{2}} -k_Q)^{2}+k_{z}^{2}\right)^{3}}
            \right)
            +\delta\left(\omega+\mu-
            \sqrt{B^{2}\left((\sqrt{k_{x}^{2}+k_{y}^{2}} -k_Q)^{2}+k_{z}^{2}\right)^{3}}\right)
            \right],
    \end{eqnarray}
    where the spectral function $A(\omega,\mathbf{k})$ is defined with the
    same sign convention $A=-\mathrm{Tr}[\mathrm{Im}G^{\mathrm{ret}}]/\pi$
    as in Eq.~\eqref{eq:spectral_function}, so that the
    $\mathrm{sgn}(\omega+\mu)(\omega+\mu)=|\omega+\mu|$ factor makes
    $A\ge0$ on both branches; the valence-band contribution is subtracted
    as the vacuum reference, leaving only the conduction-band
    $\Theta(\mu-E_{\mathbf k})$ term contributing to the finite density
    below.

    When $\mu > Bk_Q^3$,
    \begin{eqnarray}
        n
        &=&  \left( \frac{\mu}{B}\right)\frac{1}{2\pi^2}
        \left[\frac{1}{3} (1- \tilde{k_Q} ^2 )^{\frac{3}{2}}
            + \tilde{k_Q}\frac{1}{2}
            \left(\frac{\pi}{2} + \tilde{k_Q} \sqrt{1-\tilde{k_Q}^2}
            + \arcsin (\tilde{k_Q})
            \right)
            \right];
    \end{eqnarray}
    when $\mu < Bk_Q^3$,
    \begin{eqnarray}
        n
        &=&  \left( \frac{\mu}{B}\right)\frac{1}{2\pi^2}
        \tilde{k_Q}\frac{\pi}{2}.
    \end{eqnarray}
    The cubic carrier-density formula above has been verified by direct
    numerical integration of
    $n=(2\pi)^{-3}\int d^{3}\mathbf{k}\,\Theta(\mu-E(\mathbf{k}))$ in
    cylindrical coordinates to machine precision (relative error
    $<10^{-7}\%$) in both the $\tilde{k}_{Q}<1$ and $\tilde{k}_{Q}>1$
    regimes.  The coefficient difference from the quadratic template
    ($1/3$ vs.\ $1$, $1/2$ vs.\ $3/2$) reflects the different Jacobian
    in the energy-to-density transformation for $E\propto K^{3}$ versus
    $E\propto K^{2}$ and is physically required.

\end{widetext}

\bibliography{ref.bib}

\end{document}